\documentclass[journal ]{new-aiaa}
\usepackage[utf8]{inputenc}
\usepackage{textcomp}

\usepackage{graphicx}
\usepackage{amsmath}
\usepackage[version=4]{mhchem}
\usepackage{siunitx}
\usepackage{longtable,tabularx}
\usepackage{subcaption}
\title{Component-Based Reduced-Order Modeling Framework for Rocket Combustion Dynamics in Multi-Injector Configurations}

\author{Brody Gatza \footnote{Graduate Research Assistant, Department of Aerospace Engineering, AIAA Student Member} and Cheng Huang\footnote{Assistant Professor, Department of Aerospace Engineering, AIAA Member}}
\affil{University of Kansas, Lawrence, KS, 66045}

\begin{document}

\maketitle

\begin{abstract}
Even with the most advanced computational capabilities, high-fidelity (e.g., large-eddy) simulations of large-scale rocket engines remain far out of reach. In the current work, we develop and establish a component-based reduced-order modeling (CBROM) framework to enable accurate and efficient parametric modeling of large-scale rocket engines by geometrically decomposing a single domain into a combination of several representative components, including injectors, combustor and nozzle. Individual component-based reduced-order models (ROMs) are trained for each component with fabricated system-level responses enforced through carefully formulated boundary conditions during the training, which only require high-fidelity simulations of a much smaller computational domain, thereby significantly reducing the costs of ROM training. The trained component-based ROMs are then coupled together to enable full-system simulations. Specifically, we pursue an advanced adaptive ROM formulation leveraging a model-form preserving least-squares with variable transformation (MP-LSVT) projection to construct the component-based ROMs. The CBROM framework is evaluated using a seven-injector model rocket combustor configuration that exhibits self-excited combustion dynamics with distinct characteristics that vary with flow condition and geometric variations. The framework is demonstrated to provide accurate parametric predictions of the changes in dynamic behaviors, expressed in the spectra from dynamic mode decomposition (DMD) analysis and features in the time-averaged and RMS fields of target state variables.
\end{abstract}

\section{Introduction}\label{sect1}
\lettrine{I}{n} recent years, the rapid advancement in computing technology has enabled high-fidelity computational fluid dynamics (CFD) to become an important tool in modeling and understanding combustion dynamics for both conventional rocket engines \cite{urbano_exploration_2016,harvazinski_coupling_2015,harvazinski_large_2020,harvazinski_modeling_2019,schmitt_large-eddy_2017,liu_numerical_2023}, and promising next-generation rotating detonation rocket engines (RDREs) \cite{schwer_numerical_2011,hargus_air_2018,paxson_computational_2022,rong_investigation_2023,van_beck_nox_2024,prakash_numerical_2021,prakash_three-dimensional_2024}. In particular for RDREs, CFD has proven to be a vital tool for understanding the formation and stability of intersecting shock structures \cite{prakash_three-dimensional_2024,prakash_numerical_2021,rong_investigation_2023}, estimating performance and losses \cite{schwer_numerical_2011}, optimizing nozzle design \cite{paxson_computational_2022}, and studying droplet breakup in multiphase systems \cite{prakash_three-dimensional_2024}. Though powerful, these simulations incur high computational costs to resolve the complex physical interactions between turbulence, combustion, and shocks. Though the fast development of high-performance computing (HPC) technology is making these simulations more accessible, especially the utilization of graphics processing units (GPUs) \cite{ghioldi_acceleration_2023,piscaglia_gpu_2023,oconnell_gpu_nodate}, they still remain out of reach for practical rocket engine design, which is a many-query process requiring iterative evaluations of computational models to account for various design parameters (e.g., flow conditions and geometric configurations). 

To address this gap, different reduced-model methods have been pursued and developed in the literature to reduce the computational cost of these high-fidelity simulations while preserving reasonable accuracy in combustion-dynamics modeling. One group of researchers seeks to develop \emph{reduced-fidelity models (RFMs)} through simplification of physical models, common approaches of which include: (1) coarsening grid resolution \cite{zahtila_bi-fidelity_2025}, (2) using simplified chemical kinetic mechanisms \cite{li_comparative_2019,ranjan_effects_2016}, (3) using less complex turbulence models with fewer terms \cite{balabel_assessment_2011,ostlund_assessment_1999}, and (4) using fewer spatial dimensions (e.g. reducing 3D to 2D) \cite{koch_modeling_2020,frezzotti_quasi-1d_2018,harvazinski_analysis_2013,zhao_combustion_2021}. More advanced RFMs have also been developed to pursue the study of RDREs, including: (1) time-dependent boundary conditions that vary the boundary type as a simulation advances \cite{keskinoz_reduced-fidelity_2025}, (2) simplifying or excluding injector geometry \cite{schwer_numerical_2011,chu_numerical_2023}, (3) using a collection of pulse detonation tubes to represent an RDRE \cite{paxson_simple_2021}, and (4) manually specifying detonation wave parameters rather than letting the solver calculate them \cite{keskinoz_reduced-fidelity_2025}. While these methods yield computational savings, they often result in a decrease in simulation fidelity, which for propulsion simulations that are highly sensitive to the modeled dynamics, often results in data with limited applicability to the physical systems being modeled. Careful consideration must also be given to the use of RFMs to ensure that the quantities of interest are not sensitive to the simplifications.

Another class of methods pursues the use of data-driven methods to build more efficient reduced-order models (ROMs) using a small number of high-fidelity simulation data instead of relying on simplifications of the physics. Such methods can be classified as either machine learning (ML) or model-order reduction (MOR). ML seeks to build ROMs through pure data-driven black-box approaches that estimate associations between system inputs, outputs, and parameters using a limited amount of training data \cite{cherkassky_learning_2007}. Prominent examples of using ML to accelerate high-fidelity CFD include: (1) using super-resolution to improve data quality in a post-processing step to resolve fine features when using a low-fidelity mesh \cite{fukami_super-resolution_2023,hu_super-resolution-assisted_2024}, (2) enhancing turbulence modeling by predicting turbulence characteristics, calibrating constraints, and introducing new modeling terms \cite{naseem_machine_2025,zhang_machine_2015,wu_physics-informed_2018}, (3) augmenting the modeling of chemical kinetics by predicting chemical reaction rate constants and evolving chemical source terms in time \cite{li_machine_2024,berkemeier_accelerating_2023,sharma_deep_2020}, (4) training ROMs for predicting full-state information \cite{mcquarrie_data-driven_2021,hesthaven_non-intrusive_2018,guo_bayesian_2022}, (5) learning moving coordinate frames that track detonation waves \cite{mendible_data-driven_2021}, and (6) projecting ROMs onto nonlinear manifolds \cite{lee_model_2020}. Though promising, the current ML methods present fundamental limitations: (1) limited inherent physical consistency - due to their black-box nature, ML models do not always satisfy physical laws (e.g., conservation of mass). While recent success of physics-informed neural networks (PINNs) has been demonstrated in embedding physical laws into ML \cite{raissi_physics-informed_2019,zhang_crk-pinn_2024}, its applicability to complex problems such as rocket combustion remains an open question; (2) lack of generalizability \cite{zhou_machine_2022,duraisamy_perspectives_2021,ihme_combustion_2022} - the ML-based ROMs are often specially trained for a narrow range of operating conditions and present limited predictive capability beyond the training regions; and (3) high demand on training datasets \cite{caron_machine_2025,zhou_machine_2022} - training accurate ML-based ROMs requires a substantial amount of high-fidelity simulation data, incurring significant computational cost, extensive data storage, and considerable memory requirements for training.

An alternative class of data-driven methods to build ROMs is based on MOR, which seeks to derive ROMs to represent the evolution of dynamical systems based on high-dimensional nonlinear partial differential equations (PDEs), or often generally denoted as full order models (FOMs), by projecting the FOMs onto low-dimensional manifolds. The resulting projection-based ROMs inherit the majority of modeling fidelity from the FOMs while significantly reducing the computational cost. The core factor that distinguishes MOR from ML is the direct integration of PDEs (i.e., physics) into the ROM, ensuring that the MOR-ROM respects the model physics and produces physically consistent results, the importance of which has been highlighted in several studies \cite{peherstorfer_data-driven_2016,mcquarrie_data-driven_2021}. Projection-based MOR has shown success in developing ROMs for numerous aerospace applications, including hypersonic aerodynamics \cite{blonigan_model_2020}, channel flow with sheared convection layers \cite{lumley_low-dimensional_1997}, turbulent jets \cite{schmidt_guide_2020}, turbine blade flow \cite{schmidt_guide_2020}, and reacting flows \cite{buffoni_projection-based_2010,huang_model_2022,huang_analysis_2016}. Typically, the construction of projection-based ROMs consists of two stages: 1) an offline training stage that performs high-fidelity simulations of the target system at selected parameters to collect FOM data, which is used to generate low-dimensional manifolds; and 2) an online execution stage that projects the FOM onto the low-dimensional manifolds to construct and execute the ROM. Though successfully demonstrated for many complex problems, the direct application of projection-based ROMs to large-scale engineering systems, such as full-scale rocket engines, remains infeasible simply because the high-fidelity simulations of these systems are inaccessible, and therefore, FOM data is unavailable to train the ROMs. For example, a course-mesh (‘low-fidelity’) large-eddy simulation (LES) of a small-scale rocket engine (dozens of injectors) requires more than O($10^7$) CPU-hours while a fine-mesh (high fidelity) simulation would require more than O($10^9$) CPU-hours \cite{urbano_exploration_2016}. By simply scaling up the domain to a full-scale engine (with hundreds of injectors), at least one order of magnitude additional computational cost is required, leading to more than O($10^{10}$) CPU hours, which is far out of reach for modern high-performance computing capabilities.

To address this specific gap, researchers have formulated component-based, or domain decomposition methods \cite{barth_integration_2008,li_nonoverlapping_2013,choi_defining_2025}, which decompose a large-scale system into representative components with identical geometries, trains ROMs within each component, and couples the resulting component-based ROMs as a network to model the full system. In practice, domain decomposition methods have been commonly used in applications across engineering disciplines, from finite element analysis (FEA) \cite{white_reduced_2023,magoules_basics_2007}, to electromagnetism \cite{dolean_domain_2008,lu_compression_2024}, and CFD \cite{barth_integration_2008,li_nonoverlapping_2013,huang_component-based_2022,gropp_domain_1992}. For consistency, we refer to this type of approach as the component-based reduced-order modeling (CBROM) method in the current paper. CBROM methods leverage the fact that many large-scale engineering systems can be decomposed into components of identical geometric features and therefore enable the offline training of the ROMs to be performed within each individual component for multiple parameters rather than on the entire system. Such a component-based training strategy significantly reduces the cost of the offline ROM training, which makes ROM applications for large-scale systems feasible. The trained CBROMs can then be repetitively applied to model the identical components and coupled together to model different system configurations. Early success of CBROM methods has been mostly demonstrated for systems governed by linear PDEs, which include constructing a network of ROMs to model the unsteady aerodynamics of a blade row in a compressor \cite{willcox_application_2002}, to build digital twins of a 12-ft wingspan unmanned aerial vehicle (UAV) \cite{kapteyn_data-driven_2022}, and to accelerate the design of lattice-type structures \cite{mcbane_component-wise_2021}. CBROM methods have also been demonstrated for systems governed by nonlinear PDEs with representative examples including pipe flow \cite{maday_reduced-basis_2002}, rocket combustors \cite{huang_parametric_2023}, and RDREs \cite{farcas_domain_2024}. In addition, some CBROM applications adopt a hybrid ROM/FOM strategy, which deploys component-based ROMs to model parts of the domain while applying a FOM to model the remainder. Examples of this hybrid strategy include compressible nozzle flow \cite{buffoni_iterative_2009}, three-dimensional flow across a cylinder \cite{baiges_domain_2013}, and oceanic current formation \cite{ahmed_multifidelity_2021}. Recently, our group has successfully demonstrated CBROM methods on modeling either the injector-element dynamics \cite{huang_parametric_2023,huang_component-based_2022} or the downstream chamber/nozzle \cite{huang_investigations_2024} in multi-injector rocket combustors. 

Therefore, in the current work, we aim to establish and demonstrate a component-based reduced-order modeling (CBROM) framework to apply to full-scale rocket engines with the goal of enabling (1) efficient and accurate simulations of combustion dynamics, and (2) parametric predictive capability in multi-injector rocket combustors. Specifically, we decompose the combustor domain into multiple components and categorize these components into three types: (1) wall-adjacent injectors, (2) interior injectors, and (3) downstream nozzle. Then, individual component-based ROMs are trained for each component class. To mitigate the ROM training cost, the two injector ROMs are trained using reduced-geometry simulations that were crafted to emulate the injector dynamics anticipated in the full systems. The full-domain simulation is used to train the downstream nozzle ROM with the upstream injectors modeled using the injector component-based ROMs rather than the FOM to reduce the offline training cost. Specifically, we adopt an adaptive MOR formulation \cite{huang_parametric_2023} to construct predictive component-based ROMs. The resulting three component-based ROMs can then be flexibly coupled to model the full system and inherently enable parametric simulations spanning different operating conditions and geometric configurations. The remainder of the paper is outlined as follows. Section \ref{sect2} presents the FOM formulation and time discretization. Section \ref{sect3} reviews the model reduction techniques. Section \ref{sect4} presents the CBROM framework and its implementation for a multi-injector rocket combustor. Section \ref{sect5} presents the numerical results of the CBROM framework based on a seven-injector rocket combustor with a detailed assessment. Section \ref{sect6} provides concluding remarks.

\section{Full-Order Model}\label{sect2}
We first describe the full-order model (FOM) used in the numerical studies in the current work as a general dynamical system for a physical domain $\Omega$ with a boundary $\partial\Omega$,

\begin{equation}
\label{eq1}
\begin{array}{c}
\dfrac{\mathrm{d}\mathbf{q}(\mathbf{q}_p)}{\mathrm{dt}} =\mathbf{f}(\mathbf{q}_p,t) \quad \mathrm{in} \quad \Omega \\
\mathrm{with }\quad \mathbf{u}(\mathbf{q}_p) = \mathbf{u}_{BC} \quad \mathrm{on} \quad \partial\Omega \quad \mathrm{and}
\quad \mathbf{q}_p(t=0)=\mathbf{q}_p^0,
\end{array}
\end{equation}

\noindent where $\mathbf{q}_p \in \mathbb{R}^N$ is the state or solution vector with dimension ${N=N_{c} \times N_{v}}$ being the degrees of freedom (DOF) of the problem. ${N_{c}}$ is the number of elements or cells in the domain, and $N_{v}$ is the number of state variables, corresponding to the number of state equations being solved. $\mathbf{q}$, $\mathbf{f}$, and $\mathbf{u}$ are nonlinear functions of $\mathbf{q}_p$. The function $\mathbf{q}:\mathbb{R}^{N} \rightarrow \mathbb{R}^{N}$ represents the transformation to the conservative state variables. $\mathbf{f}: \mathbb{R}^N \rightarrow \mathbb{R}^N$ produces the source terms, surface fluxes, and body forces that result from transforming a geometry into a discrete domain. $\mathbf{u}: \mathbb{R}^N \rightarrow \mathbb{R}^{(N_{BC} )}$ represents the boundary conditions on $\partial\Omega$, and $\mathbf{u}_{BC} \in \mathbb{R}^{(N_{BC} )}$ denotes the states to be satisfied at the boundary conditions, where $N_{BC}$ is the number of cells located along the domain boundary. The initial condition at time $(t=0)$ is denoted by $\mathbf{q}_p^0 \in \mathbb{R}^N$. For all the numerical studies performed in this paper, we introduce a second-order implicit scheme for time discretization
\begin{equation}
\label{eq2}
\begin{array}{c}
{\mathbf{r}(\mathbf{q}_p^n) \triangleq \frac{3}{2} \mathbf{q}(\mathbf{q}_p^n) - 2 \mathbf{q}(\mathbf{q}_p^{n-1})+\frac{1}{2}\mathbf{q}(\mathbf{q}_p^{n-2})-\Delta \mathrm{t}\mathbf{f}(\mathbf{q}_p^n,t^n)=0} \\
{\mathrm{with} \quad \mathbf{u}(\mathbf{q}_p^n)=\mathbf{u}_{BC}^n\quad \mathrm{on} \quad \partial\Omega},
\end{array}
\end{equation}

\noindent where $\mathbf{r}: \mathbb{R}^N\rightarrow\ \mathbb{R}^N$ is defined as the FOM equation residual, $\Delta{t} \in \mathbb{R}^+$ is the physical time step, and the superscript $n$ denotes the iteration number. The states $\mathbf{q}_p^n$ are solved iteratively at each time step such that $\mathbf{r}\left(\mathbf{q}_p^n\right)=0$.

\section{Reduced-Order Model}\label{sect3}
Due to the nonlinear and non-stationary nature of combustion dynamics in rocket engines, conventional ROM formulations often fail to provide accurate predictions of combustion simulations due to the slow decay of the Kolmogorov N-width \cite{greif_decay_2019}, that results from the complex dynamics. To address this gap, we leverage an adaptive ROM formulation using the methodology proposed by Huang and Duraisamy \cite{huang_predictive_2023} to construct the component-based ROMs in the CBROM framework. This formulation seeks to update and tailor the low-dimensional subspace to the evolving physics of a problem, enabling accurate prediction of nonlinear and non-stationary dynamics. It has proven to be effective for modeling complex combustion dynamics in rocket engines \cite{huang_parametric_2023,huang_component-based_2022} and RDREs \cite{camacho_investigations_2025}. 

\subsection{Construction of the Low-Dimensional Subspace for Solution Variables}
In the current work, we construct the low-dimensional subspace $\mathbf{V}_p $, to approximate the full states, $\mathbf{q}_p$, using the proper orthogonal decomposition (POD) \cite{lumley_low-dimensional_1997,berkooz_proper_1993}. The basis is used to approximate the state vector $\tilde{\mathbf{q}}_p$ through

\begin{equation}
\label{eq3}
{\tilde{\mathbf{q}}_p = \mathbf{q}_{p,\mathrm{ref}}} + \mathbf{H}^{-1}\mathbf{V}_p\mathbf{q}_r,
\end{equation}

\noindent where $\mathbf{q}_r \in \mathbb{R}^{n_p}$ is the reduced state and $\mathbf{V}_p \in \mathbb{R}^{N \times n_p}$ is the low-dimensional subspace (or trial basis) with $n_p$ representing the number of POD modes. $\mathbf{q}_{p,\mathrm{ref}} \in \mathbb{R}^{N}$ represents a reference state incorporated to prevent bias in basis computation where the time-averaged FOM solution is used as the reference state - i.e., $\mathbf{q}_{p,\mathrm{ref}} = \frac{1}{t_1-t_0}\int_{t_0}^{t_1}\mathbf{q}_p(t) \mathrm{d}t$. In addition, a scaling matrix $\mathbf{H} \in \mathbb{R}^{N \times N}$ is introduced to Eq.~\ref{eq3} to ensure that the variables corresponding to different physical quantities have similar orders of magnitude in the basis computation. Specifically, we choose to use the $L^2$-norm of the full states for the scaling matrix following the work of Lumley and Poje \cite{lumley_low-dimensional_1997}
\begin{equation}
\label{eq:H}
\mathbf{H} = diag(\mathbf{H}_1,...,\mathbf{H}_i,...,\mathbf{H}_{N_c}),
\end{equation}
where $\mathbf{H}_i = diag(\phi_{1,norm}^{-1},...,\phi_{v,norm}^{-1},...\phi_{N_v,norm}^{-1})$ and $v$ represents the $v^\mathrm{th}$ state variable. The $L^2$-norm is computed using
\begin{equation}
\label{eq:L2norm}
\phi_{v,norm} = \frac{1}{t_1-t_0}\int_{t_0}^{t_1}\frac{1}{\Omega}\int_{\Omega}\phi_v'^2(\mathbf{x},t)\mathrm{d}\mathbf{x}\mathrm{d}t
\end{equation}

\subsection{Least-Squares with Variable Transformation}
To guarantee the construction of a stable ROM, model-form preserving least-squares with variable transformation (MP-LSVT) is used \cite{huang_model_2022}. Using the trial ROM basis, MP-LSVT is leveraged to minimize the residual of the discretized FOM with respect to the reduced state. The minimization problem takes the form

\begin{equation}
\label{eq4}
\begin{array}{c}
\mathbf{q}_r^n \triangleq \underset{\mathbf{q}_r \in \mathbb{R}^{n_p}}{\mathrm{arg\  min}} \|\mathbf{Pr}(\tilde{\mathbf{q}}_p)\|_2^2, \\
\mathrm{with} \quad \mathbf{u}(\tilde{\mathbf{q}}_p^n) \quad \mathrm{on} \quad \partial\Omega \quad \mathrm{and}\quad \tilde{\mathbf{q}}_p^0=\mathbf{V}_p(\mathbf{V}_p)^\mathrm{T}\mathbf{q}_p^0 \quad \mathrm{or} \quad \tilde{\mathbf{q}}_p^0=\mathbf{q}_p^0,
\end{array}
\end{equation}

\noindent where $\mathbf{P} \in \mathbb{R}^{N \times N}$ is a scaling matrix of the same form as $\mathbf{H}$, but with the $L^2$-norm being computed based on the conservative state values $\mathbf{q}(\mathbf{q}_p)$. The approximated solution values and the initial condition can be represented by the projected FOM states, $\mathbf{V}_p(\mathbf{V}_p)^T\mathbf{q}_p$, or the FOM state, $\mathbf{q}_p$, itself. The boundary conditions of the MP-LSVT formulation are identical to those of the FOM, ensuring that they are satisfied by the ROM. Using the Petrov-Galerkin projection (LSPG) \cite{carlberg_galerkin_2017}, the system is projected to a lower dimension following 
\begin{equation}
\label{eq5}
\begin{array}{c}
(\mathbf{W}_p^n)^T\mathbf{Pr}(\tilde{\mathbf{q}}_p^n) = 0 \\
\mathrm{with} \quad \mathbf{u}(\tilde{\mathbf{q}}_p^n) = \mathbf{u}_{BC}^n \quad \mathrm{on} \quad \partial\Omega \quad \mathrm{and} \quad \tilde{\mathbf{q}}_p^0=\mathbf{V}_p(\mathbf{V}_p)^T\mathbf{q}_p^0 \quad \mathrm{or} \quad \tilde{\mathbf{q}}_p^0=\mathbf{q}_p^0,
\end{array}
\end{equation}
\noindent where the test basis $\mathbf{W}_p$ takes the form
\begin{equation}
\label{eq6}
\begin{array}{c}
\mathbf{W}_p^n=\frac{\partial\mathbf{Pr}(\tilde{\mathbf{q}}_p^n)}{\partial\mathbf{q}_r^n} = \mathbf{P}(\frac{3}{2}\tilde{\boldsymbol{\Gamma}}^n-\Delta t\tilde{\mathbf{J}}^n\tilde{\boldsymbol{\Gamma}}^n)\mathbf{H}^{-1}\mathbf{V}_p, \\
\mathrm{with} \quad \tilde{\boldsymbol{\Gamma}}^n=[\frac{\partial\mathbf{q}}{\partial\mathbf{q}_p}]^n_{q_p=\tilde{q}_p} \quad \mathrm{and} \quad \tilde{\mathbf{J}}^n=[\frac{\partial\mathbf{f}}{\partial\mathbf{q}}]^n_{q_p=\tilde{q}_p}. 
\end{array}
\end{equation}
\subsection{Hyper-Reduction}
While MP-LSVT reduces the dimension of the problem, evaluation of the non-linear residual remains a computational bottleneck. To reduce the evaluations, of the nonlinear residual, the discrete empirical interpolation method (DEIM) \cite{chaturantabut_nonlinear_2010} is used to approximate the full-field residual based on sparsely sampled evaluations using
\begin{equation}
\label{eq7}
\bar{\mathbf{r}}\approx\mathbf{U}(\mathbf{S}^T\mathbf{U})^+\mathbf{S}^T\mathbf{r},
\end{equation}
\noindent where $\bar{\mathbf{r}}$ is the approximated residual and $\mathbf{U} \in \mathbb{R}^{N \times n_d}$ is a basis used to approximate the residual that can be constructed using POD, similar to the trial basis $\mathbf{V}_p$  where $n_d$ is the number of temporal POD modes. However, it has been found that using the trial basis in place of the residual basis produces accurate approximations of the residual while reducing computational overhead \cite{huang_model_2022}. $\mathbf{S} \in \mathbb{S}^{N \times n_s}$ is a selection operator that contains the sampling points with $n_s$ denoting the number of sampling points. Applying Eq. \ref{eq7} to approximate the residual in Eq. \eqref{eq4} yields
\begin{equation}
\label{eq8}
\tilde{\mathbf{q}}_p^n \triangleq \underset{\tilde{\mathbf{q}}_p^n \in \mathrm{Range}(\mathbf{V}_p)}{\mathrm{arg \ min}} \| \mathbf{U}(\mathbf{S}^T\mathbf{U})^+\mathbf{S}^T\mathbf{Pr}(\tilde{\mathbf{q}}_p^n)\|_2^2,
\end{equation}
\noindent and results in the test basis
\begin{equation}
\label{eq9}
\overline{\mathbf{W}}_p^n = \frac{\partial\mathbf{U}(\mathbf{S}^T\mathbf{U})^+\mathbf{S}^T\mathbf{Pr}(\tilde{\mathbf{q}}_p^n)}{\partial\mathbf{q}_r^n} = \mathbf{U}(\mathbf{S}^T\mathbf{U})^+\mathbf{S}^T\frac{\partial\mathbf{Pr}(\tilde{\mathbf{q}}_p^n)}{\partial\mathbf{q}_r^n} = \mathbf{U}(\mathbf{S}^T\mathbf{U})^+\mathbf{S}^T\mathbf{W}_p^n.
\end{equation}
This reduces the evaluation of the test basis to $n_s$ rows, reducing the number of computations and results in the hyper-reduced MP-LSVT ROM formulation
\begin{equation}
\label{eq10}
(\mathbf{S}^T\mathbf{W}_p^n)^T[(\mathbf{S}^T\mathbf{U})^+]^T(\mathbf{S}^T\mathbf{U})^+\mathbf{S}^T\mathbf{Pr}(\tilde{\mathbf{q}}_p^n)=\mathbf{0}.
\end{equation}
The term $[(\mathbf{S}^T\mathbf{U})^+]^T(\mathbf{S}^T\mathbf{U})^+ \in \mathbb{R}^{n_s \times n_s}$ may be precomputed; however, it may be memory-prohibitive, and as such, only $(\mathbf{S}^T\mathbf{U})^+ \in \mathbb{R}^{n_d \times n_s}$ is precomputed.

\subsection{Adaptive ROM Formulation}
To overcome the Kolmogorov barrier and enhance the ROM’s predictive capabilities, the adaptive ROM framework proposed by Huang and Duraisamy \cite{huang_predictive_2023} is used. The adaptive MOR technique builds on the computational savings of MP-LSVT and hyper-reduction by updating the trial basis and sampling points during the online simulation. Adaptation is enabled by the minimization problem in Eq. \ref{eq11}, which feeds new information to the ROM, facilitating accurate prediction of combustion dynamics.
\begin{equation}
\label{eq11}
\{\mathbf{q}_r^n,\mathbf{V}_p^n,\mathbf{S}^n\} \triangleq \underset{\mathbf{q}_r \in \mathbb{R}^{n_p},\ \mathbf{V}_p^n \in \mathbb{R}^{N \times n_p},\ \mathbf{S}^n \in \mathbb{S}^{N \times {n_s}}}{\mathrm{arg \ min}}\|\mathbf{V}_p^n[(\mathbf{S}^n)^T\mathbf{V}_p^n]^+(\mathbf{S}^n)^T\mathbf{Pr}(\tilde{\mathbf{q}}_p^n)\|_2^2
\end{equation}
Solving Eq. \ref{eq11} directly poses a large computational cost; rather, it is decoupled into two minimization problems to update the trial basis and the sampling points. First, $\mathbf{q}_r^n$ is selected as the solution to Eq. \eqref{eq5}, satisfying the minimization problem in Eq. \ref{eq4}. Second, the trial basis is updated through the minimization problem

\begin{equation}
\label{eq12}
\mathbf{V}_p^n \triangleq \underset{\mathbf{V}_p \in \mathbb{R}^{N \times n_p}}{\mathrm{arg \ min}} \|\mathbf{Pr}(\tilde{\mathbf{q}}_p^n)\|_2^2,
\end{equation}
which can be solved exactly via

\begin{equation}
\label{eq13}
\begin{array}{c}
\mathbf{V}_p^n = \mathbf{V}_p^{n-1} + \delta\mathbf{V}_p, \\
\mathrm{with} \quad \delta\mathbf{V}_p = \frac{(\hat{\mathbf{q}}_p^n-\tilde{\mathbf{q}}_p^n)(\mathbf{q}_r^n)^T}{\|\mathbf{q}_r^n\|_2^2},
\end{array}
\end{equation}

\noindent where $\hat{\mathbf{q}}_p^n \in \mathbb{R}^N$ represents the full-state information evaluated from the FOM equation residual using

\begin{equation}
\label{eq14}
\mathbf{r}(\hat{\mathbf{q}}_p^n) \triangleq \frac{3}{2}\mathbf{q}(\hat{\mathbf{q}}_p^n) - 2\mathbf{q}(\tilde{\mathbf{q}}_p^{n-1}) + \frac{1}{2}\mathbf{q}(\tilde{\mathbf{q}}_p^{n-2}) - \Delta t\mathbf{f}(\hat{\mathbf{q}}_p^n,t^n) = 0.
\end{equation}

This method is used to update the basis every $z_b$ time steps which is determined empirically. Third, the locations of the sampling points are updated every $z_s$ $(z_s > z_b)$ time steps by solving the minimization problem
\begin{equation}
\label{eq15}
\mathbf{S}^n \triangleq \underset{\mathbf{S}^* \in \mathbb{S}^{N \times n_s}}{\mathrm{arg \ min}} \|(\hat{\mathbf{q}}_p^n - \mathbf{q}_{p,\mathrm{ref}}) - \mathbf{V}_p^n[(\mathbf{S}^*)^T\mathbf{V}_p^n]^+(\mathbf{S}^*)^T\mathbf{H}(\hat{\mathbf{q}}_p^n - \mathbf{q}_{p,\mathrm{ref}})\|_2^2.
\end{equation}
Updating the basis and sampling points are computationally intensive steps, and the frequency of the updates must be carefully selected to balance computational efficiency and ROM accuracy. To reduce the update cost, the full basis is only updated when the sampling points are updated; otherwise, the basis is only updated at the sampled points using 
\begin{equation}
\label{eq16}
\mathbf{S}^{n-1}\mathbf{V}_p^n = \mathbf{S}^{n-1}\mathbf{V}_p^{n-1} + \mathbf{S}^{n-1}\delta\mathbf{V}_p,
\end{equation}
which only requires full-state information to be evaluated at the sampled points
\begin{equation}
\label{eq17}
\mathbf{S}^{n-1}\mathbf{r}(\hat{\mathbf{q}}_p^n) = \mathbf{S}^{n-1}[\frac{3}{2}\mathbf{q}(\hat{\mathbf{q}}_p^n) - 2\mathbf{q}(\tilde{\mathbf{q}}_p^{n-1}) + \frac{1}{2}\mathbf{q}(\tilde{\mathbf{q}}_p^{n-2}) - \Delta t\mathbf{f}(\hat{\mathbf{q}}_p^n,t^n)] = 0.
\end{equation}
This reduces the computational cost of the adaptive ROM formulation while maintaining its predictive capabilities. While this method incurs higher costs than a static basis ROM, the benefits it provides have made it the clear choice for this work. 

\section{Component-Based Reduced-Order Modeling Framework}\label{sect4}

\begin{figure}[hbt!]
    \centering
    \includegraphics[width=1.0\linewidth, trim=70 250 70 250, clip]{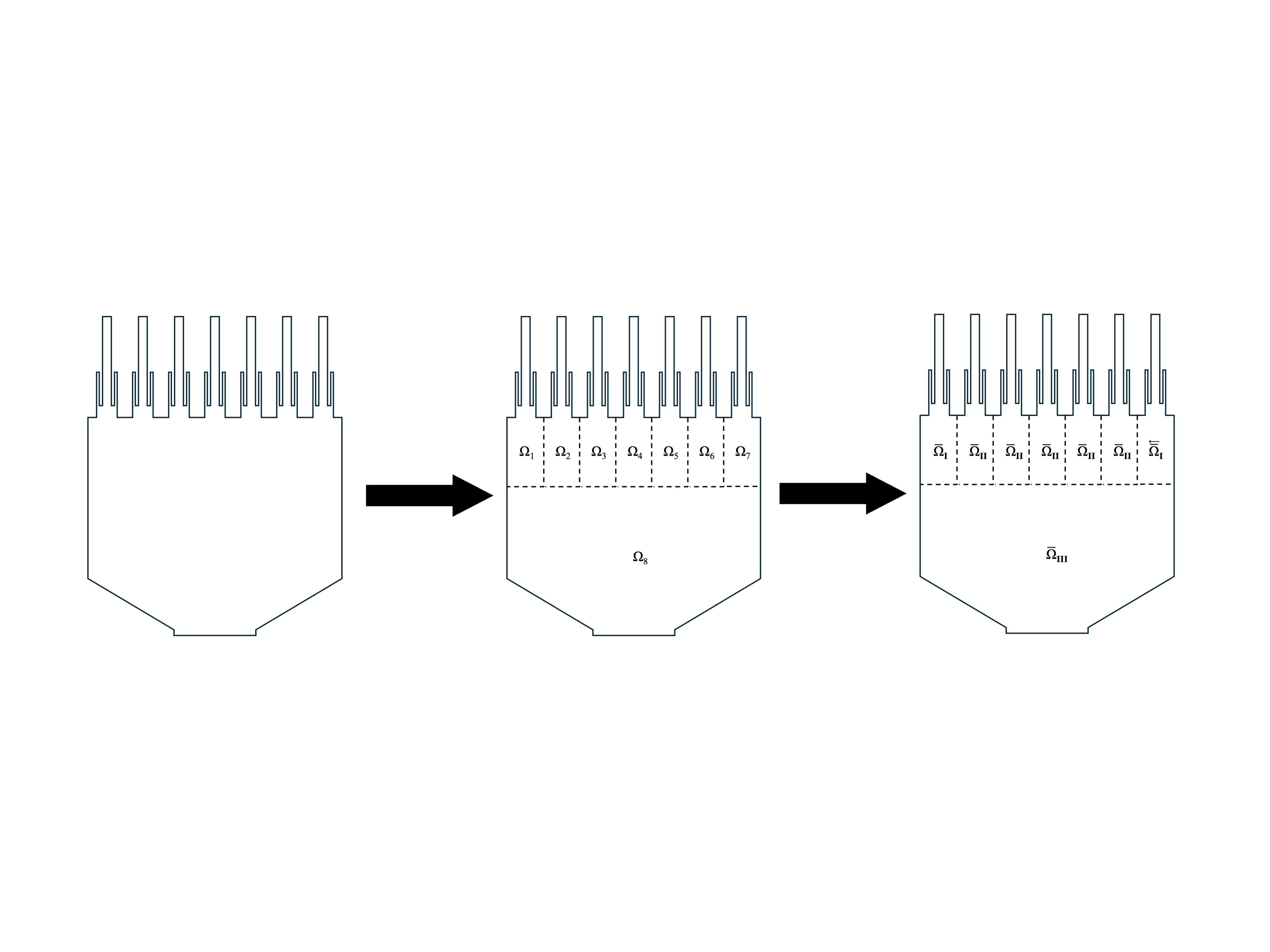}
    \caption{Component-based representation of a multi-injector rocket combustor}
    \label{fig1}
\end{figure}

In this section, we introduce the component-based reduced-order modeling (CBROM) framework for simulations of large-scale rocket engines. The framework leverages the domain decomposition method and seeks to reduce the computational and memory overhead of ROM construction by decomposing the domain of a full system into a collection of components with identical geometric features and constructing individual ROMs within each component. This CBROM framework is crucial for modeling large-scale problems, as constructing a single global ROM for the full system directly can be unattainable due to limitations in data storage and memory for essential operations such as basis generation. We use a multi-injector rocket combustor as shown in Fig. \ref{fig1} to illustrate the domain decomposition method. The combustor is comprised of seven identical injector elements that feed propellants into the combustion chamber, where they mix and react before accelerating down the convergent nozzle. It can be easily seen that the multi-injector rocket combustor can aptly be decomposed into eight components: one for each of the seven injectors $(\Omega_{1-7})$ and one containing the downstream combustor and nozzle $(\Omega_8)$. Moreover, the interior injector components $(\Omega_{2-6})$ share identical geometric features and the wall injector components $(\Omega_{1})$ and $(\Omega_{7})$ are geometrically symmetric. Therefore, the multi-injector rocket configuration in Fig. \ref{fig1} can be simplified as a combination of three reference components: $(\bar{\Omega}_\text{I})$ -- wall injector component, $(\bar{\Omega}_\text{II})$ -- interior injector component, and $(\bar{\Omega}_\text{III})$ -- downstream combustor and nozzle component (simply denoted as downstream component in the current paper for compactness). This component-based representation enables independent training of ROMs for each reference component respectively, and allows flexible integration of these ROMs for full-system simulations without accessing the expensive high-fidelity FOM of the full geometry. The mathematical formulation of the domain decomposition method is presented as follows 
\begin{equation}
\label{eq18}
\begin{array}{c}
\mathbf{B}_k\mathbf{r}(\bar{\mathbf{q}}_{p,k}^n) = \mathbf{0}, \\
\mathrm{with} \quad \mathbf{u}_k(\bar{\mathbf{q}}_{p,k}^n) = \mathbf{u}_{BC,k}^n \quad \mathrm{on} \quad \partial\Omega_k, \\
\mathrm{and} \quad \mathbf{v}_{km}(\bar{\mathbf{q}}_{p,k}^n) = \mathbf{v}_{km}(\bar{\mathbf{q}}_{p,m}^n) \quad \mathrm{on} \quad \partial\Omega_{km},
\end{array}
\end{equation}
\noindent where $\mathbf{B}_k \in \mathbb{R}^{n_{B,k} \times N_k}$ is a matrix that selects whether a component adopts FOM or ROM and $k$ is the component number. The DOF of component $k$ is represented by $N_k = N_{c,k} \times N_{v}$, with $N_{c,k}$ representing the number of cells in the subdomain $\Omega_k$, and $\bar{\mathbf{q}}_{p,k}^n \in \mathbb{R}^{N_k}$ the state vector for the component. The nonlinear function $\mathbf{u}_k : \mathbb{R}^N \rightarrow \mathbb{R}^{N_{BC,k}}$ represents the physical boundary conditions prescribed to be satisfied as $\mathbf{u}_{BC,k}^n$ on the boundary $\partial\Omega_k$, with $N_{BC,k} = N_{c,BC,k} \times N_{v}$ and $N_{c,BC,k}$ as the number of cells adjacent to the boundary. In addition, the nonlinear function $\mathbf{v}_{km} : \mathbb{R}^N \rightarrow \mathbb{R}^{N_{km}}$ represents the conditions at the interface $\partial\Omega_{km}$ of two adjacent components, $k$ and $m$, where $N_{km} = N_{c,km} \times N_{v}$ and $N_{c,km}$ the number of cells next to an interface boundary in an individual component. With this general domain decomposition formulation, individual CBROMs can then be trained for each representative component, which are then coupled together to enable full-system modeling.

\subsection{Component-Based ROM Training}\label{sect4A}
Based on the component-based representation introduced in Fig. \ref{fig1}, we introduce a component-based ROM training strategy in this section, which eliminates the need for high-fidelity FOM data of the full system. Instead, the strategy seeks to perform high-fidelity FOM simulations on carefully designed reduced-geometry configurations that produce dynamics mimicking the full system with a small fraction of the computational cost of simulating the full-system. Individual component-based ROMs are then developed for each reference component identified in Fig. \ref{fig1}. To achieve this, we consider two component-based ROM training strategies to account for the variations in dynamics between components: (1) reduced-geometry-based training for injector components ($\bar{\Omega}_\text{I}$ and $\bar{\Omega}_\text{II}$), and (2) full-geometry-based training for the downstream component $(\bar{\Omega}_\text{III})$.

\subsubsection{Reduced-geometry-based training strategy for injector-component ROMs}\label{sect4A1}
\begin{figure}[hbt!]
    \centering
    \includegraphics[width=0.65\linewidth, trim=250 160 200 50, clip]{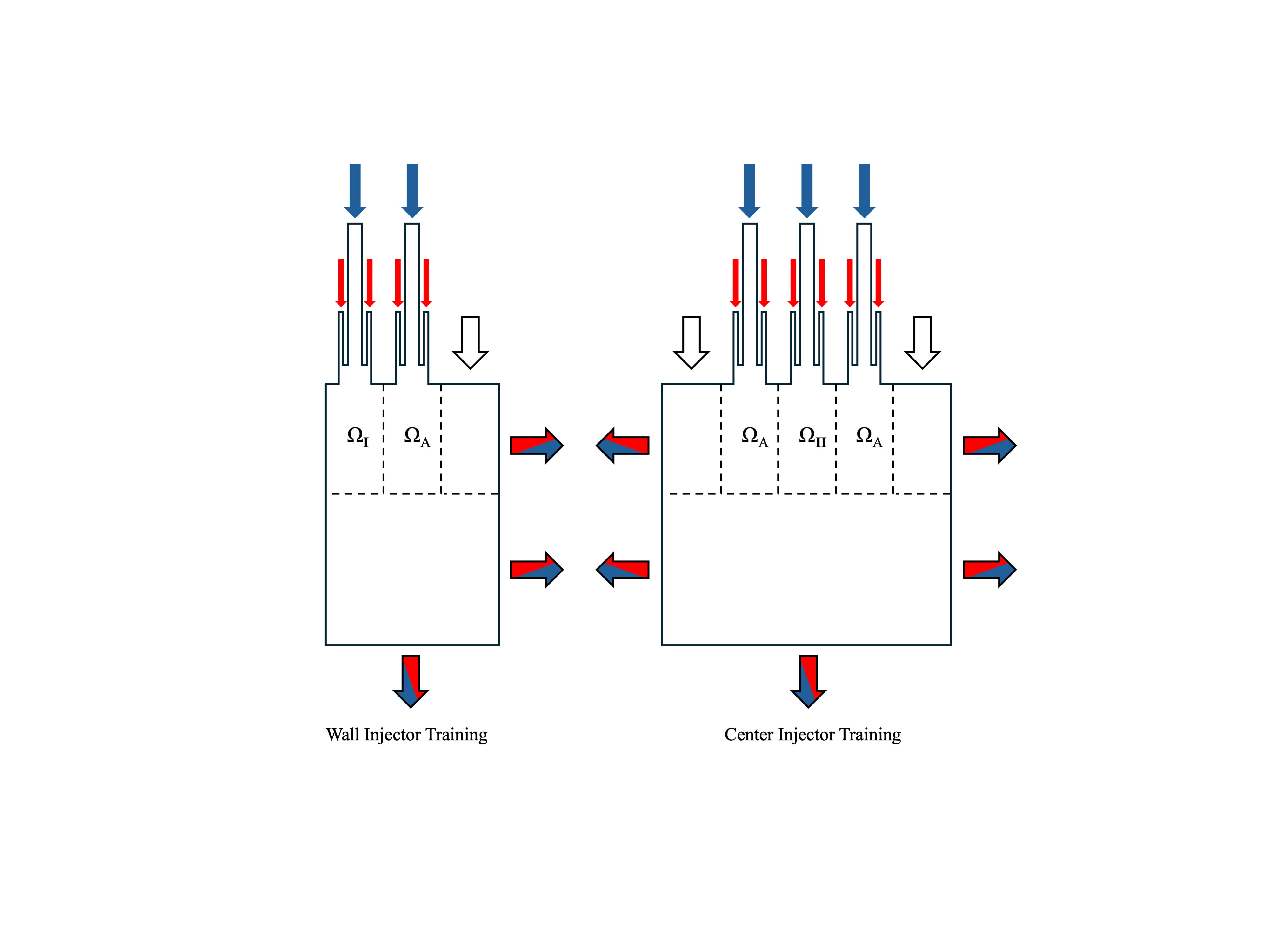}
    \caption{Reduced-geometry-based training strategy for the injector-component ROM}
    \label{fig2}
\end{figure}

The training configurations for the two injector-component ROMs are shown in Fig. \ref{fig2}. Each training configuration consists of the reference injector components (i.e., $\Omega_\text{I}$ for the wall-injector component and $\Omega_\text{II}$ for the interior-injector component) and auxiliary injector components (one for the wall-injector component and two for the interior-injector component). The auxiliary injector components incorporate the strong interactions between injectors, which are expected in the full system \emph{otherwise} it is difficult to account for such interactions by imposing boundary conditions due to the complexity of the dynamics at the injector interfaces. In addition, following the work by Huang et al. \cite{huang_component-based_2022}, buffer regions are included surrounding the injector components with exponentially stretched meshes, which prove to be necessary to represent the undamped large-scale motions (e.g., vortex shedding) anticipated in the full system. Including these buffer regions also ensures that the small-scale motions (e.g., chemical reaction) are damped and do not interfere with the non-reflective boundary conditions set at the downstream and the side ends. The inlet conditions for the injectors (indicated by the red and blue arrows) are identical to the full system (details in Sec. \ref{sect5}), while the inlet conditions (indicated by the hollow arrows) for the buffer regions are set up to mimic the mean flow of the injector components. Non-reflective boundary conditions are specified at the downstream end and the side ends (indicated by the mixed-color arrows) in both training configurations to maintain a nominal pressure consistent with the full system and to ensure that the imposed boundary conditions do not imprint the dimensions of the individual component upon the resulting combustion dynamics for ROM training. High-fidelity simulations on these two reduced-geometry training configurations are initialized and performed until the combustion dynamics reach stationary states. Then, from the snapshots, the subset of data corresponding to the two reference injector components ($\bar{\Omega}_\text{I}$ and $\bar{\Omega}_\text{II}$) are extracted to construct the two injector-component ROMs.

\subsubsection{Full-geometry-based training strategy for the downstream-component ROM}\label{sect4A2}
\begin{figure}[hbt!]
    \centering
    \includegraphics[width=0.45\linewidth, trim=300 130 300 130, clip]{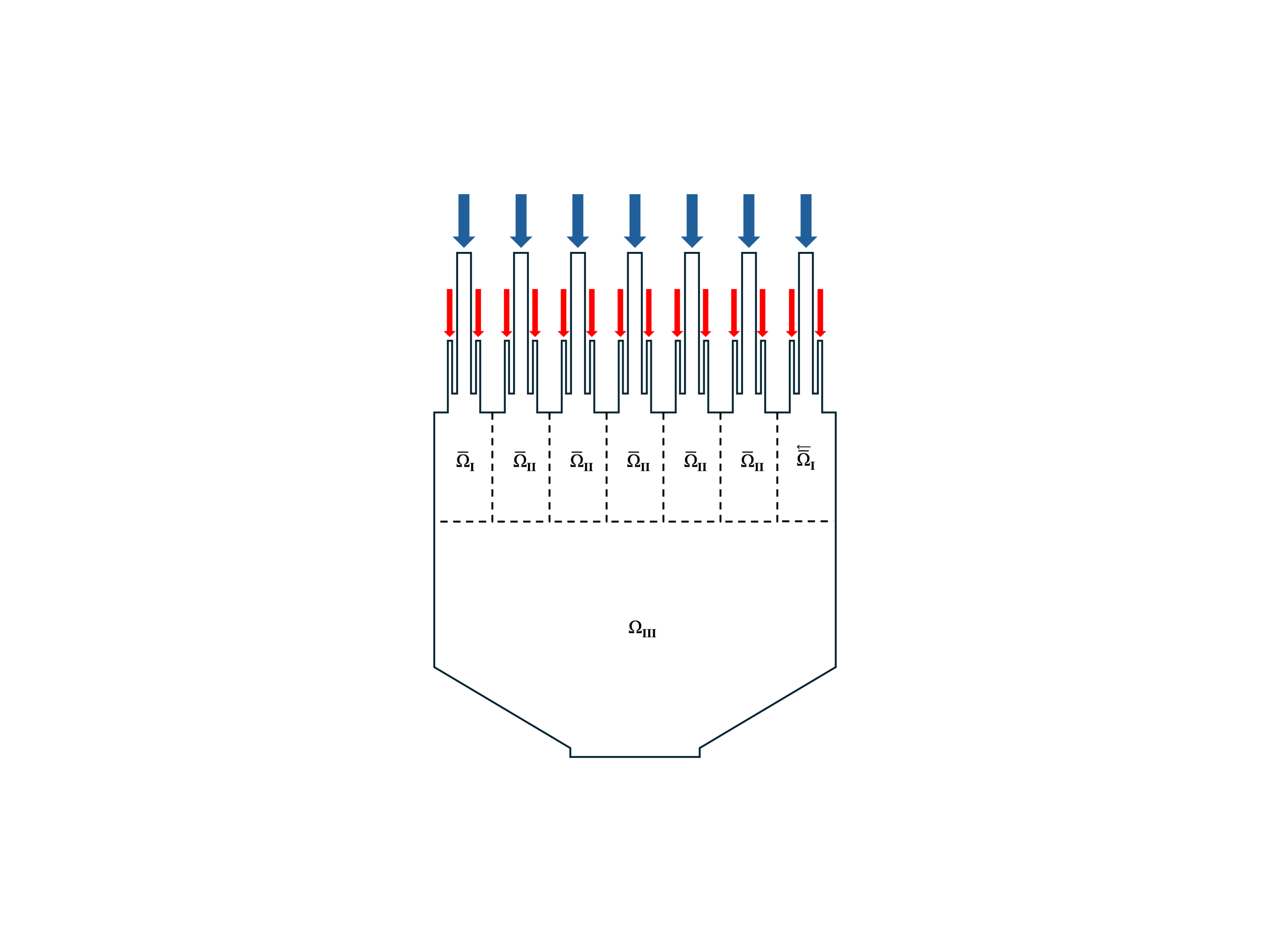}
    \caption{Full-geometry-based training strategy for the downstream-component ROM}
    \label{fig3}
\end{figure}

Due to the complexity of the dynamics arising from the upstream injectors, accurate replication of the combustion dynamics in the downstream component requires detailed modeling of the upstream dynamics, which cannot be achieved using the reduced-geometry training strategy as illustrated by Huang et al. \cite{huang_investigations_2024}. Instead, we adopt a full-geometry-based training strategy for the downstream-component ROM as shown in Fig. \ref{fig3} with the upstream injectors modeled using the injector-component ROMs obtained in Sec. \ref{sect4A1} and the downstream combustor and nozzle modeled using the FOM. Specifically, the wall-injector ROM $(\bar{\Omega}_\text{I})$ is deployed to model the two injectors on the side $(\Omega_{1,7})$ while the  interior-injector ROM $(\bar{\Omega}_\text{II})$ is used to model all five interior injectors $(\Omega_{2-5})$. Similar to the injector-component ROM training, only snapshot data corresponding to the portion of the downstream combustor and nozzle ($\bar{\Omega}_\text{II}$) are extracted to construct the basis for the downstream-component ROM.

\subsection{Component-Based ROM Integration and Deployment in Full-System Simulations}\label{cbrom_integration}
Once the component-based ROMs are trained following the strategies in Sec. \ref{sect4A}, we introduce effective methods to integrate and deploy these ROMs in the CBROM framework to ensure accurate information exchange between components and enable flexible full-system simulations in the current section.

\subsubsection{Component-based ROM coupling}\label{cbrom_coupling}
To ensure accurate predictions of combustion dynamics in full systems arising from the component interactions, it is crucial to ensure that essential information is precisely exchanged between components. To achieve this, we adopt a direct flux matching method via ghost cell assignment \cite{huang_component-based_2022} to couple two components, $\Omega_{k}$ and $\Omega_{m}$, at the interface, $\partial\Omega_{km}$, with no overlap
\begin{equation}
\label{eq6}
\mathbf{v}_{km}(\bar{\mathbf{q}}_{p,k}^n,\bar{\mathbf{q}}_{p,m}^n) = \mathbf{v}_{km}(\bar{\mathbf{q}}_{p,m}^n,\bar{\mathbf{q}}_{p,k}^n) \quad \mathrm{on} \quad \partial\Omega_{km},
\end{equation}
where information is passed between components with the goal of matching the interface condition, $\mathbf{v}_{km}$ (computed both from $\Omega_{k}$ and $\Omega_{m}$), at the component interface. When computing the the state vector, $\bar{\mathbf{q}}_{p,k}$, at time step $n$ for a cells near $\partial\Omega_{km}$ within $\Omega_k$, the state vector, $\bar{\mathbf{q}}_{p,m}$, at the adjacent cell is treated as a ghost cell and assigned by the corresponding neighboring component, $\Omega_{m}$. The combination of $\bar{\mathbf{q}}_{p,k}$ and $\bar{\mathbf{q}}_{p,m}$ is then used to calculate the interface condition $\mathbf{v}_{km}(\bar{\mathbf{q}}_{p,k}^n,\bar{\mathbf{q}}_{p,m}^n)$, and vice
versa, thus guaranteeing the interface condition state is matched as posed in Eq. \ref{eq6}. Specifically, we set the interface condition state function $\mathbf{v}_{km}$ to be the numerical fluxes (both inviscid and viscous) to better suit the finite volume scheme of the numerical solver used for the current work \cite{harvazinski_coupling_2015}. The major benefit of the direct-flux-matching interfacing method is that it inherently accounts for changes in flow characteristics at the interface and therefore important phenomena such as reverse flows are naturally supported. More importantly it makes the training of the component-based ROMs relatively independent of their coupling with other components in the framework, which allows for more flexibility in the ROM training strategy.

\subsubsection{Component-based ROM deployment for geometric variations}\label{cbrom_geom}
\begin{figure}[hbt!]
    \centering
    \includegraphics[width=0.35\linewidth, trim=80 50 80 70, clip]{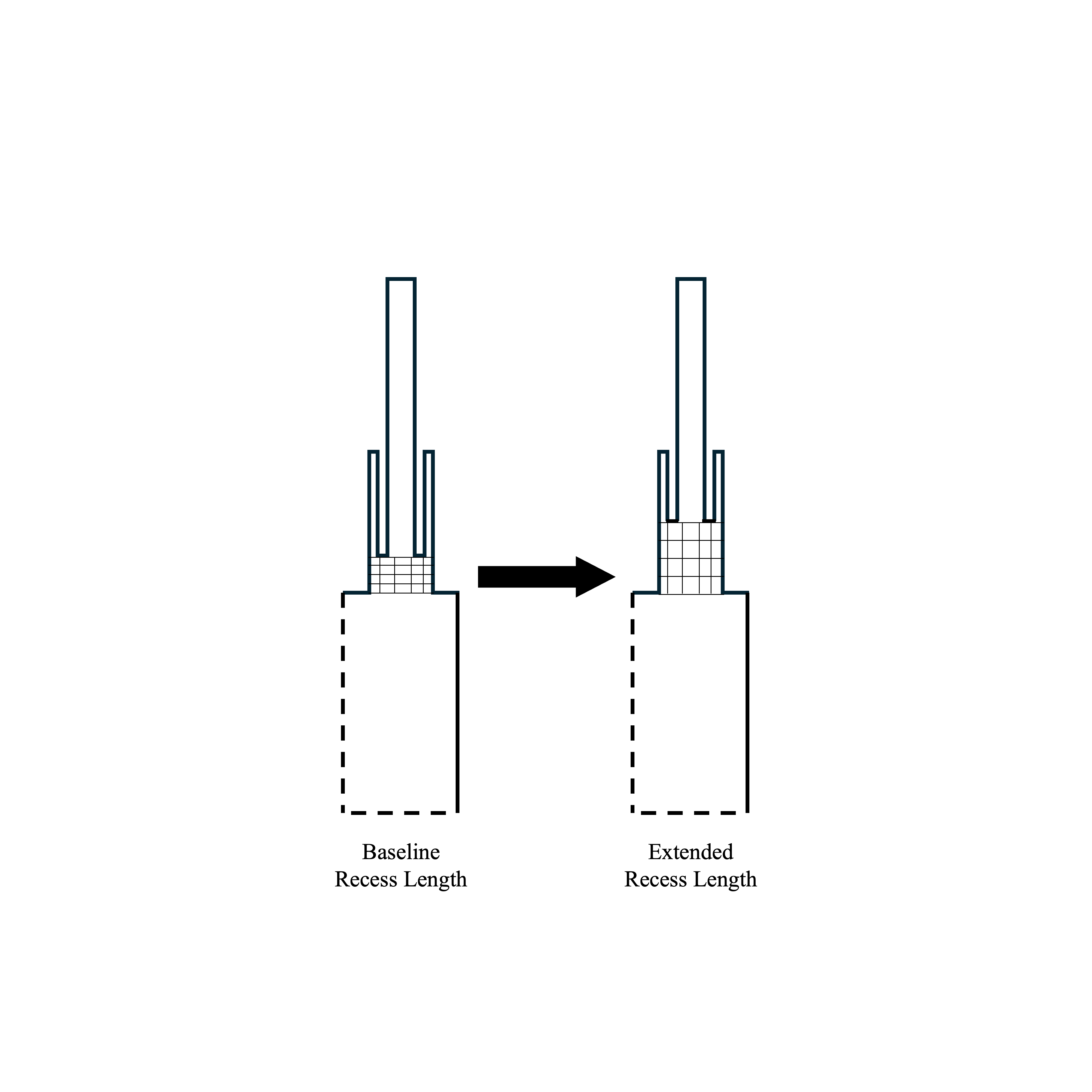}
    \caption{Illustration of ROM basis mapping for geometric variations}
    \label{geom_inj}
\end{figure}

\begin{figure}[hbt!]
    \centering
    \begin{subfigure}{0.3\textwidth}
        \centering
        \includegraphics[height=0.3\textheight, trim=2000 500 4000 500, clip]{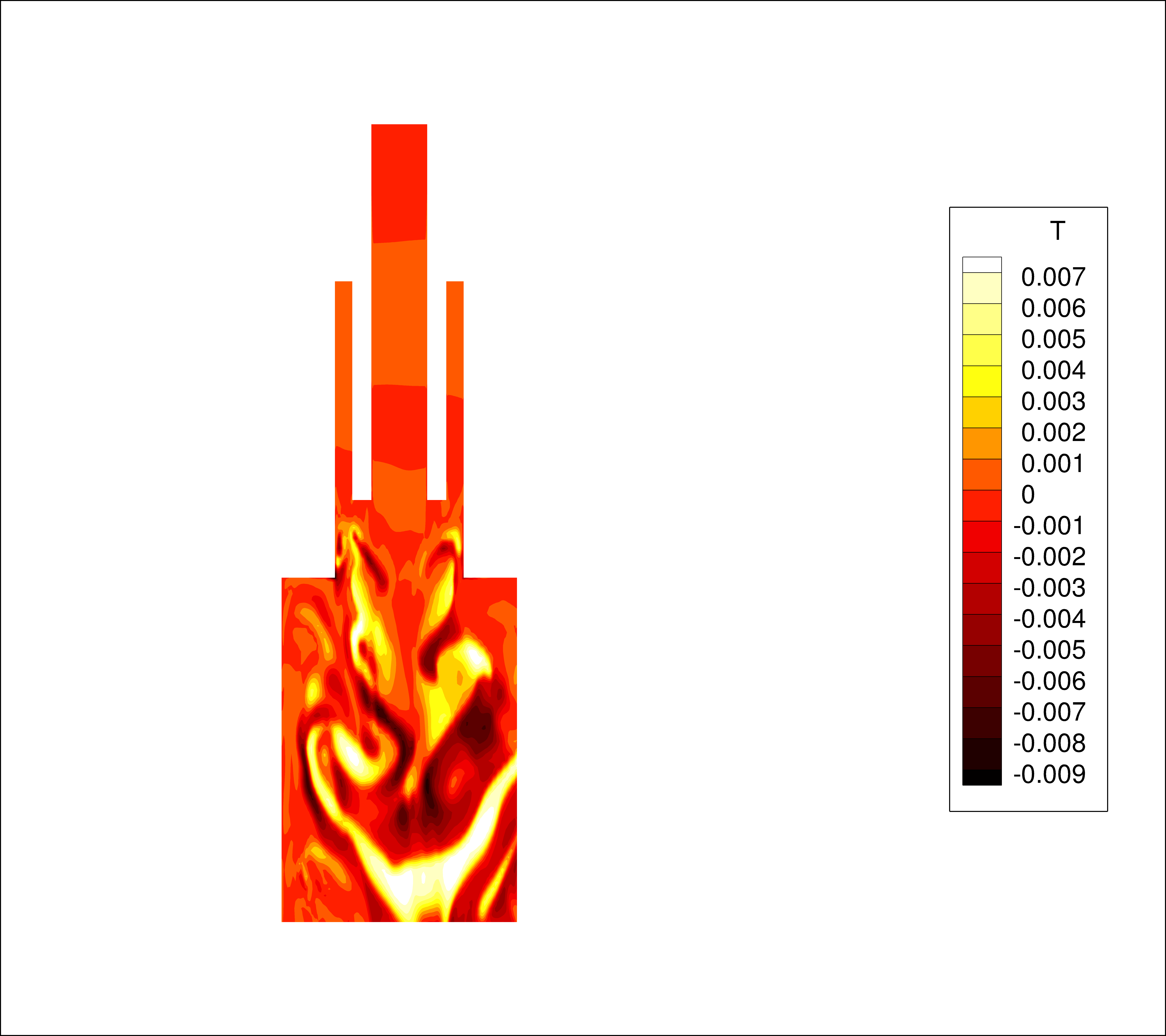}
        \caption{Baseline Recess Length}
    \end{subfigure}
    \begin{subfigure}{0.3\textwidth}
        \centering
        \includegraphics[height=0.3\textheight, trim=2000 500 4000 500, clip]{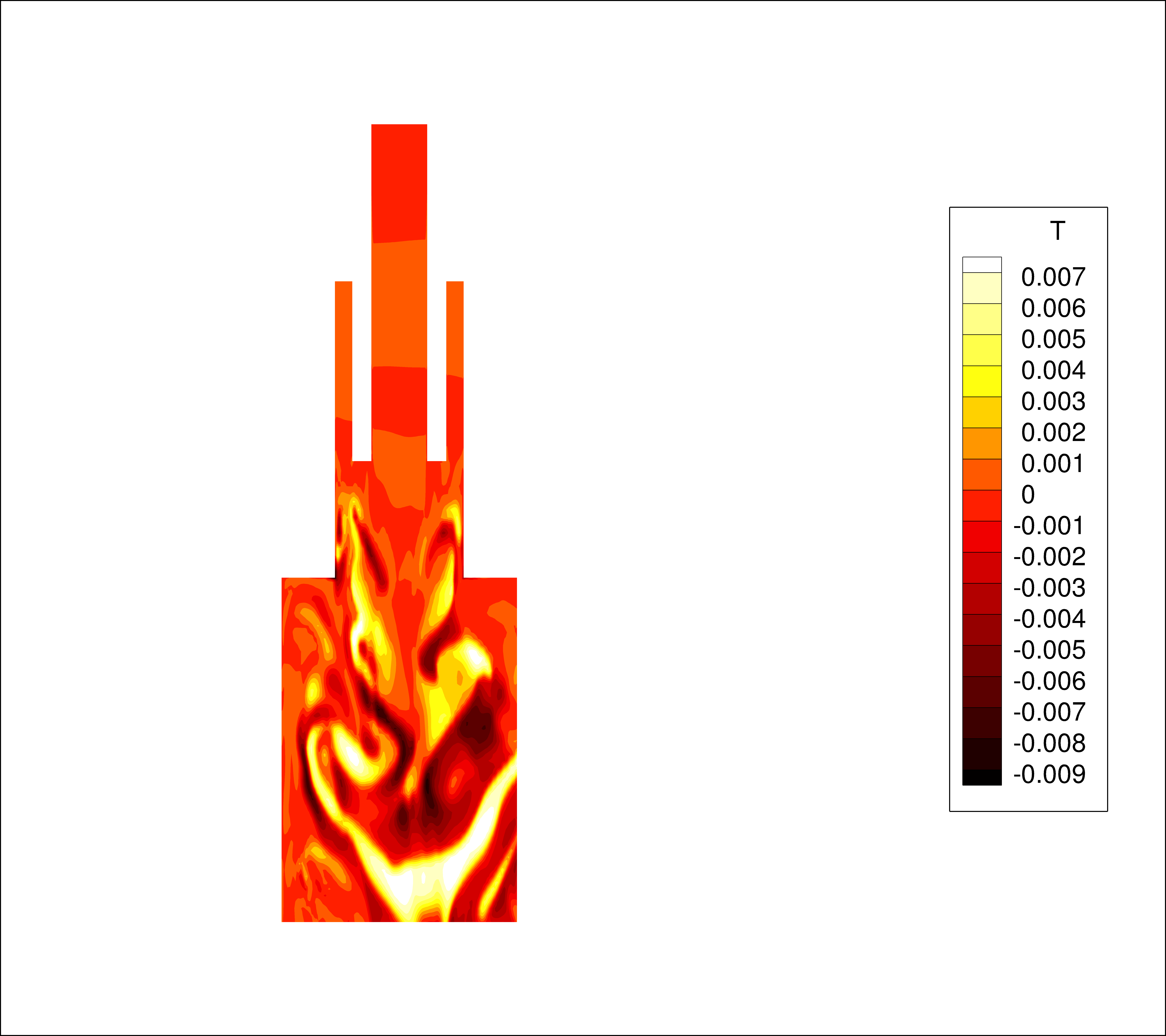}
        \caption{Extended Recess Length}
    \end{subfigure}
    \subfloat{\includegraphics[height=0.3\textheight, trim=6400 1000 300 1100, clip]{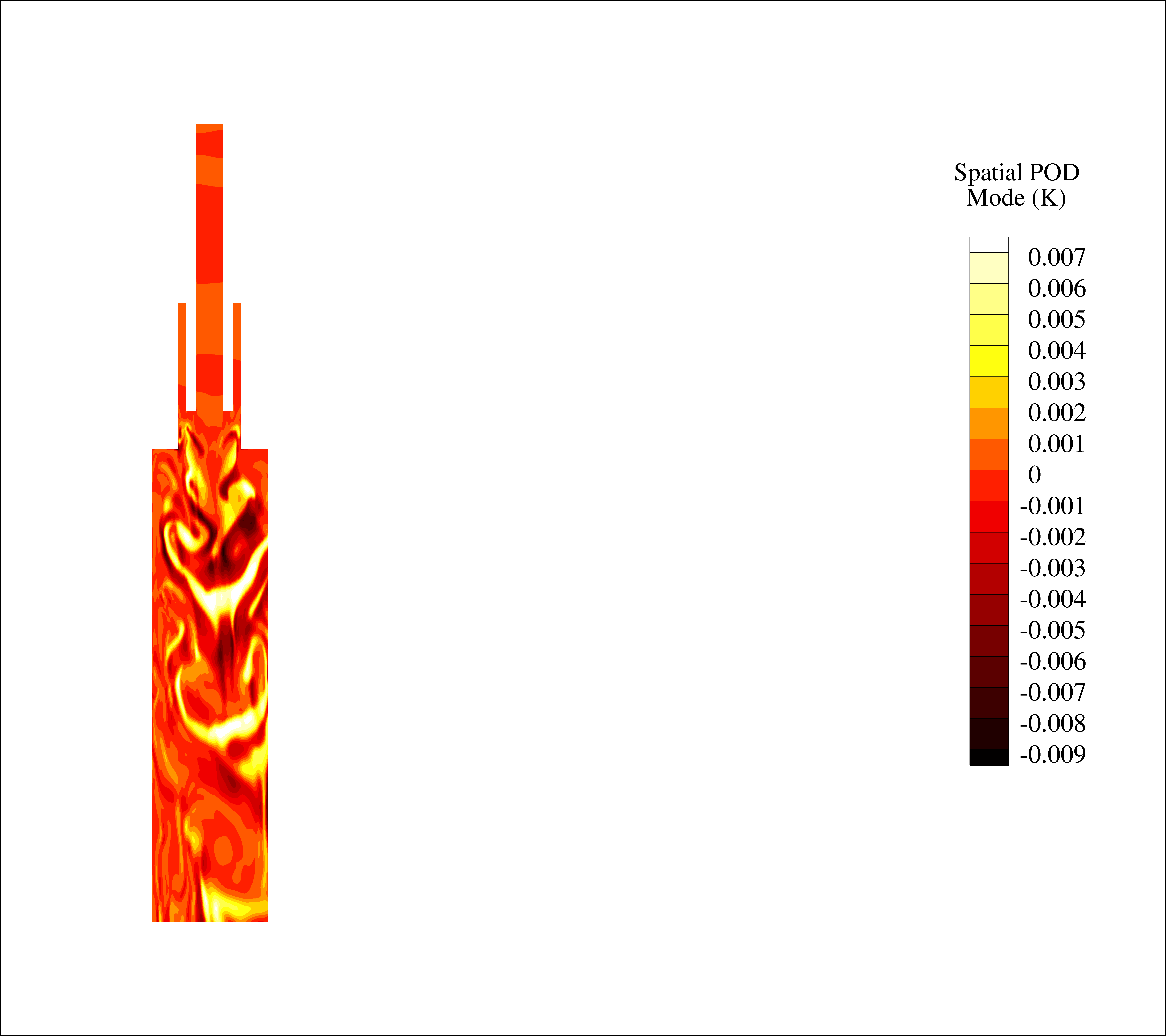}}
    \caption{Transformation of the first spatial POD mode of temperature from the baseline injector mesh to the extended recess length injector mesh}
    \label{geom_inj_POD}
\end{figure}

In addition to enabling FOM-inaccessible full-system simulations via flexible coupling of component-based ROMs, another intriguing capability of the CBROM framework is the component-level modularity, which allows individual components to be replaced with alternative geometries. This capability enables systematic investigations into the effects of geometric variations in injector or nozzle design on the full-system performance without the need to remesh the entire computational domain, which is often resource-consuming. In addition, when replacing a component with a geometrically identical variant, an existing ROM basis can be reused for the modified component. Furthermore, leveraging the adaptive ROM formulation in the CBROM framework, the ROM basis inherently adapts to the dynamics in the altered geometry, thereby eliminating the need to generate separate high-fidelity training data for each geometric variation in a parametric design study. To illustrate the capability to account for geometric variations, we take a single wall-injector component as an example in Fig. \ref{geom_inj}, where the recess length of the injector is doubled with a simplified depiction of the mesh within the injector recess. By preserving the mesh topology within the component for different recess lengths, the ROM basis for the baseline recess length injector can be directly mapped to the injector configuration with the extended recess length. Then, subsequent basis adaptation is executed to accurately capture the changes in the dynamics in the new geometry. The result of this process is shown for the first spatial POD mode of temperature in Fig.\ref{geom_inj_POD}, which has a zoomed-in view of the region around the injector recess. Within the recess, we observe that the mode features are stretched as they map to the larger mesh, while the remainder of the mapping is identical between the baseline and the modified geometry.

\section{Numerical Results and Analysis}\label{sect5}
To assess the capabilities of the component-based reduced-order modeling (CBROM) framework, we consider a 2D seven-injector rocket combustor \cite{huang_parametric_2023,huang_investigations_2024} (as shown in Fig. \ref{fig1}) derived based on a 3D configuration \cite{shipley_multi-injector_2014}, which exhibits distinct self-excited transverse combustion instability under parametric variations of both operating conditions and geometries. Each injector element consists of a shear coaxial configuration with two fuel inlets surrounding one oxidizer inlet. As shown in the previous studies \cite{huang_parametric_2023,huang_investigations_2024}, this seven-injector configuration inherently produces unstable combustion dynamics, featured with highly non-linear interactions between acoustics, turbulence, and chemical reactions that are challenging for ROM development and valuable for detailed evaluation of the CBROM framework's performance. Specifically, we consider three sets of test cases based on this configuration, including: (1) a baseline case at the nominal operation condition, (2) two cases with different fuel injectors shutdown to assess the framework’s modeling capability to capture abrupt parametric variations in operating conditions, and (3) one case with elongated recess lengths of the wall injectors to evaluate the framework's capability in modeling geometric variations. The in-house CFD code, the General Mesh and Equation Solver (GEMS), is used to conduct the FOM and ROM simulations by solving a coupled system consisting of the conservative mass, momentum, energy equations along with species transport.

\subsection{Numerical Setup and Results for the FOM 2D Seven-Injector Rocket Combustor Configuration}\label{sect5A}
In this section, we introduce the baseline geometric configuration and operating conditions of the 2D seven-injector rocket combustor following Fig. \ref{fig3}. First, a baseline case is established at a nominal condition with each fuel injector (indicated by the red arrows) feeding pure gaseous methane ($100\%$ $\text{CH}_\text{4}$) at 300K with a mass flow rate of $0.46$  kg/s and each oxidizer injector (indicated by the blue arrows) supplying a gaseous mixture of $58\%$ $\text{H}_\text{2}\text{O}$ and $42\%$ $\text{O}_\text{2}$ at 660K with a mass flow rate of $5.4$  kg/s. The downstream nozzle exit is choked to sustain acoustic waves in the combustor. The baseline operating condition is maintained with an adiabatic flame temperature of approximately 2700 K, and a mean chamber pressure of 770 kPa. Combustion is represented by the flamelet progress variable (FPV) model \cite{coclite_smld_2015} with GRI-1.2 \cite{frenklach_gri-mech_1995} chemical kinetics that contains 32 species and 177 chemical reactions. The mesh for the full domain contains 370K cells, with 37K cells per injector component and 110K cells in the downstream component. The FOM is solved using the second-order accurate backwards difference formula and dual time-stepping, with a
constant physical time step size of 0.1 $\mathrm{\mu s}$. Second, two parametric cases are created by abruptly setting the mass flow rates of selected fuel injectors to zero. These two cases are designed to emulate injector shutdown scenarios in rocket engines, which may arise from blockages or valve failures. In the first parametric case, the fuel injector in the center injector component ($\Omega_4$) is shut down. In the second parametric case, the fuel injectors of both the center ($\Omega_4$) and the right wall injector components ($\Omega_{7}$) are shut down. In addition, to better represent realistic scenarios in rocket engine failures, both parametric cases are initialized using the statistically stationary baseline solution, leading to transient flow adjustments as the system adjusts to the new inlet conditions. For consistent evaluation of important statistical quantities, all test cases are simulated for 20 ms to ensure sufficient information is included for accurate analysis of these quantities.

\begin{figure}[hbt!]
    \centering
    \subfloat{\includegraphics[width=0.48\linewidth, trim=15 10 15 25, clip]{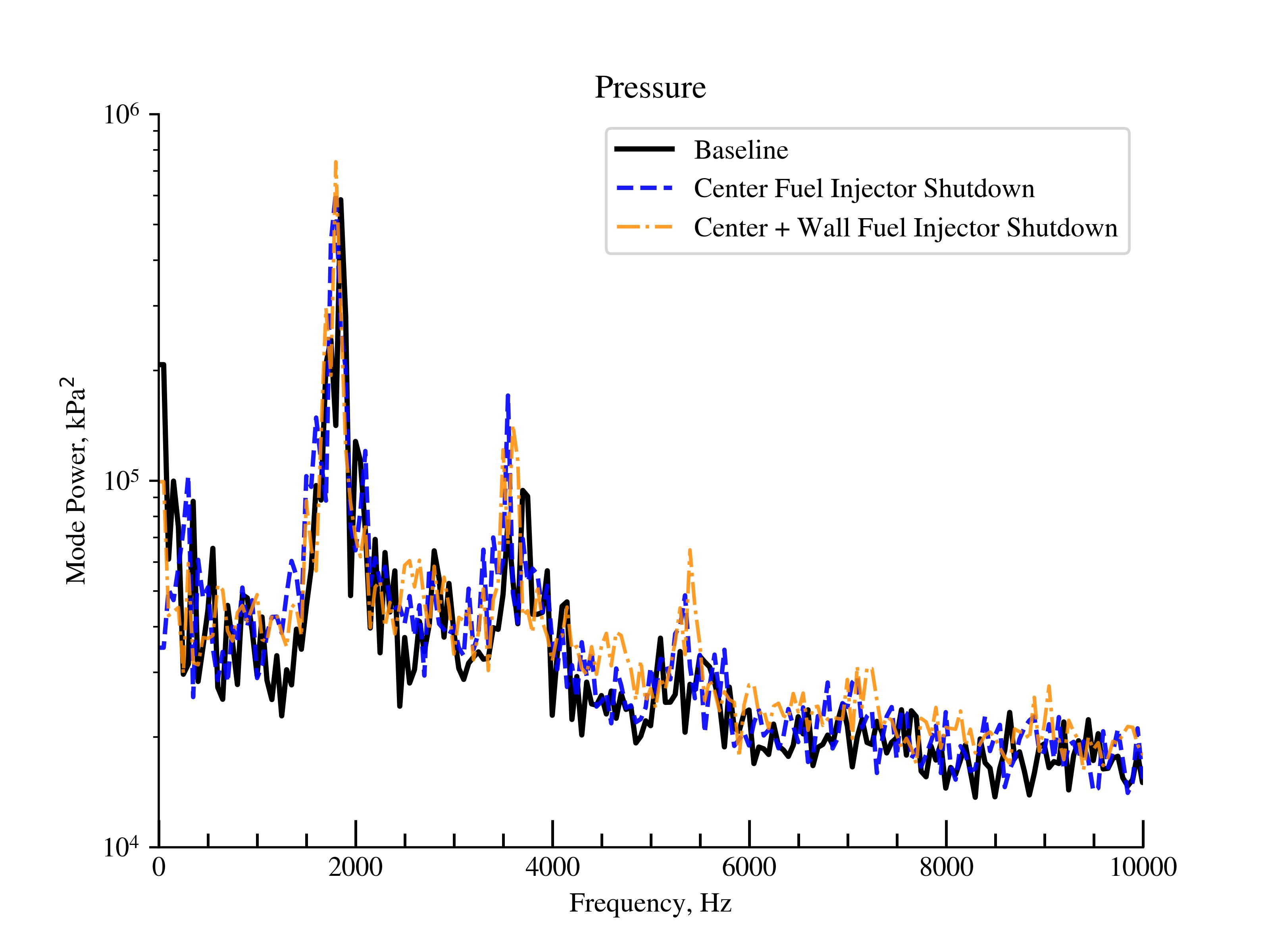}}
    \hfill
    \subfloat{\includegraphics[width=0.48\linewidth, trim=15 10 15 25, clip]{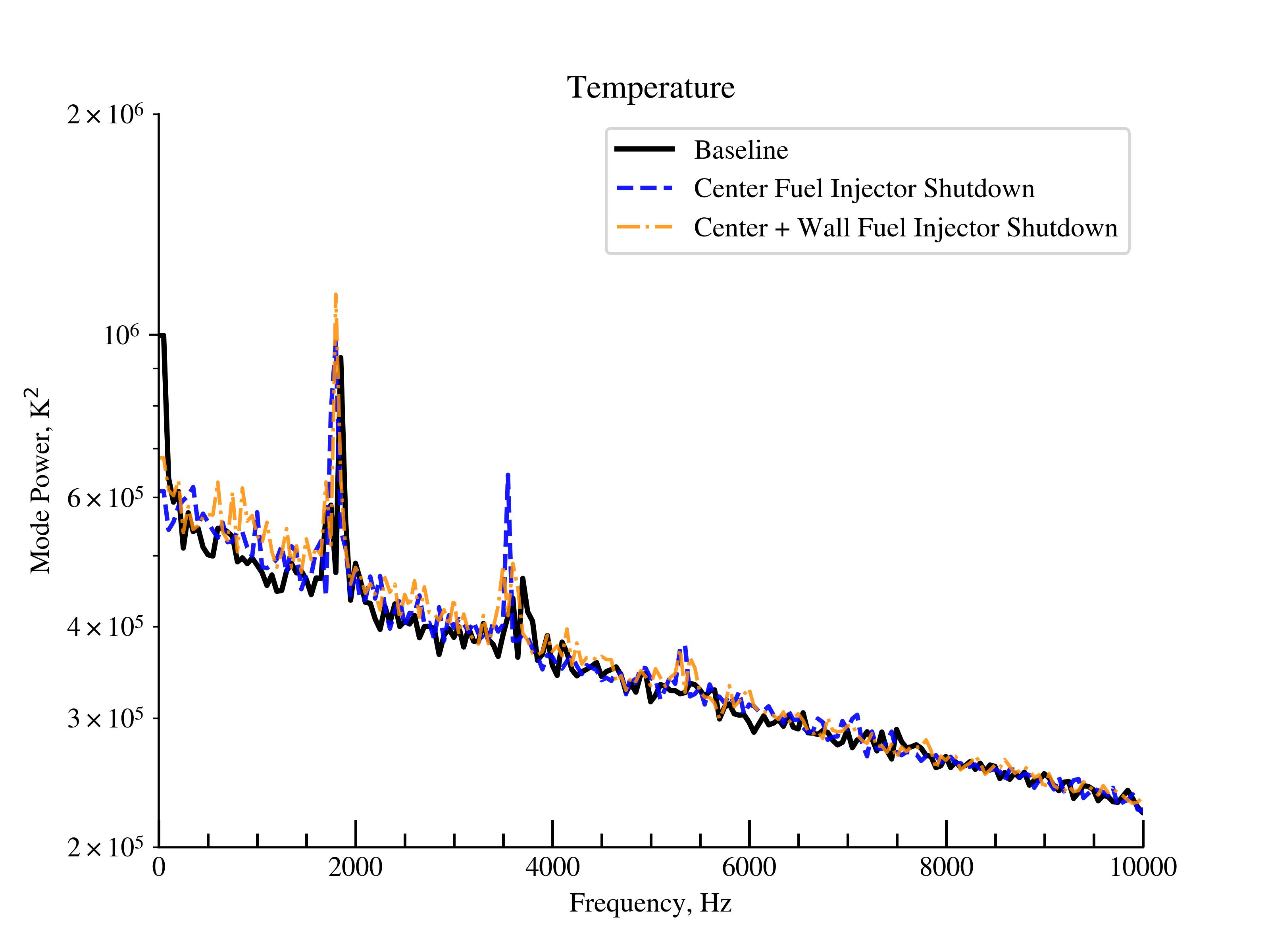}}
    \caption{Comparisons of DMD spectra for the three test problems of the 2D seven-injector rocket combustor configuration with various operating conditions}
    \label{FOM_DMD}
\end{figure}

Next, the resulting FOM solutions of the three test cases are examined and compared to assess the impacts of fuel-injector shutdown on rocket combustion dynamics. Dynamic mode decomposition (DMD) analysis \cite{huang_analysis_2016} is applied to the FOM solutions and the DMD spectra for pressure and temperature are compared in Fig.~\ref{FOM_DMD}. Overall, both pressure and temperature spectra exhibit distinct frequency peaks corresponding to the dominant transverse acoustic modes. However, with fuel injectors shut down, the dominant acoustic modes are observed to shift to lower frequencies and higher amplitudes compared to the baseline in both pressure and temperature spectra. In addition, a third prominent acoustic mode is established at approximately 5500Hz with fuel injectors shut down, which is not observed in the baseline case.

\begin{figure}[hbt!]
    \centering
    \subfloat[Baseline]{{\includegraphics[height=6cm, trim=950 850 2850 800, clip]{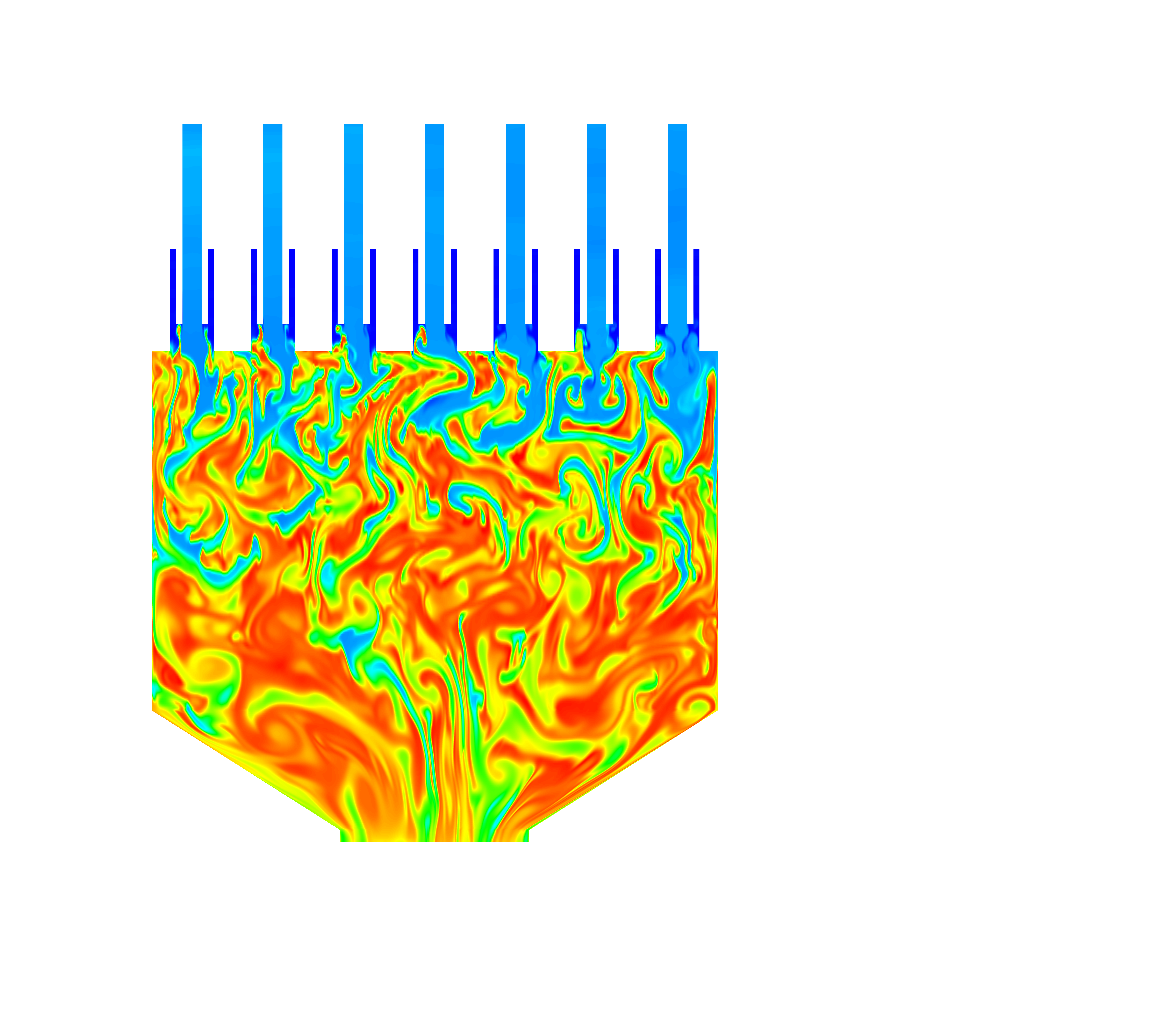}}}
    \hfill
    \subfloat[Center Fuel Injector Shutdown]{\includegraphics[height=6cm, trim=950 850 2850 800, clip]{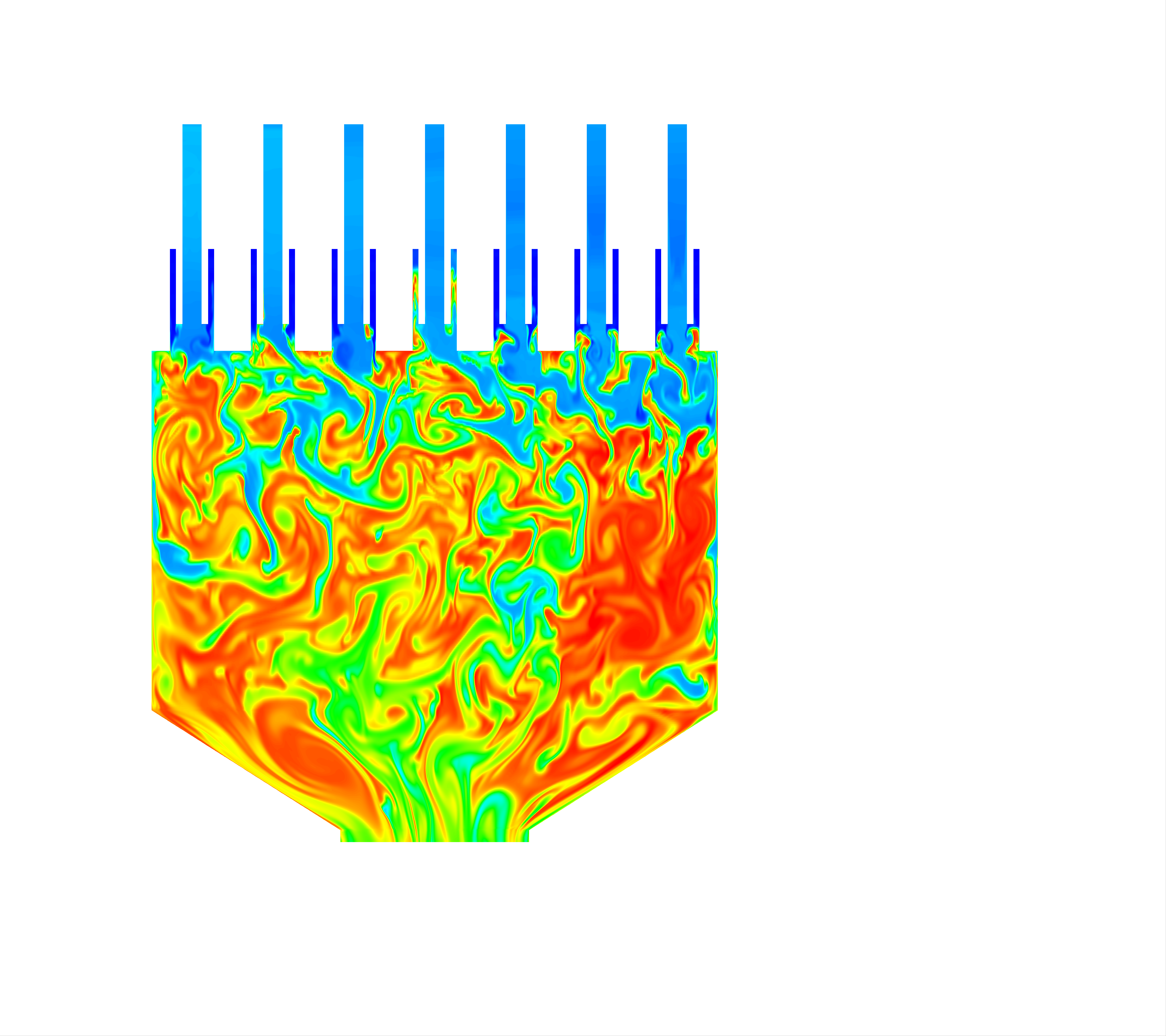}}
    \hfill
    \subfloat[Center and Right-wall Fuel Injector Shutdown]{\includegraphics[height=6cm, trim=950 850 1000 800, clip]{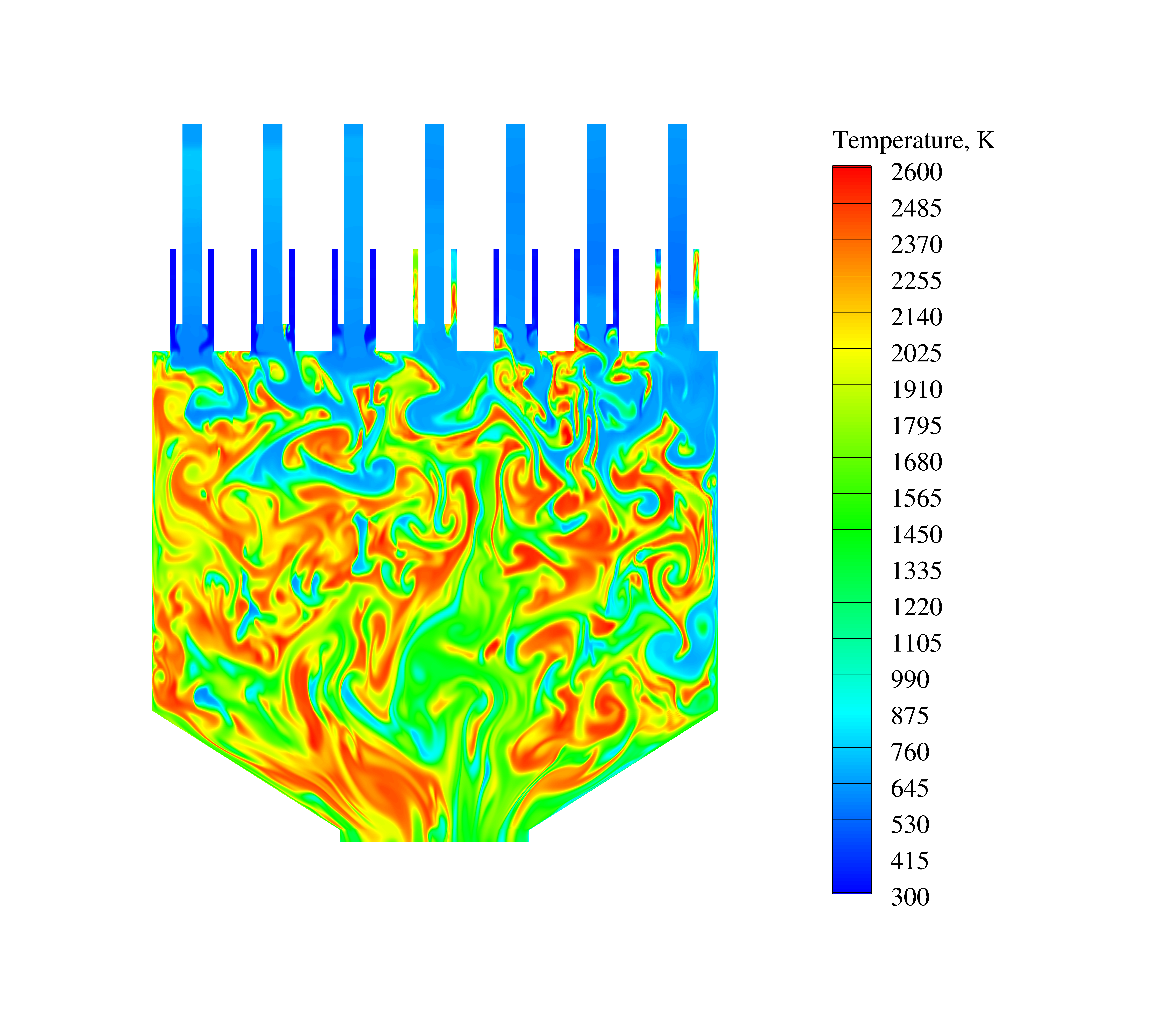}}
    \caption{Comparisons of instantaneous temperature fields for the three test problems of the 2D seven-injector rocket combustor configuration with various operating conditions}
    \label{FOM_Tfield}
\end{figure}

Moreover, instantaneous temperature fields are compared in Fig.~\ref{FOM_Tfield} at 20 ms for the three cases. All cases exhibit strong interactions between acoustics, turbulence, and combustion while the combustion dynamics behave quite differently by varying the operating conditions. Specifically, by shutting down the fuel injectors, high-temperature products enter and recirculate into the fuel injectors with no incoming flow, which results in substantial asymmetric dynamical behaviors compared to the baseline case, leading to higher oscillating amplitudes as observed in DMD spectra (Fig.~\ref{FOM_DMD}). More importantly, as the number of shut down fuel injectors increases, cold propellants penetrate farther downstream, which delays and suppress the overall combustion, leading to decreased mean combustor temperature, and consequently, reduced frequencies of the acoustic modes as observed in DMD spectra (Fig.~\ref{FOM_DMD}).

\subsection{Numerical Setup of CBROM Framework}\label{sect5B}
Following the methodology in Sec. \ref{sect4}, three component-based ROMs are trained, including: two for wall and interior injectors respectively, and one for the downstream combustor and nozzle. The training of injector-component ROMs adopts the reduced-geometry-based strategies in Fig. \ref{fig2} with the inflow conditions for the injector components used in the training domains matching those of the nominal operating conditions for each injector in the full seven-injector simulations. The inflows of the buffer regions (indicated by the white arrows) are set to replicate the mean flow of the injector components, which consist of a mixture of $29\%$ $\text{H}_2\text{O}$ and $71\%$ $\text{C}\text{O}_2$ entering the training domain at 250 m/s and 2400 K. All the outlets are specified with non-reflective boundary conditions. High-fidelity simulations of the two training domains are performed using the FOM and are advanced until the dynamics reach statistically stationary states. The simulations are then continued for an additional 1,000 time steps (corresponding to 0.1 ms of dynamics), the resulting data of which is then collected within the reference components, $\bar{\Omega}_\text{I}$ and $\bar{\Omega}_\text{II}$, and used to construct ROM bases. The training of the downstream-component ROM follows the full-geometry-based strategy in Fig. \ref{fig3} with the injector components modeled using ROMs and the downstream component modeled using the FOM. The same nominal boundary conditions are applied as the baseline case, and a solution snapshot from the baseline case is used to initialize the training simulation for the downstream-component ROM. Similar to the injector-component ROM training, the high-fidelity simulation is advanced for 1,000 time steps (0.1 ms of dynamics), the resulting data of which is collected and utilized to construct the ROM basis for the downstream component. 

Specifically, when constructing the three component-based ROMs, the training data is down-sampled every 100 time steps, leading to 10 evenly spaced snapshots. This strategy increases the temporal separation between successive snapshots, allowing the flow field to evolve more substantially between samples. As a result, the POD basis captures a broader range of dynamical behavior, reducing the risk of overfitting to slowly varying or highly correlated states. All three component-based ROMs are constructed using the first 7 POD modes (covering > 99.99$\%$ of the total energy) to eliminate possible influence of the noisy low-energy modes that can contaminate the resulting ROMs' performance. In addition, $1\%$ of the total cells are sampled for hyper-reduction to construct the adaptive ROM within each component, the basis and sampling points of which are updated every time step $(z_b=1)$ and every 5 time steps $(z_s=5)$ respectively. The three resulting component-based ROMs are then deployed and coupled in the CBROM framework to enable the full seven-injector simulations established in Sec. \ref{sect5A}. For consistent comparison, the CBROM calculation of each test case uses the same initial condition as the corresponding FOM calculation and advances for 20 ms to ensure sufficient information is included for detailed statistical analysis.

\subsection{Performance of the CBROM Framework}\label{sect5C}
In this section, we assess the modeling capabilities of the CBROM framework based on the true FOM simulation results (e.g., in Figs. \ref{FOM_DMD} and \ref{FOM_Tfield}), focusing on the framework's parametric predictive capabilities for: (1) variations of operating conditions, and (2) variations of injector geometries. Given the complexity of combustion dynamics in the seven-injector configuration (\ref{FOM_Tfield}), direct comparison of the solutions between the FOM and the CBROM framework (e.g., $\text{L}_\text{2}$-norm errors or flow-field visualization) can lead to biased evaluations, thereby providing an inaccurate assessment of the framework's performance. Therefore, to consistently evaluate the performance of the CBROM framework, we focus our assessment on comparing key statistical quantities of interest (QoIs) that are of the most value in engineering design. Specifically, three quantities are computed and compared between the FOM and the CBROM framework to obtain systematic and comprehensive evaluations of the framework's performance, which include the time-averaged flow fields, root-mean-square (RMS) flow fields, and frequency spectra from DMD analysis (as illustrated in Fig. \ref{FOM_DMD}). 

\subsubsection{Evaluations of parametric predictive capabilities on variations of operating conditions}\label{sect5C2}
\begin{figure}[hbt!]
\centering
\begin{subfigure}{0.8\textwidth}
    \centering
    \includegraphics[width=0.48\linewidth, trim=20 10 25 25, clip]{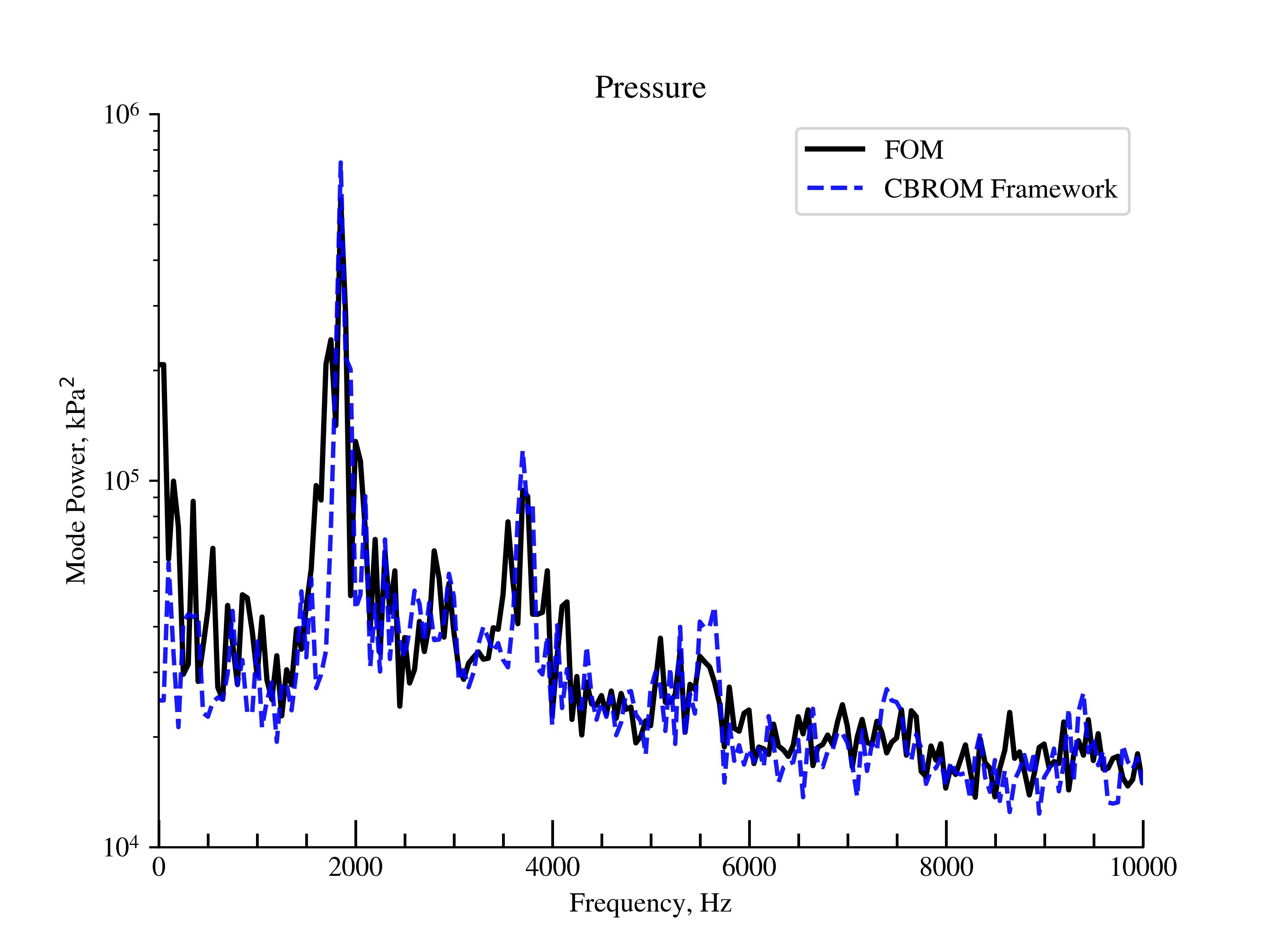}
    \hfill
    \includegraphics[width=0.48\linewidth, trim=20 10 25 25, clip]{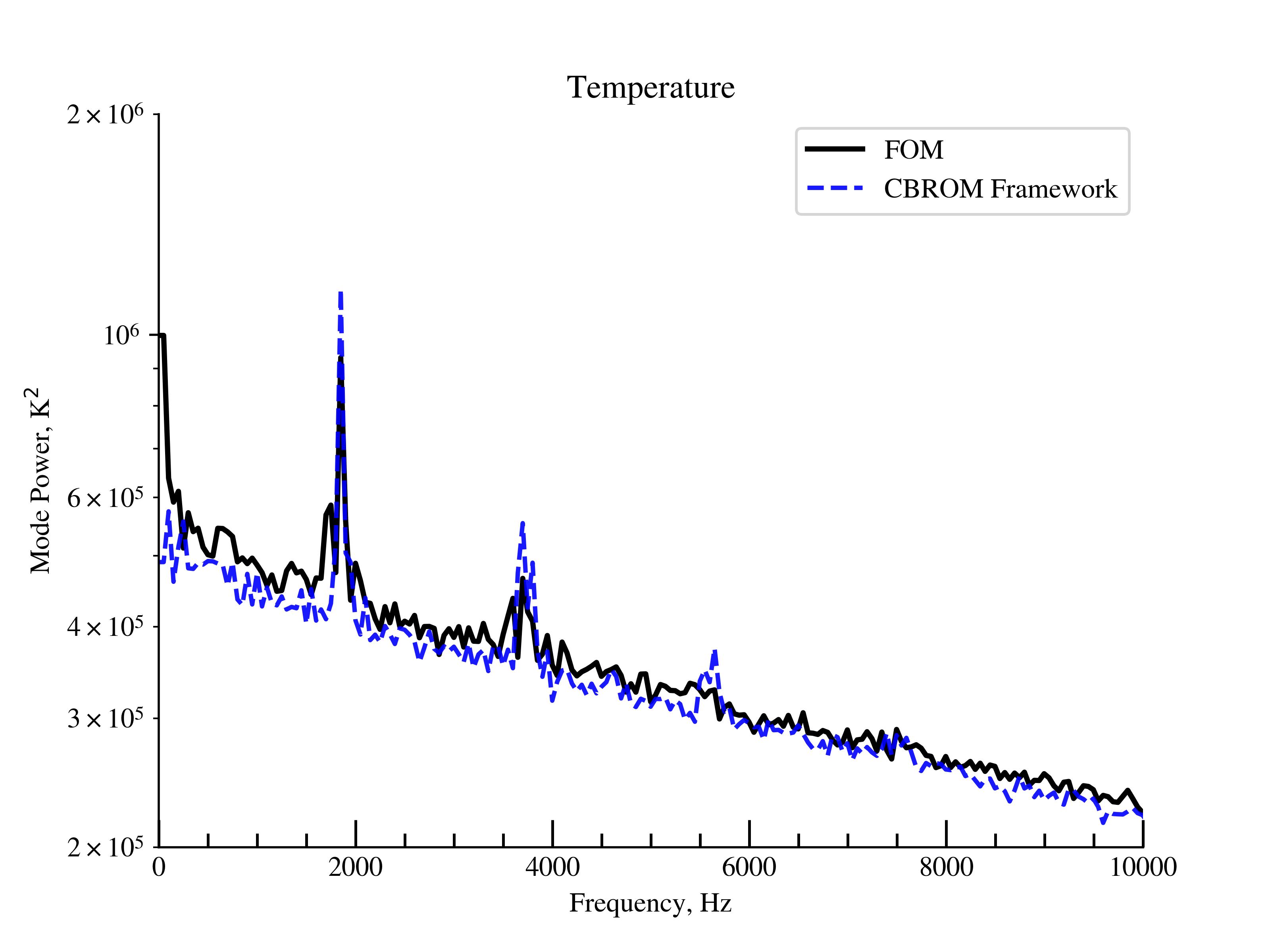}
    \caption{Baseline test case}
    \label{baseline_DMD}
\end{subfigure}
\medskip
\begin{subfigure}{0.8\textwidth}
    \centering
    \includegraphics[width=0.48\linewidth, trim=20 10 25 25, clip]{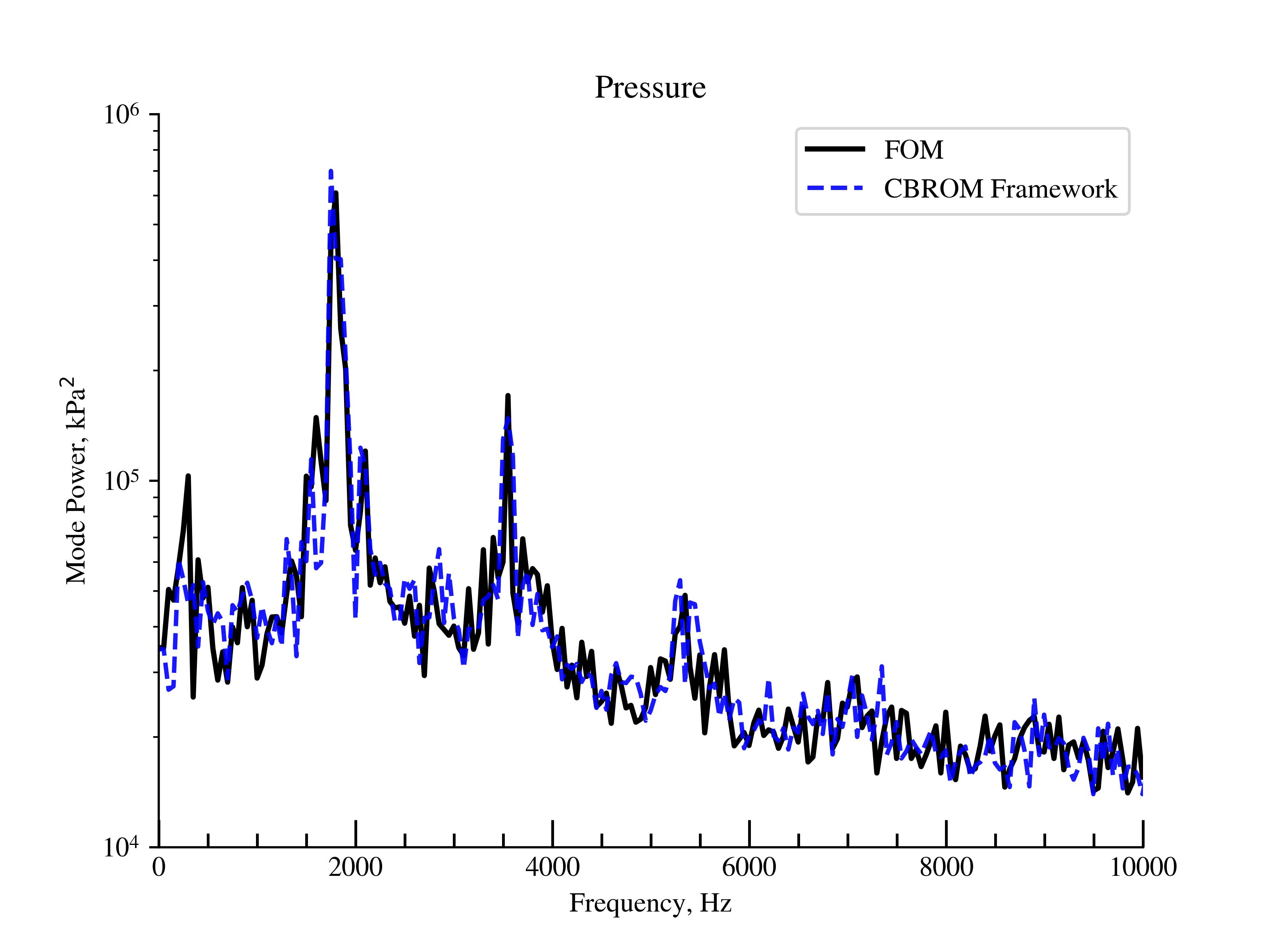}
    \hfill
    \includegraphics[width=0.48\linewidth, trim=20 10 25 25, clip]{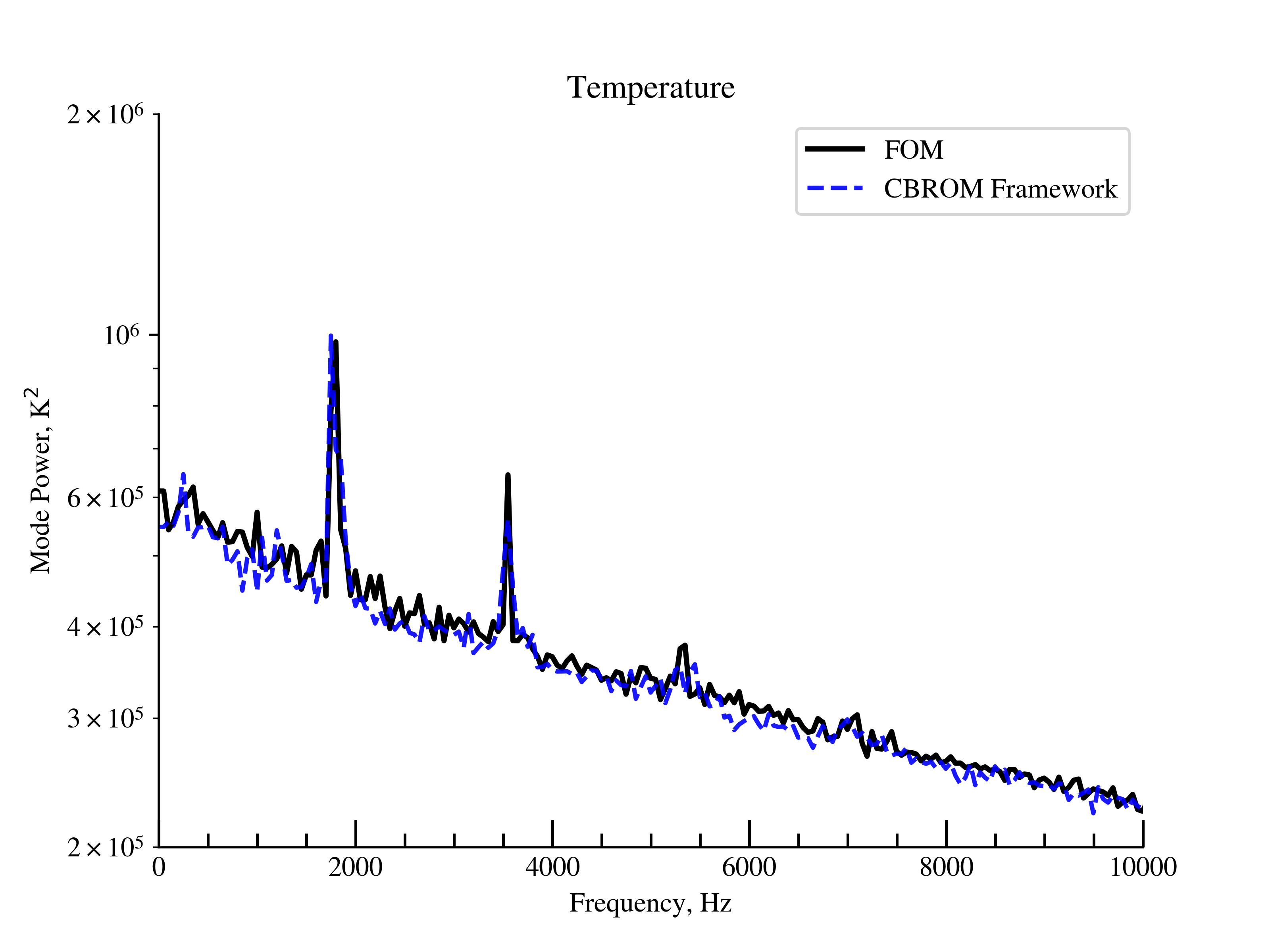}
    \caption{Parametric test case with center fuel injector shutdown}
    \label{m_DMD}
\end{subfigure}
\medskip
\begin{subfigure}{0.8\textwidth}
    \centering
    \includegraphics[width=0.48\linewidth, trim=20 10 25 25, clip]{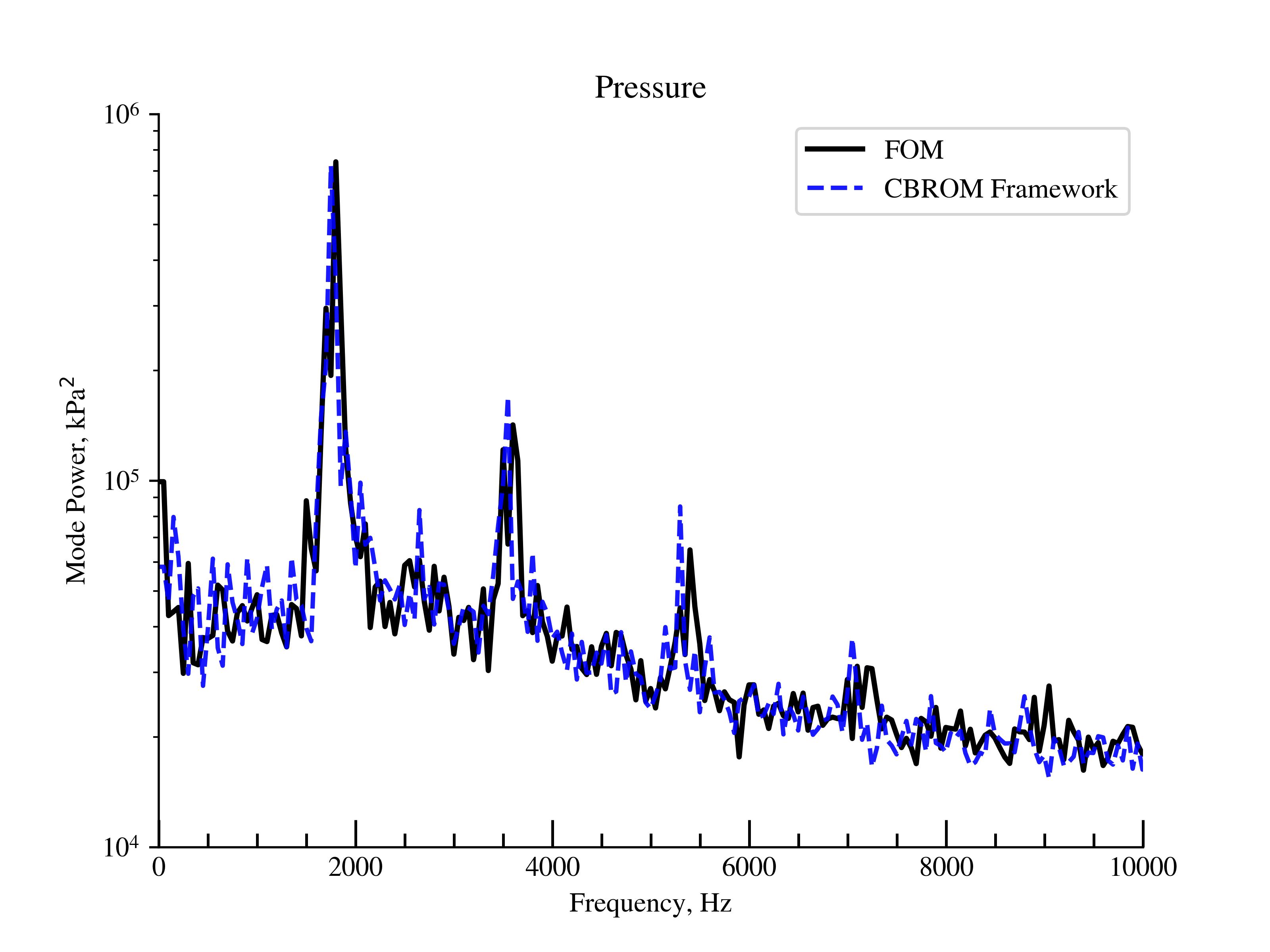}
    \hfill
    \includegraphics[width=0.48\linewidth, trim=20 10 25 25, clip]{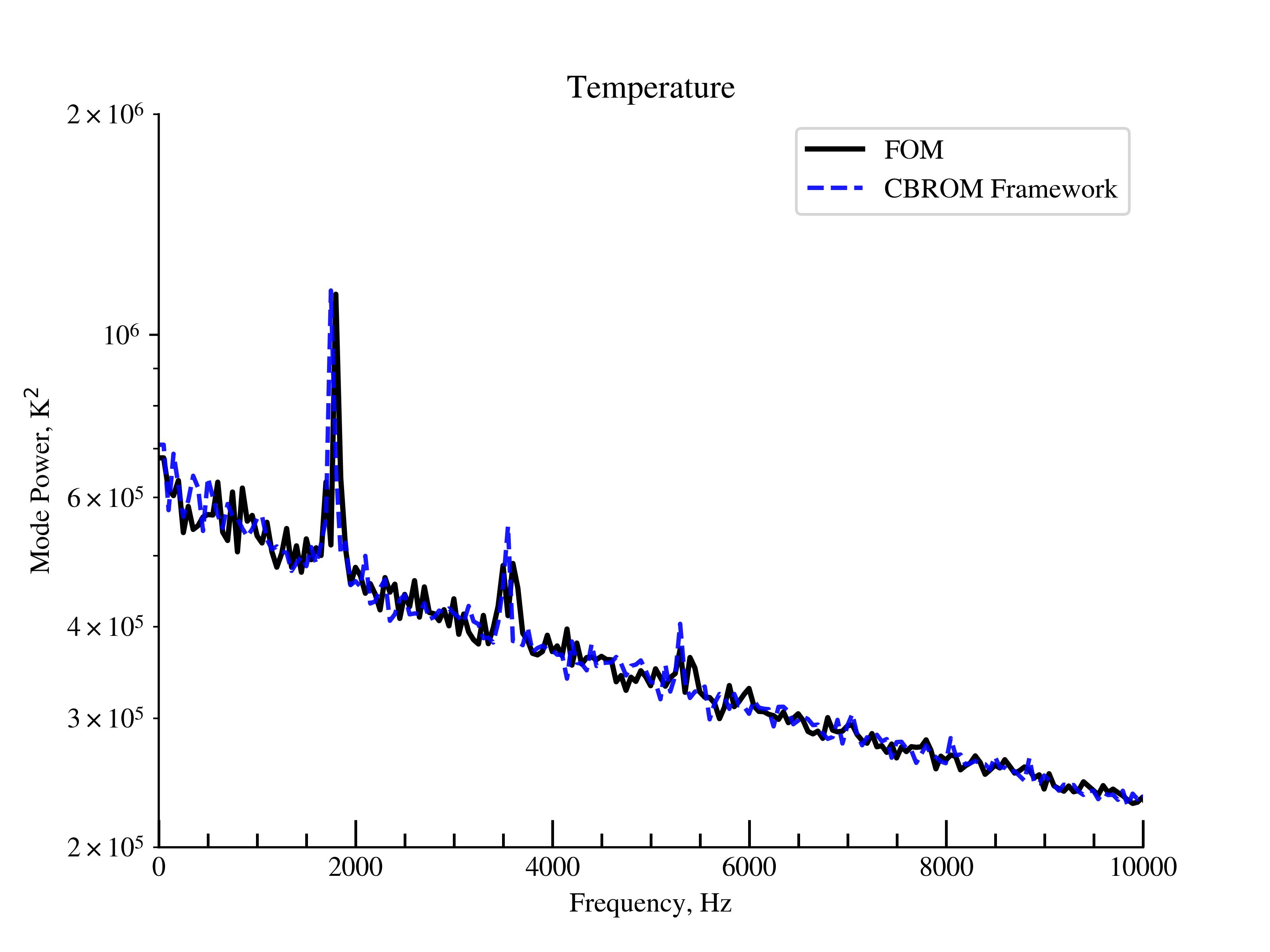}
    \caption{Parametric test case with center and right-wall fuel injectors shutdown}
    \label{mr_DMD}
\end{subfigure}
\caption{Comparisons of DMD spectra between FOM and CBROM framework for different operating conditions}
\label{DMD_para}
\end{figure}

The first evaluation focuses on assessing the CBROM framework's modeling capabilities of predicting the changes in dynamics due to variations of operating conditions based on the three test cases introduced in Sec. \ref{sect5A}. Figure~\ref{DMD_para} compares the DMD spectra of both pressure and temperature, which shows excellent agreement between the FOM and the CBROM framework for all three tests cases, especially in capturing the changes in frequencies and magnitudes of dominant acoustic modes due to variations of operating conditions. Specifically, the CBROM framework successfully predicts the frequency decrease and magnitude increase of dominant acoustic modes as an increasing number of fuel injectors are shut down, consistent with the observations of the FOM results in Fig. \ref{FOM_DMD}. In addition, the CBROM framework accurately captures the emergence of higher acoustic modes at different operating conditions. For example, compared to the baseline test case, the third acoustic mode becomes more distinct with the center fuel injector shut down and the fourth acoustic mode appears as an additional dominant frequency when the center and right-wall fuel injectors are shut down. 

\begin{figure}[hbt!]
\centering
\begin{subfigure}{0.8\textwidth}
    \centering
    \begin{minipage}{0.48\linewidth}
        \centering
        \includegraphics[height=5.8cm, trim=0 0 2300 0, clip]{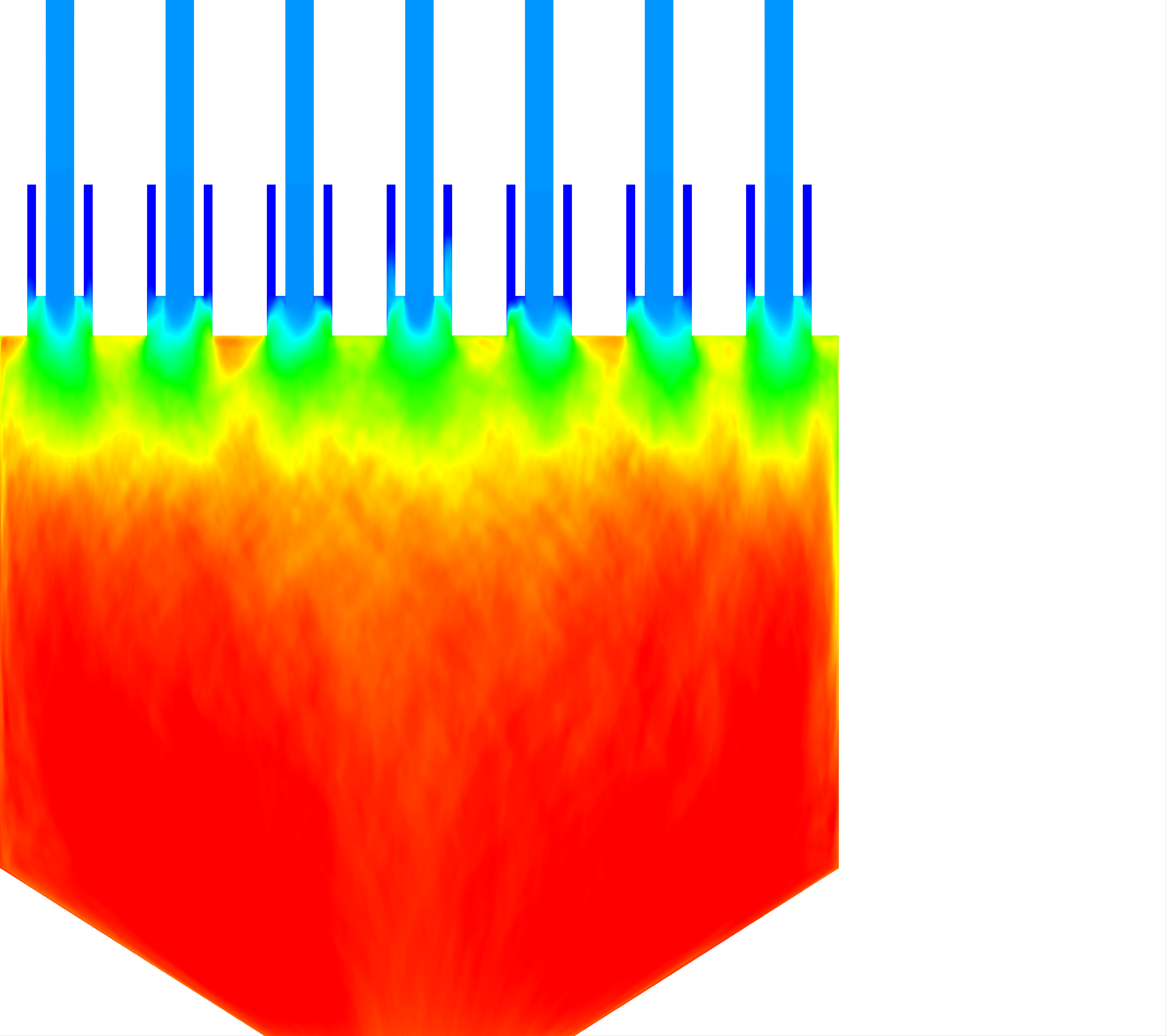}\\
        \small\bfseries FOM
    \end{minipage}
    \hfill
    \begin{minipage}{0.48\linewidth}
        \centering
        \includegraphics[height=5.8cm, trim=0 0 0 0, clip]{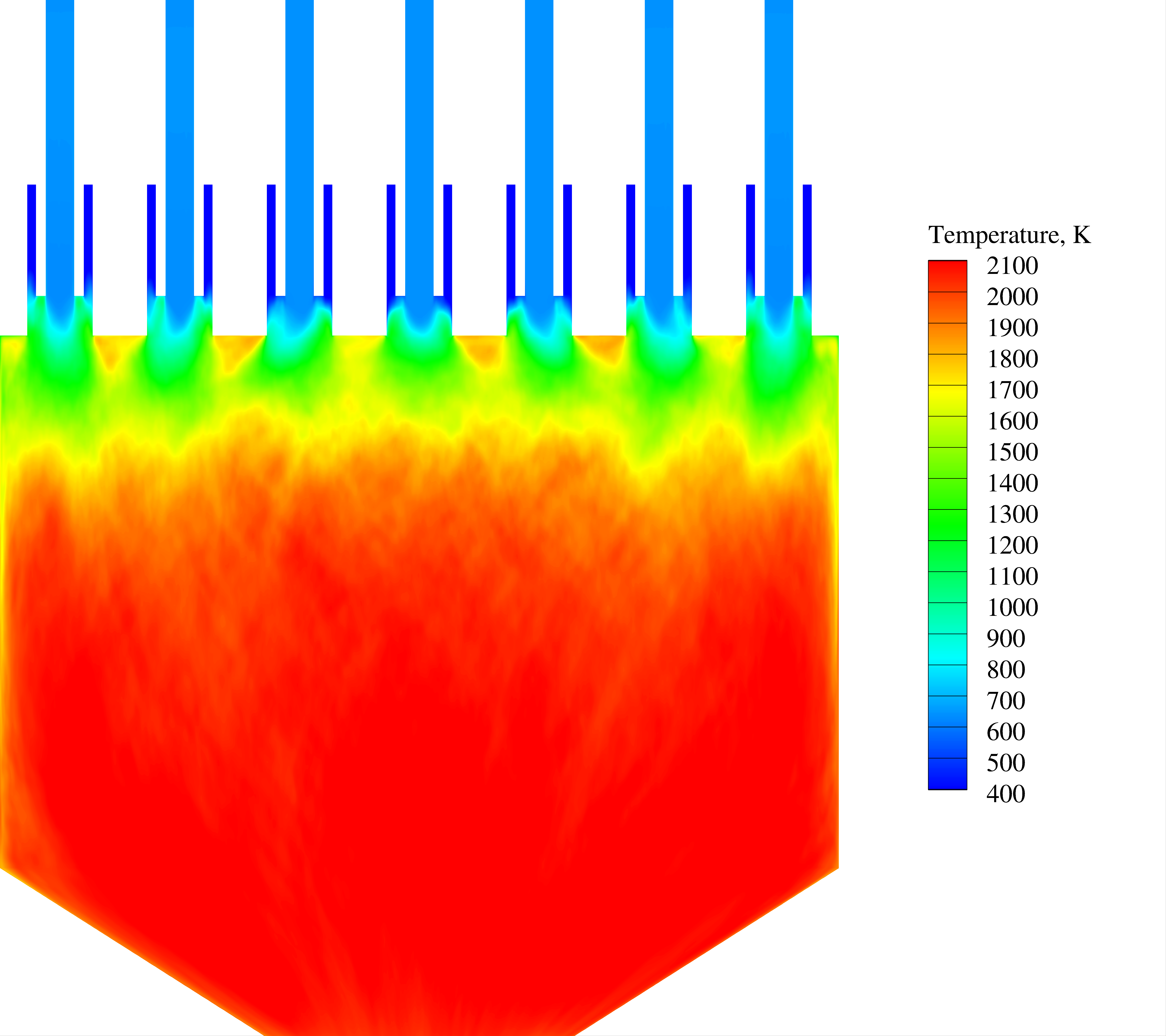}\\
        \hspace*{-1.4cm}\small\bfseries CBROM Framework
    \end{minipage}
    \caption{Baseline test case}
    \label{baseline_mean_para}
\end{subfigure}
\medskip
\begin{subfigure}{0.8\textwidth}
    \centering
    \begin{minipage}{0.48\linewidth}
        \centering
        \includegraphics[height=5.8cm, trim=0 0 2300 0, clip]{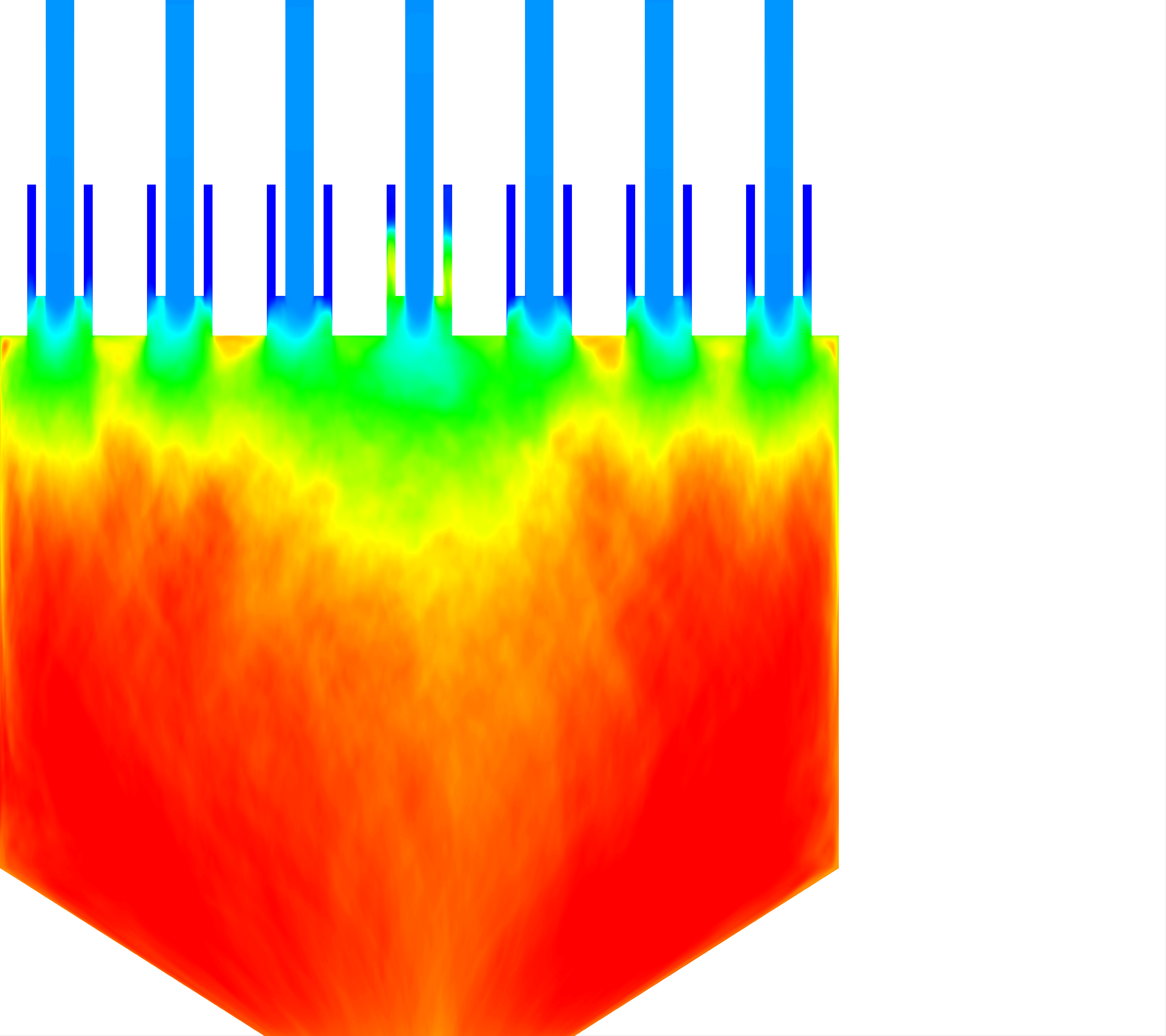}\\
        \small\bfseries FOM
    \end{minipage}
    \hfill
    \begin{minipage}{0.48\linewidth}
        \centering
        \includegraphics[height=5.8cm, trim=0 0 0 0, clip]{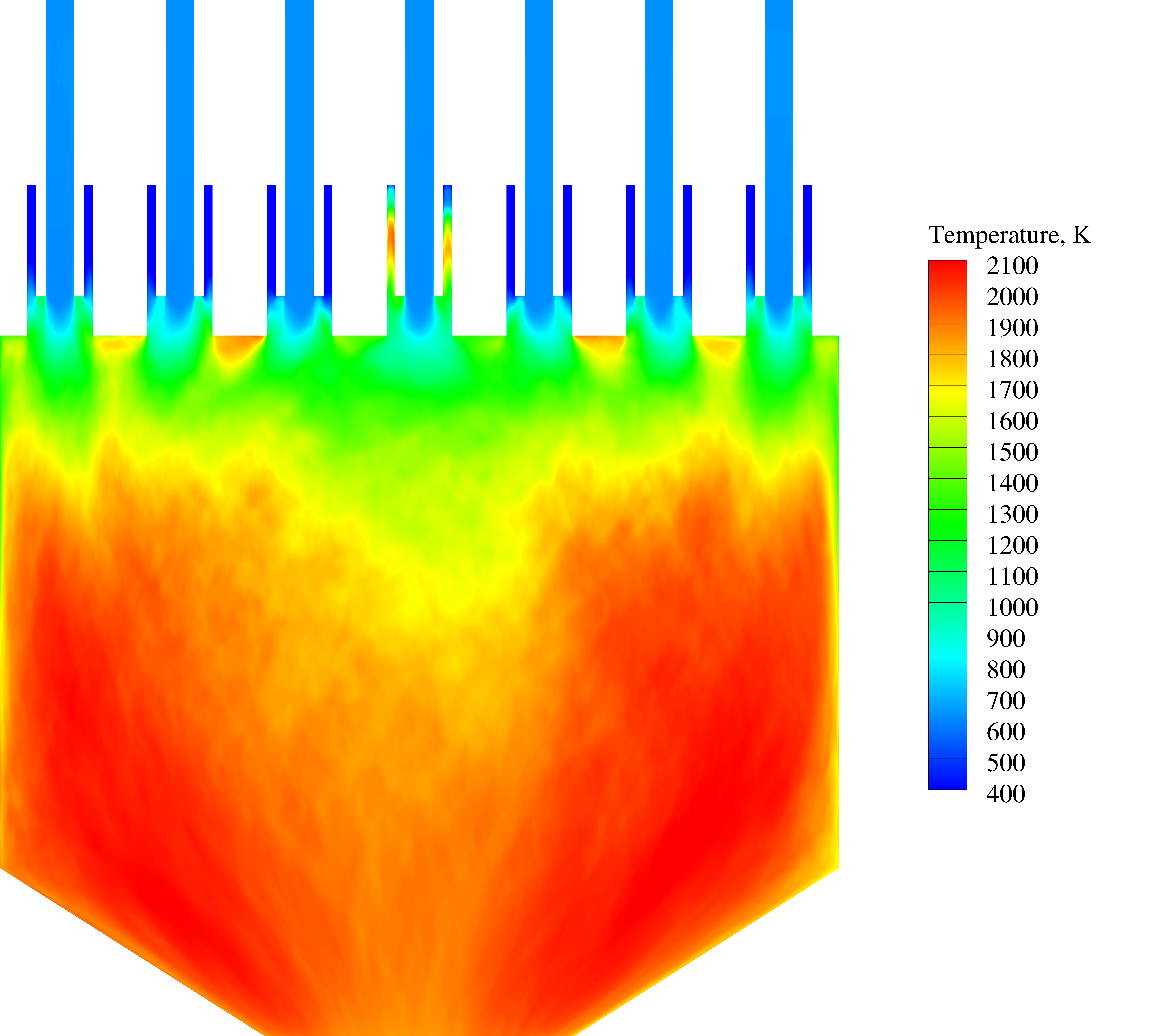}\\
        \hspace*{-1.4cm}\small\bfseries CBROM Framework
    \end{minipage}
    \caption{Parametric test case with center fuel injector shutdown}
\end{subfigure}
\medskip
\begin{subfigure}{0.8\textwidth}
    \centering
    \begin{minipage}{0.48\linewidth}
        \centering
        \includegraphics[height=5.8cm, trim=0 0 2300 0, clip]{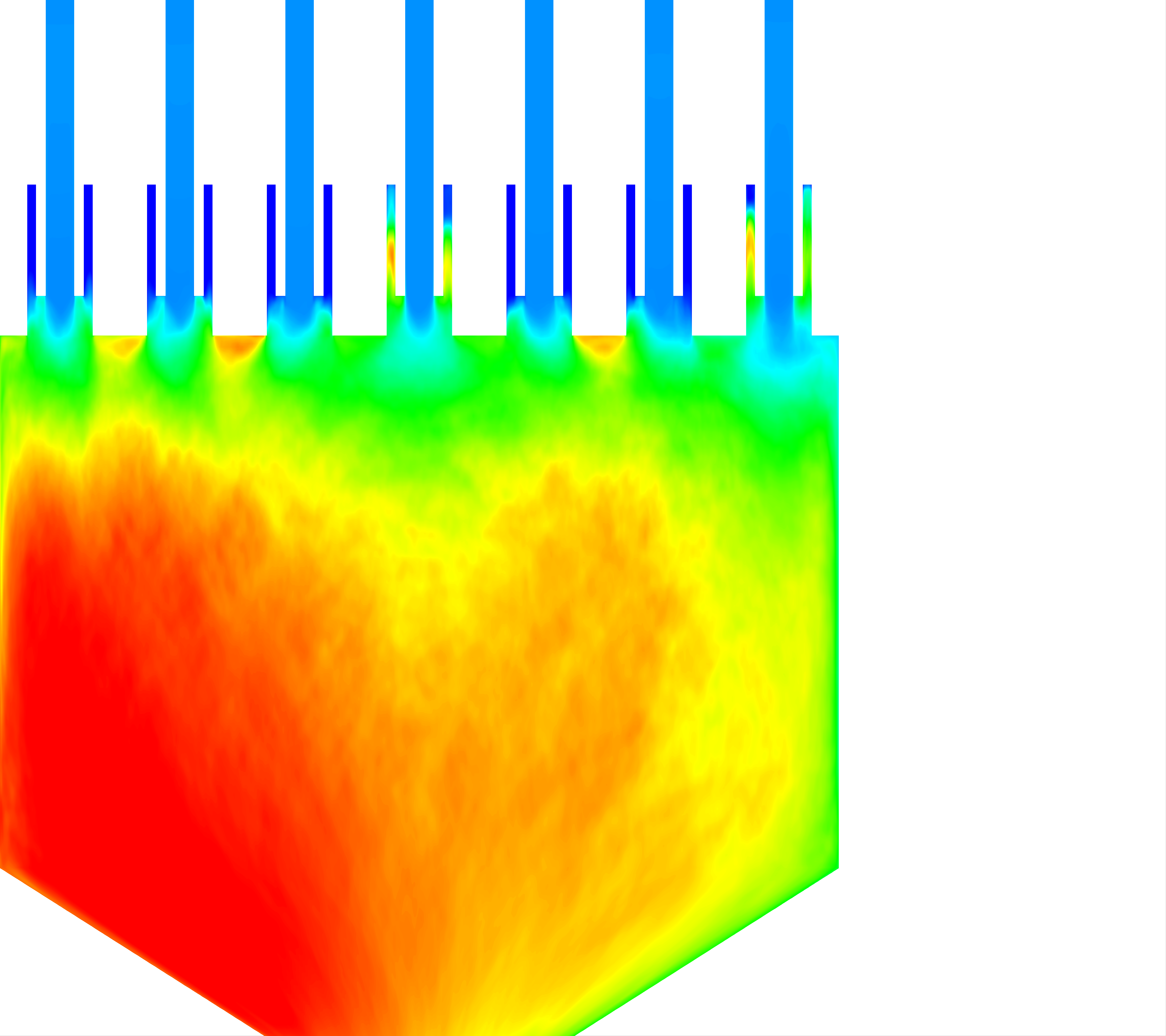}\\
        \small\bfseries FOM
    \end{minipage}
    \hfill
    \begin{minipage}{0.48\linewidth}
        \centering
        \includegraphics[height=5.8cm, trim=0 0 0 0, clip]{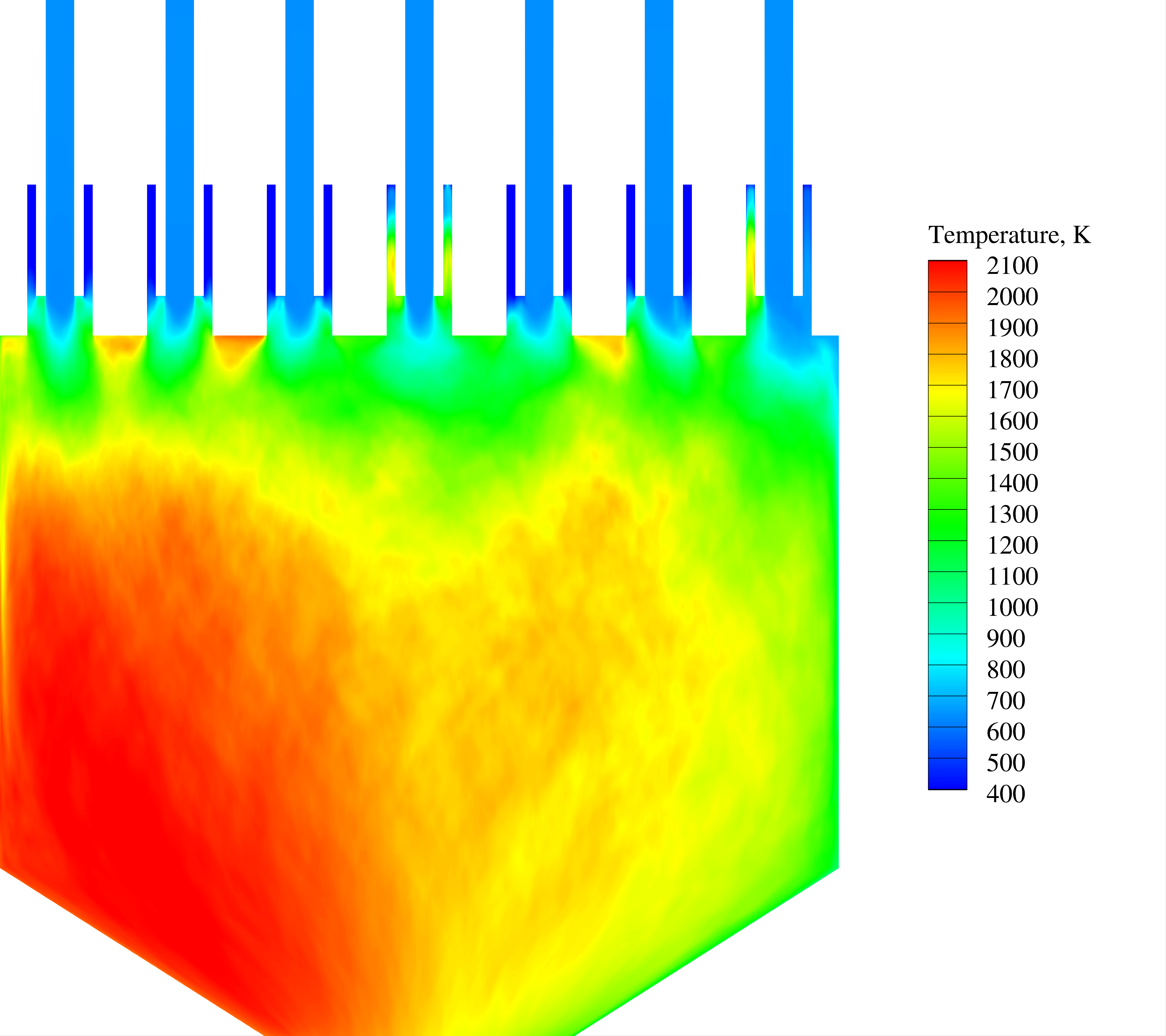}\\
        \hspace*{-1.4cm}\small\bfseries CBROM Framework
    \end{minipage}
    \caption{Parametric test case with center and right-wall fuel injectors shutdown}
\end{subfigure}
\caption{Comparisons of time-averaged temperature fields between FOM and CBROM framework for different operating conditions}
\label{mean_para}
\end{figure}

\begin{figure}[hbt!]
\centering
\begin{subfigure}{0.8\textwidth}
    \centering
    \begin{minipage}{0.48\linewidth}
        \centering
        \includegraphics[height=5.8cm, trim=0 0 2300 0, clip]{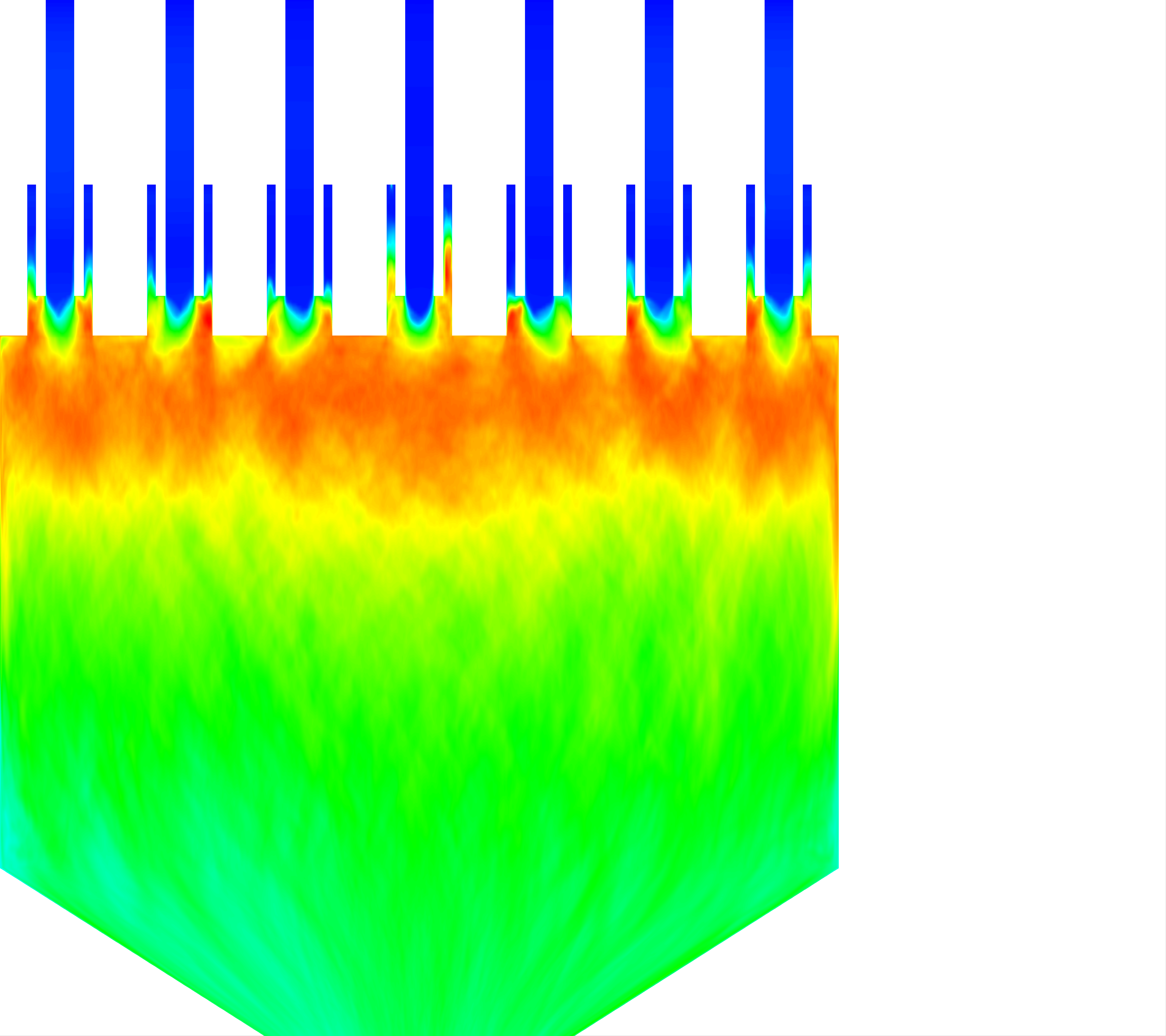}\\
        \small\bfseries FOM
    \end{minipage}
    \hfill
    \begin{minipage}{0.48\linewidth}
        \centering
        \includegraphics[height=5.8cm, trim=0 0 0 0, clip]{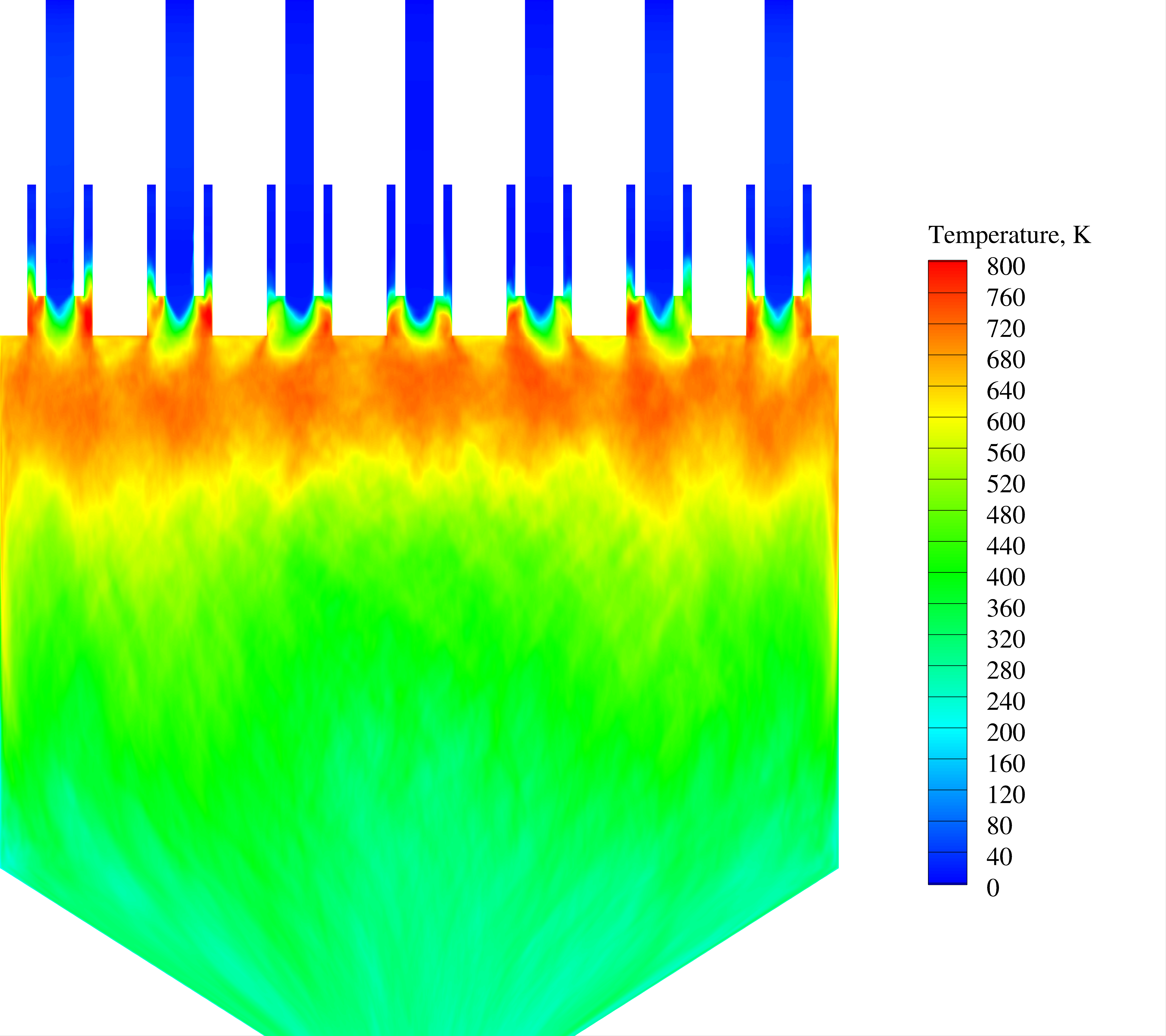}\\
        \hspace*{-1.4cm}\small\bfseries CBROM Framework
    \end{minipage}
    \caption{Baseline test case}
    \label{baseline_rms_para}
\end{subfigure}
\medskip
\begin{subfigure}{0.8\textwidth}
    \centering
    \begin{minipage}{0.48\linewidth}
        \centering
        \includegraphics[height=5.8cm, trim=0 0 2300 0, clip]{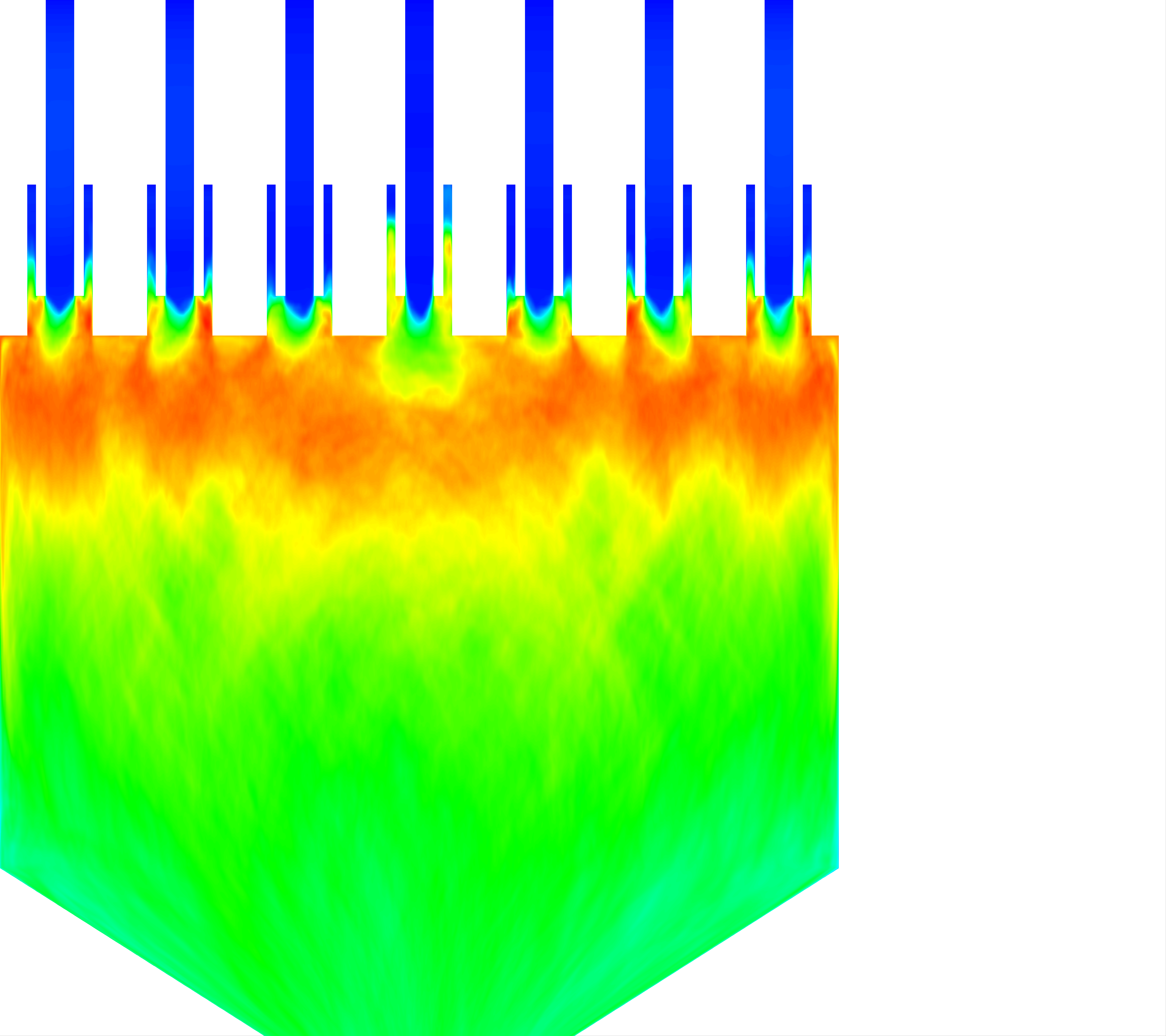}\\
        \small\bfseries FOM
    \end{minipage}
    \hfill
    \begin{minipage}{0.48\linewidth}
        \centering
        \includegraphics[height=5.8cm, trim=0 0 0 0, clip]{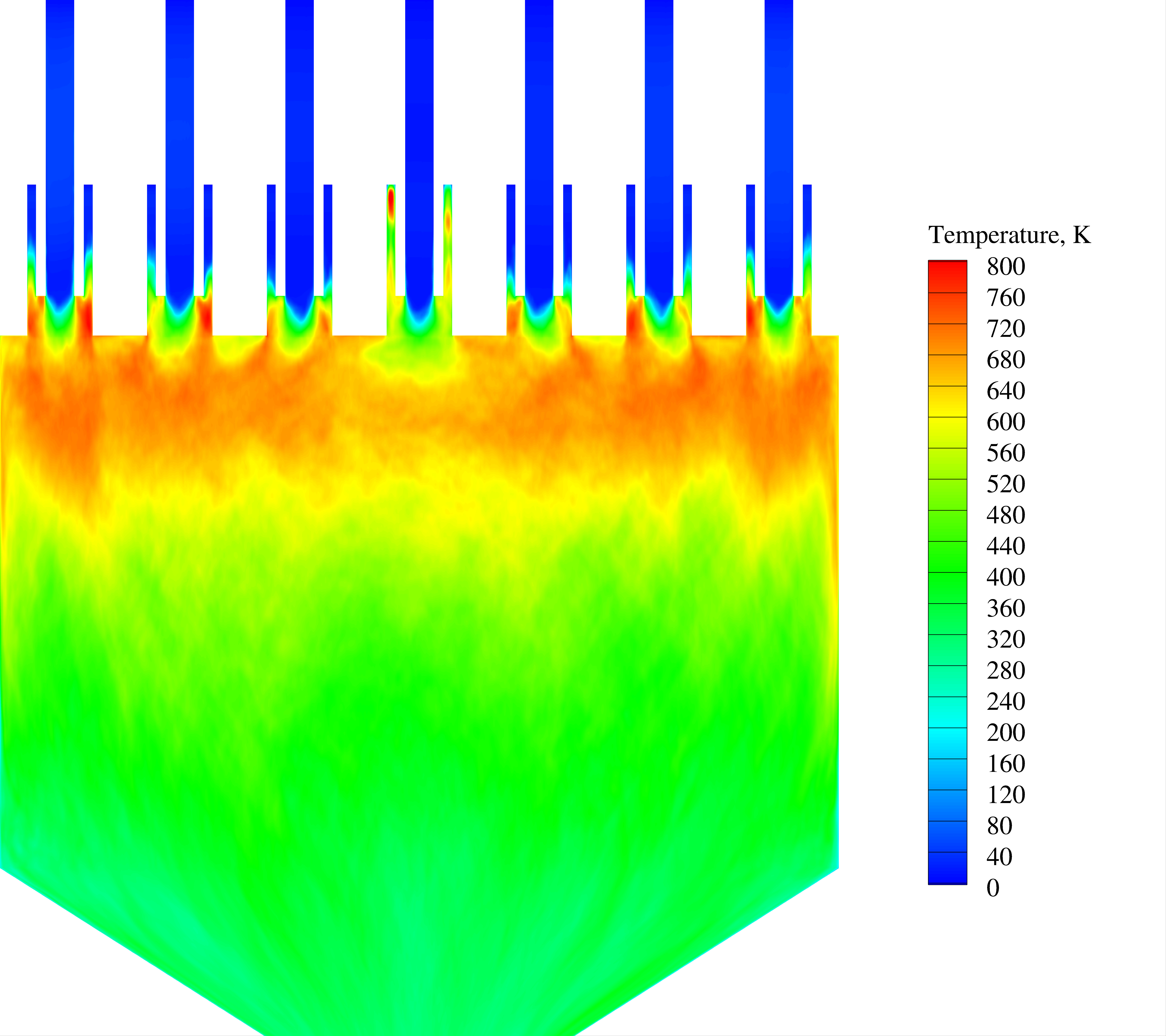}\\
        \hspace*{-1.4cm}\small\bfseries CBROM Framework
    \end{minipage}
    \caption{Parametric test case with center fuel injector shutdown}
\end{subfigure}
\medskip
\begin{subfigure}{0.8\textwidth}
    \centering
    \begin{minipage}{0.48\linewidth}
        \centering
        \includegraphics[height=5.8cm, trim=0 0 2300 0, clip]{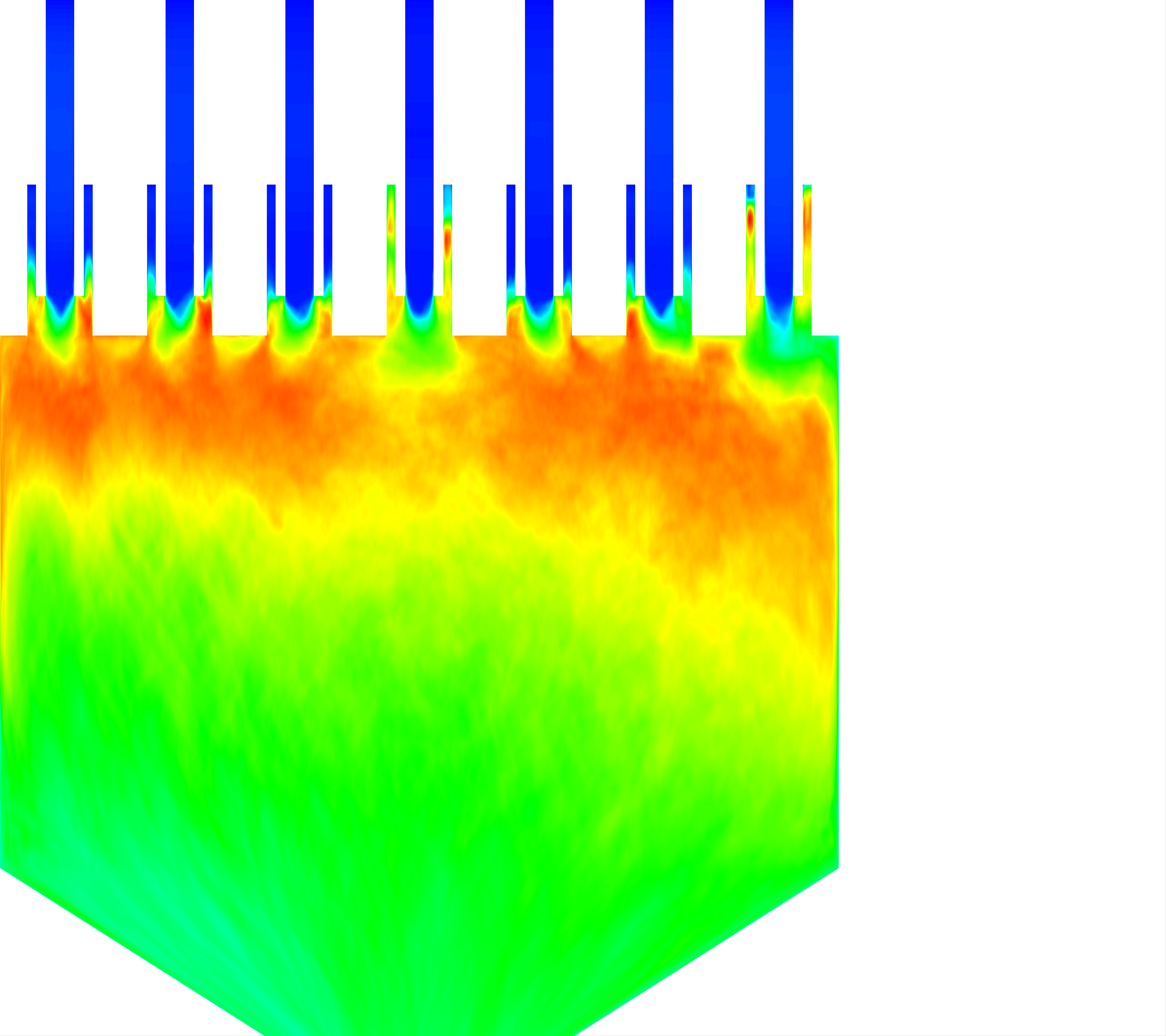}\\
        \small\bfseries FOM
    \end{minipage}
    \hfill
    \begin{minipage}{0.48\linewidth}
        \centering
        \includegraphics[height=5.8cm, trim=0 0 0 0, clip]{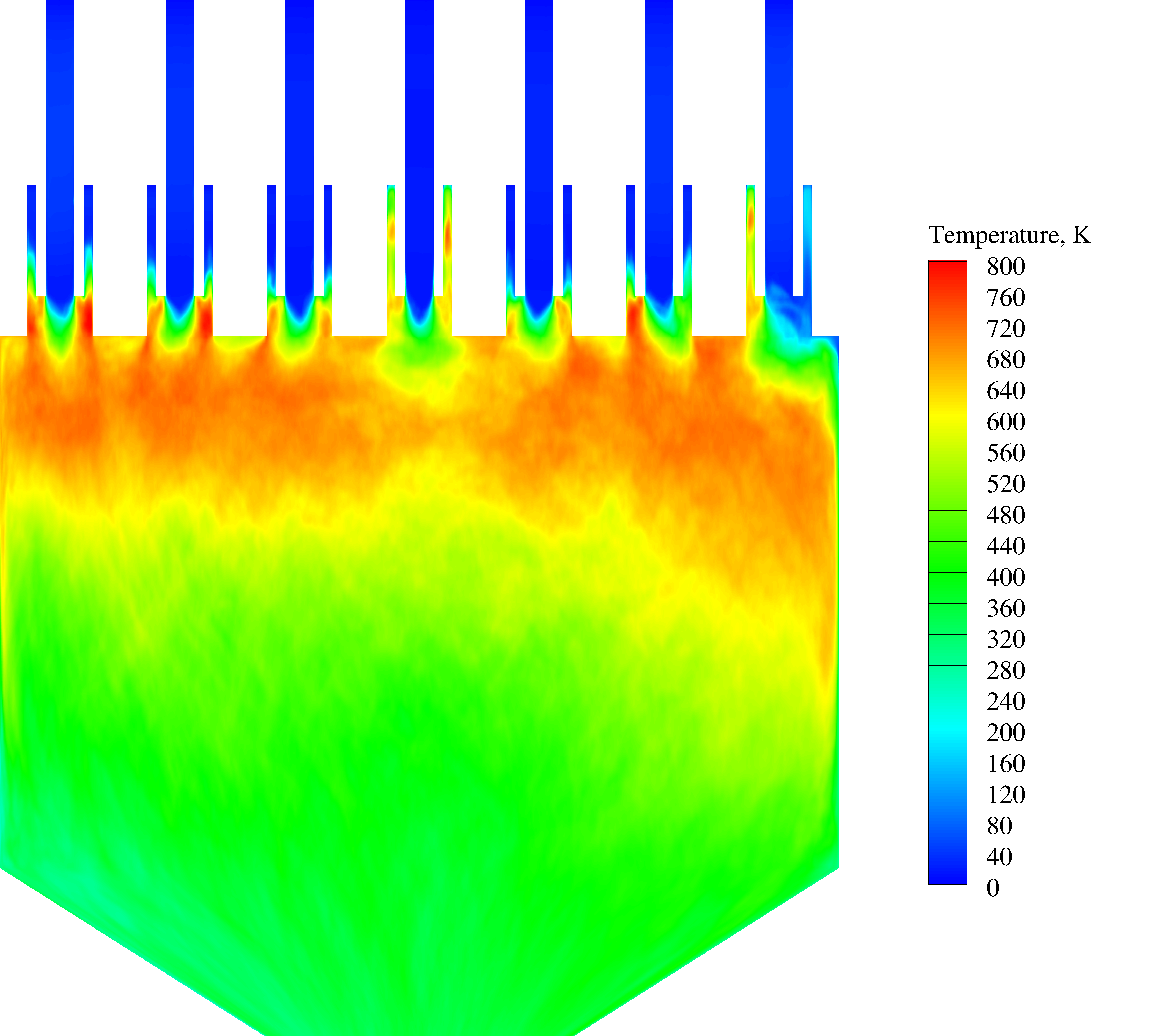}\\
        \hspace*{-1.4cm}\small\bfseries CBROM Framework
    \end{minipage}
    \caption{Parametric test case with center and right-wall fuel injectors shutdown}
\end{subfigure}
\caption{Comparisons of RMS temperature fields between FOM and CBROM framework for different operating conditions}
\label{rms_para}
\end{figure}

Having successfully demonstrated the ability of the CBROM framework in capturing changes in DMD spectra arising from variations of operating conditions, its predictive capabilities are further assessed based on two other statistical QoIs, time-averaged and root-mean-square (RMS) fields of the state variables, which serve as crucial determining factors in many engineering applications. Specifically, these two QoIs are computed using 20 ms of solution data for both the FOM and the CBROM framework to ensure that sufficient information is included for an accurate representation of statistical quantities. The time-averaged temperature fields are compared for the three cases in Fig.~\ref{mean_para}, which readily exhibit good agreement between solutions from the FOM and the CBROM framework. Specifically, the CBROM framework accurately predicts the penetration lengths of cold reactants into the combustor as well as the high-temperature pockets between injectors, alluding to the presence of recirculation zones where fresh propellant mixing is limited due to strong combustion instability. More importantly, the framework successfully captures the changes in time-averaged temperature fields of the full system with fuel injectors shutdown. For example, with no incoming flow through the center fuel injector, hot combustion gases are allowed to recirculate and remain within the injector recess and passage, which leads to intermediate time-averaged temperature within the center fuel injector. In addition,  the lack of active combustion in the center region results in deeper penetration of cold reactants downstream, ultimately reducing the average temperature in the combustor relative to the baseline case. With an additional right-wall fuel injector shutdown, the changes in combustion dynamics become more pronounced. A distinct asymmetry in the temperature distribution can be readily observed in Fig.~\ref{mean_para}, due to weakened combustion near the right wall. This leads to a further decrease in the overall temperature of the combustor, consistent with the trends observed in the instantaneous fields in Fig.~\ref{FOM_Tfield}. Figure~\ref{rms_para} compares the RMS temperature fields for the three cases to characterize the intensity of temperature fluctuations throughout the combustor. Again, the CBROM framework's predictions closely match the FOM results for all three test case. In the baseline case, the framework accurately predicts the high-intensity regions near injectors, corresponding to areas of most active mixing and combustion. Moreover, the CBROM framework successfully captures the changes in RMS temperature fields with variations in operating conditions. When the center fuel injector is shut down, less mixing occurs in the center and leads to lower RMS intensity. In contrast, when both center and right wall fuel injectors are turned off, substantial temperature fluctuations develop on the right side of the combustor, producing asymmetric patterns consistent with those observed in the time-averaged fields (Fig.~\ref{mean_para}).

\subsubsection{Evaluations of parametric predictive capabilities on variations of injector geometries}\label{sect5C3}

\begin{figure}[hbt!]
    \centering
    \includegraphics[width=0.75\linewidth, trim=290 450 290 300, clip]{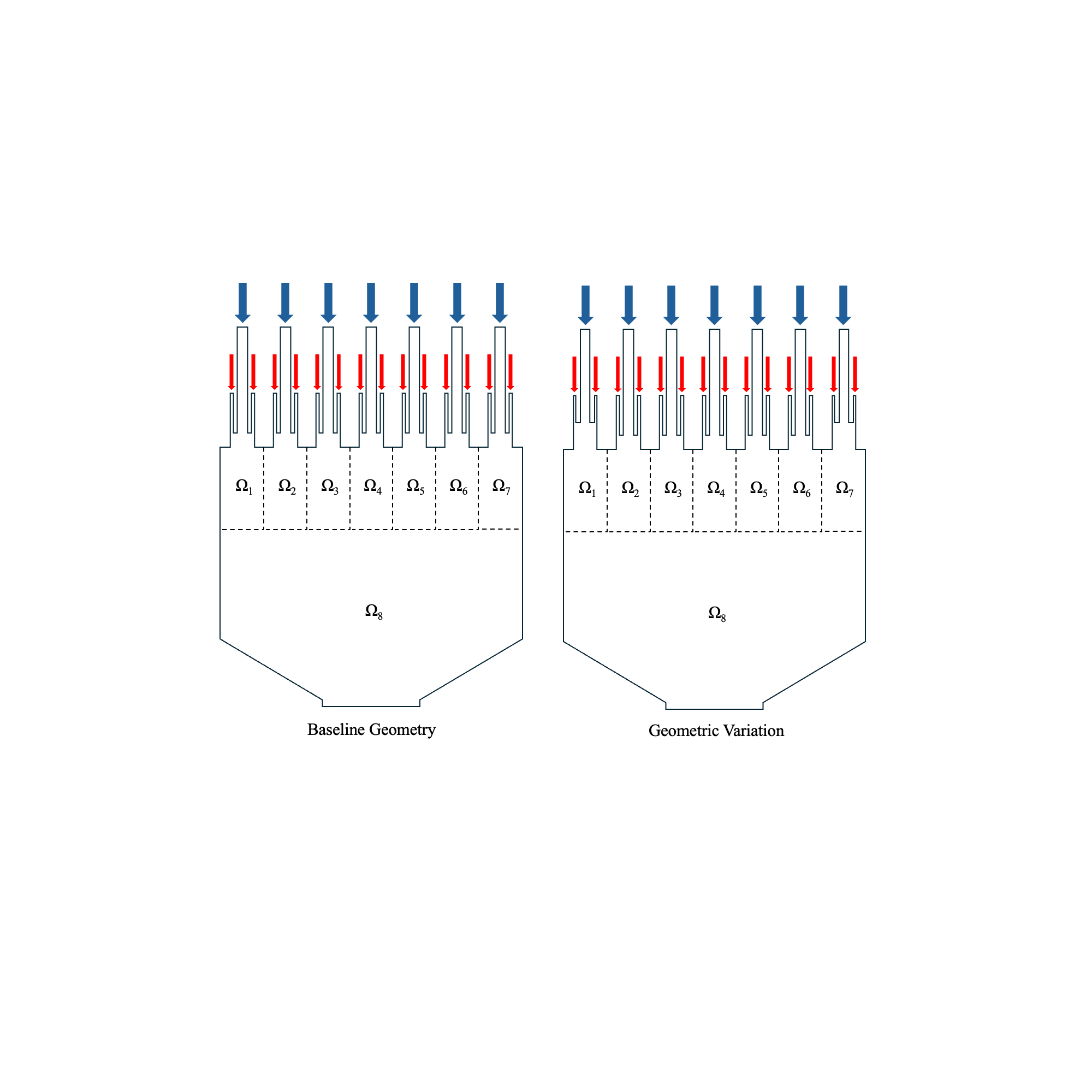}
    \caption{Computational configurations of the baseline (left) and geometrically varied (right) 2D seven-injector rocket combustor}
    \label{fig4}
\end{figure}

The second evaluation focuses on assessing the CBROM framework's predictive capabilities for combustion dynamics under injector geometry variations. Specifically, we establish a geometric test case with the recess lengths of the two wall injectors elongated by $50\%$ as shown in Fig. \ref{fig4}. The geometric test case operates under the same nominal conditions as the baseline case in Sec. \ref{sect5B}. Following the ROM deployment strategy in Sec. \ref{cbrom_geom}, it adopts the same injector mesh topology as the baseline case to allow reuse of the already-constructed component-based ROM through direct mapping of POD basis from the reference component to the altered geometry as illustrated in Fig. \ref{geom_inj}, thereby eliminating the need to establish a new training configuration to retrain a component-based ROM.

\begin{figure}[hbt!]
\centering
\begin{subfigure}{0.8\textwidth}
    \centering
    \begin{minipage}{0.48\linewidth}
        \centering
        \includegraphics[height=5.8cm, trim=0 0 2300 0, clip]{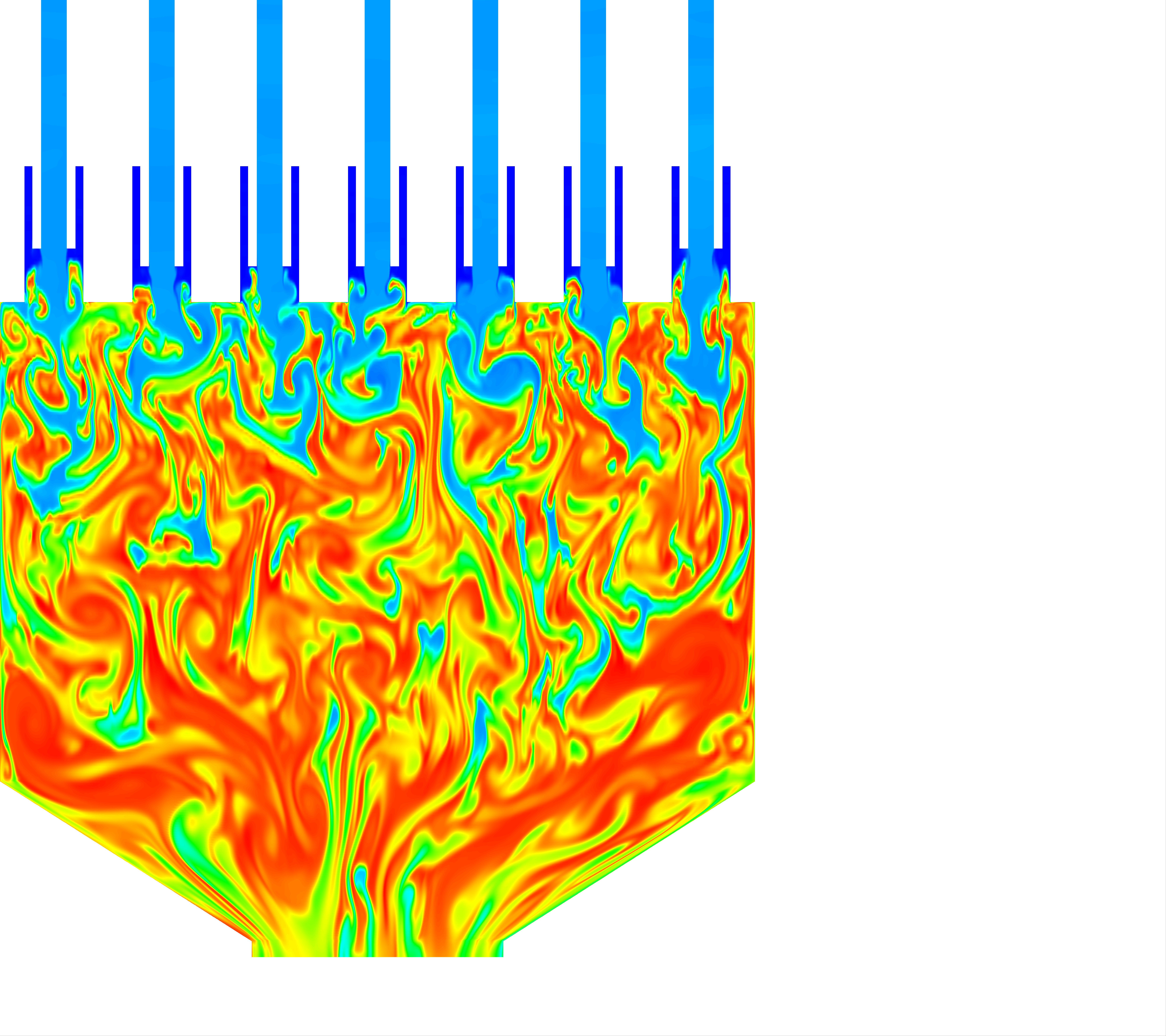}\\
        \small\bfseries FOM
    \end{minipage}
    \hfill
    \begin{minipage}{0.48\linewidth}
        \centering
        \includegraphics[height=5.8cm, trim=0 0 0 0, clip]{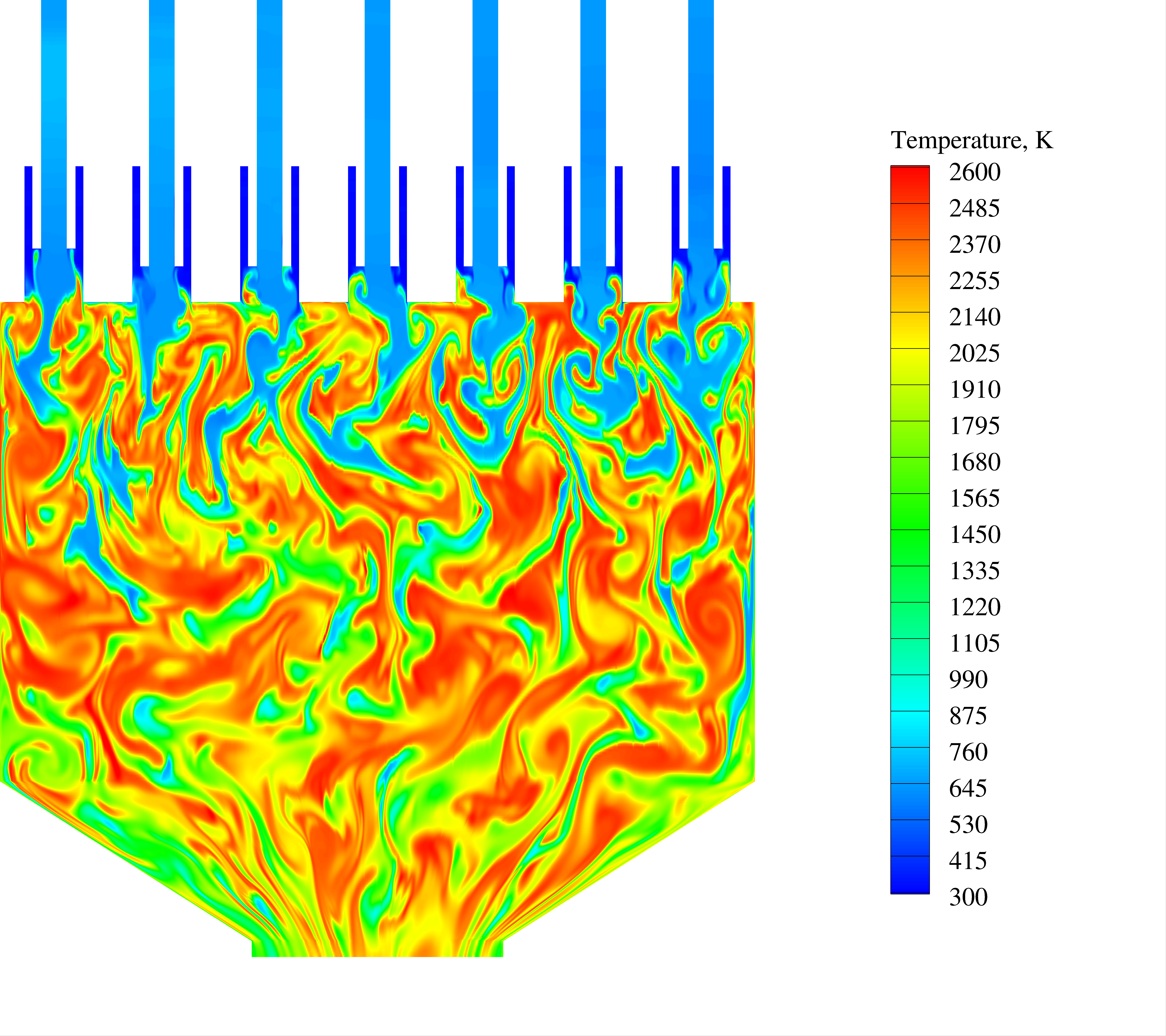}\\
        \hspace*{-1.4cm}\small\bfseries CBROM Framework
    \end{minipage}
\end{subfigure}
\caption{Comparisons of instantaneous temperature fields between FOM and CBROM framework for injector-geometry variations}
\label{bwlr_T}
\end{figure}

\begin{figure}[hbt!]
    \centering
    \subfloat{\includegraphics[width=0.48\linewidth, trim=15 10 15 25, clip]{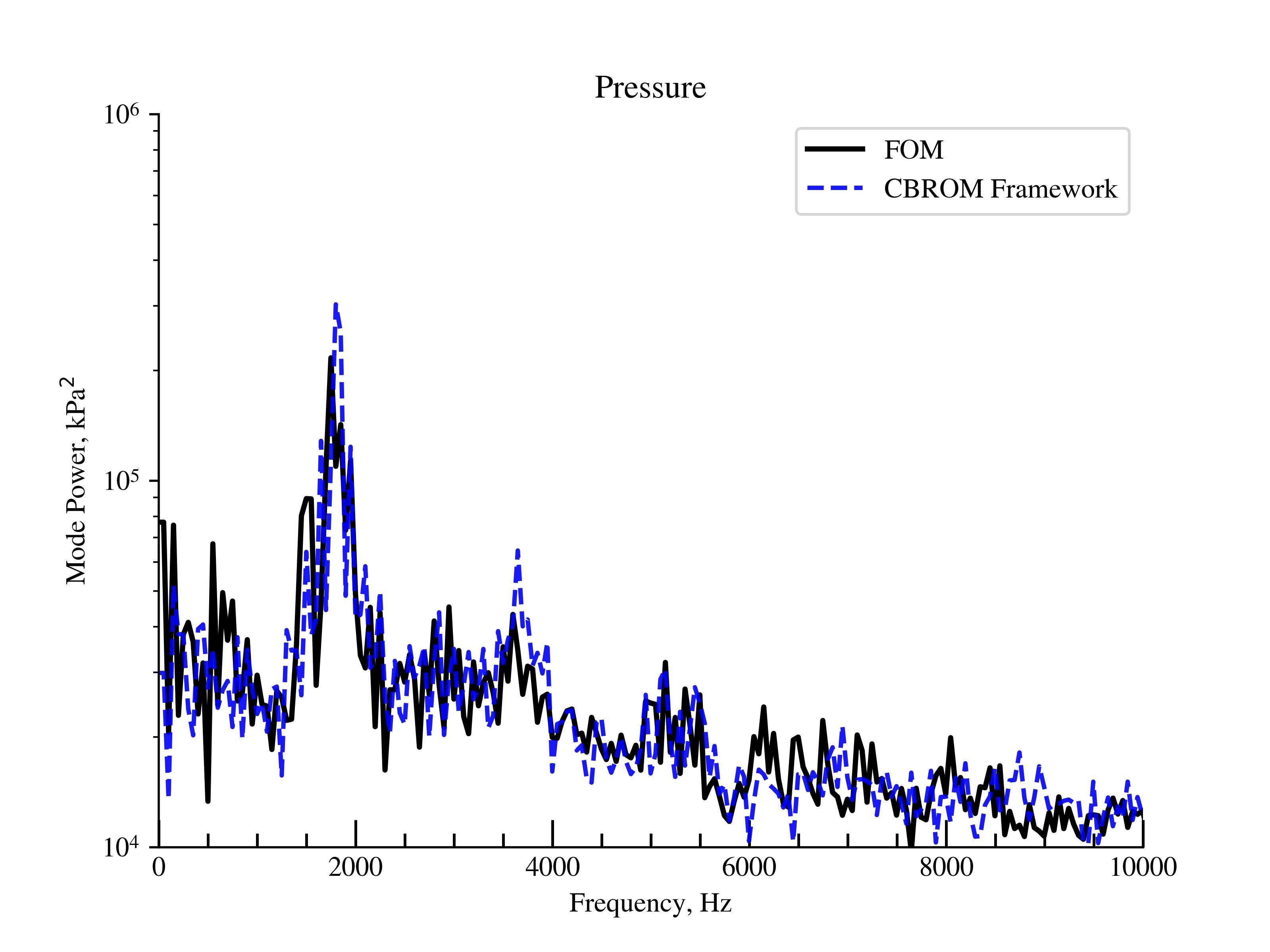}}
    \hfill
    \subfloat{\includegraphics[width=0.48\linewidth, trim=15 10 15 25, clip]{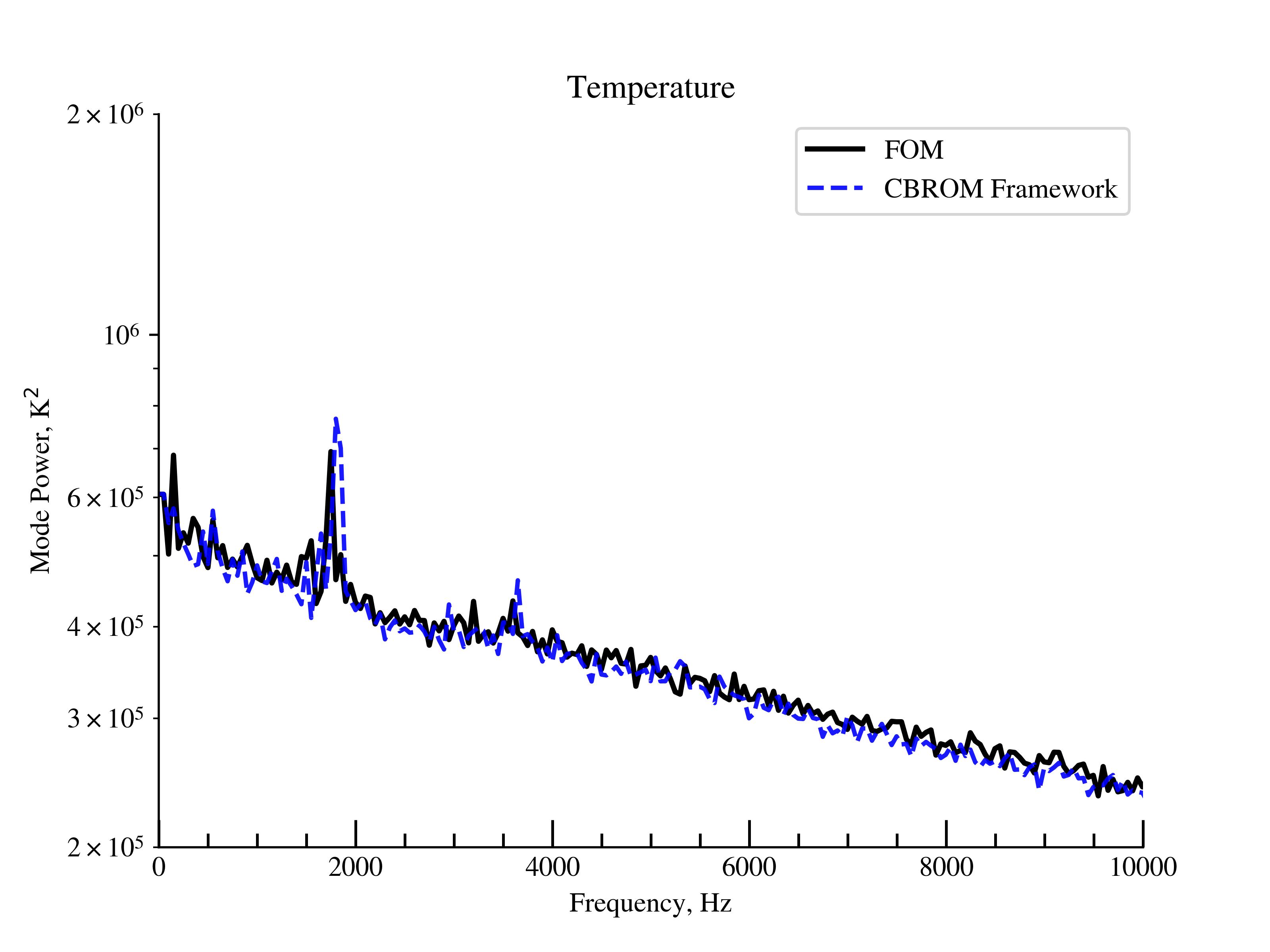}}
    \caption{Comparisons of DMD spectra between FOM and CBROM framework for injector-geometry variations}
    \label{geom_DMD}
\end{figure}

\begin{figure}[hbt!]
\centering
\begin{subfigure}{0.8\textwidth}
    \centering
    \begin{minipage}{0.48\linewidth}
        \centering
        \includegraphics[height=5.8cm, trim=0 0 2300 0, clip]{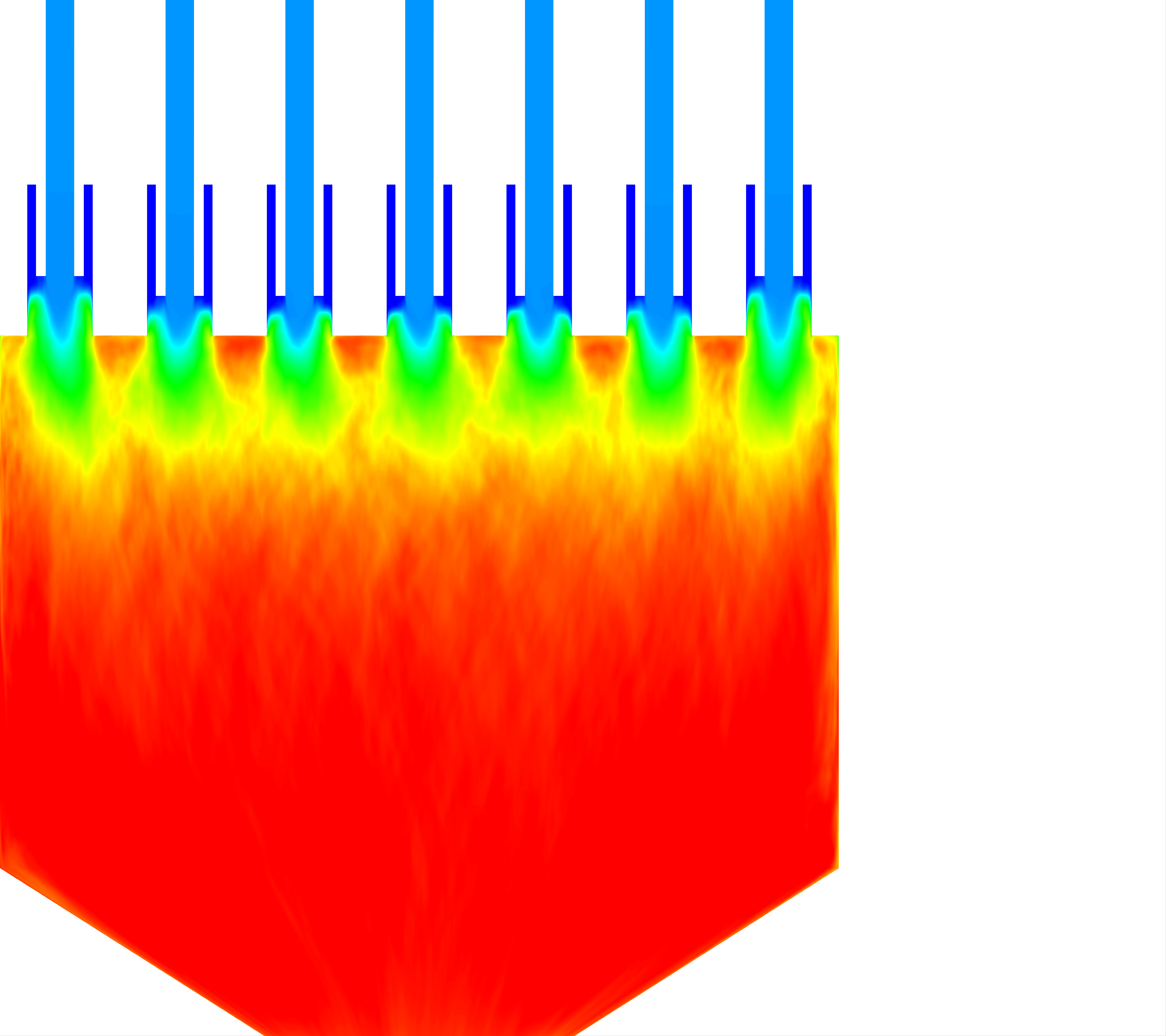}\\
        \small\bfseries FOM
    \end{minipage}
    \hfill
    \begin{minipage}{0.48\linewidth}
        \centering
        \includegraphics[height=5.8cm, trim=0 0 0 0, clip]{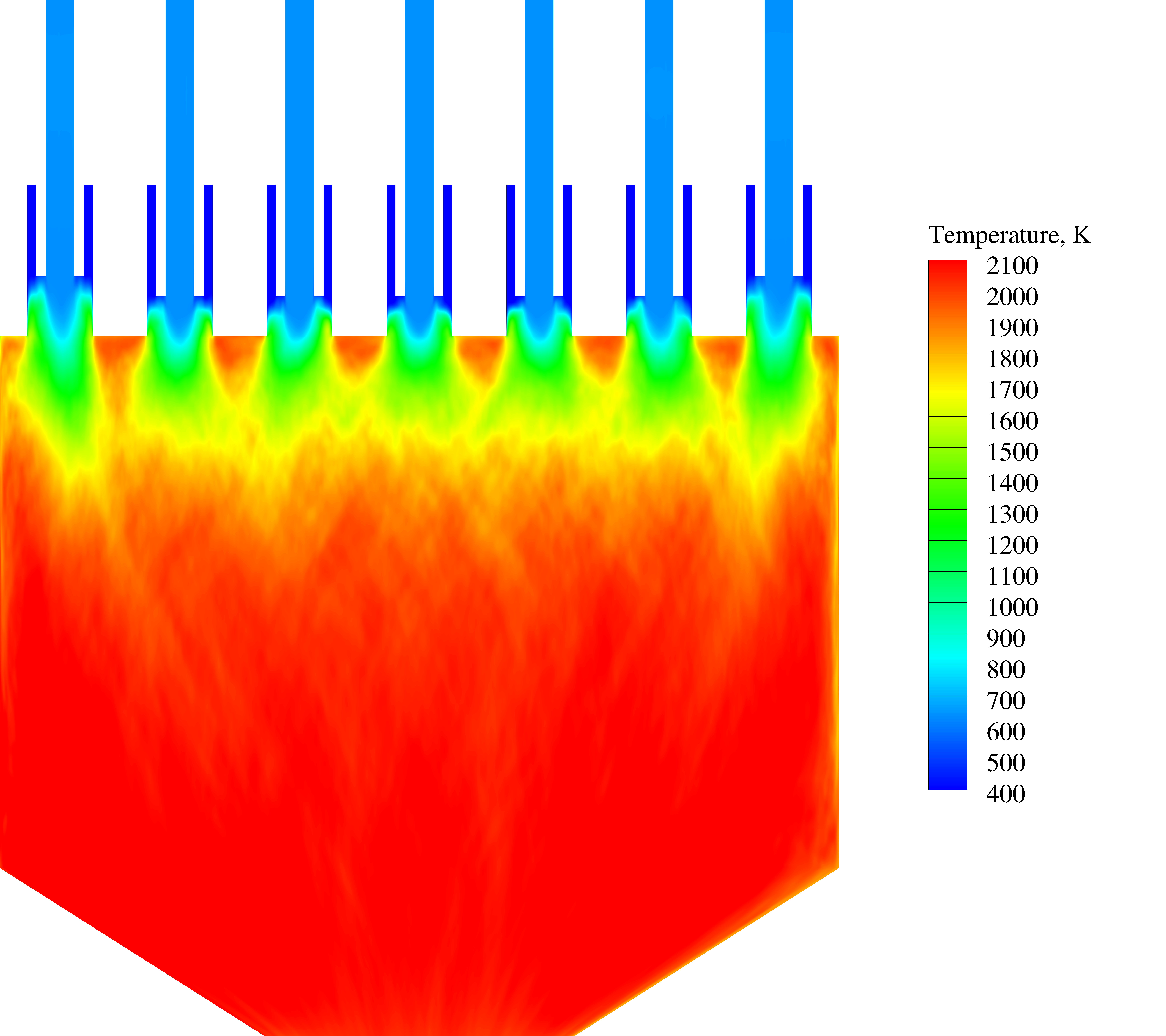}\\
        \hspace*{-1.4cm}\small\bfseries CBROM Framework
    \end{minipage}
\end{subfigure}
\caption{Comparisons of time-averaged temperature fields between FOM and CBROM framework for injector-geometry variations}
\label{mean_geom}
\end{figure}

\begin{figure}[hbt!]
\centering
\begin{subfigure}{0.8\textwidth}
    \centering
    \begin{minipage}{0.48\linewidth}
        \centering
        \includegraphics[height=5.8cm, trim=0 0 2300 0, clip]{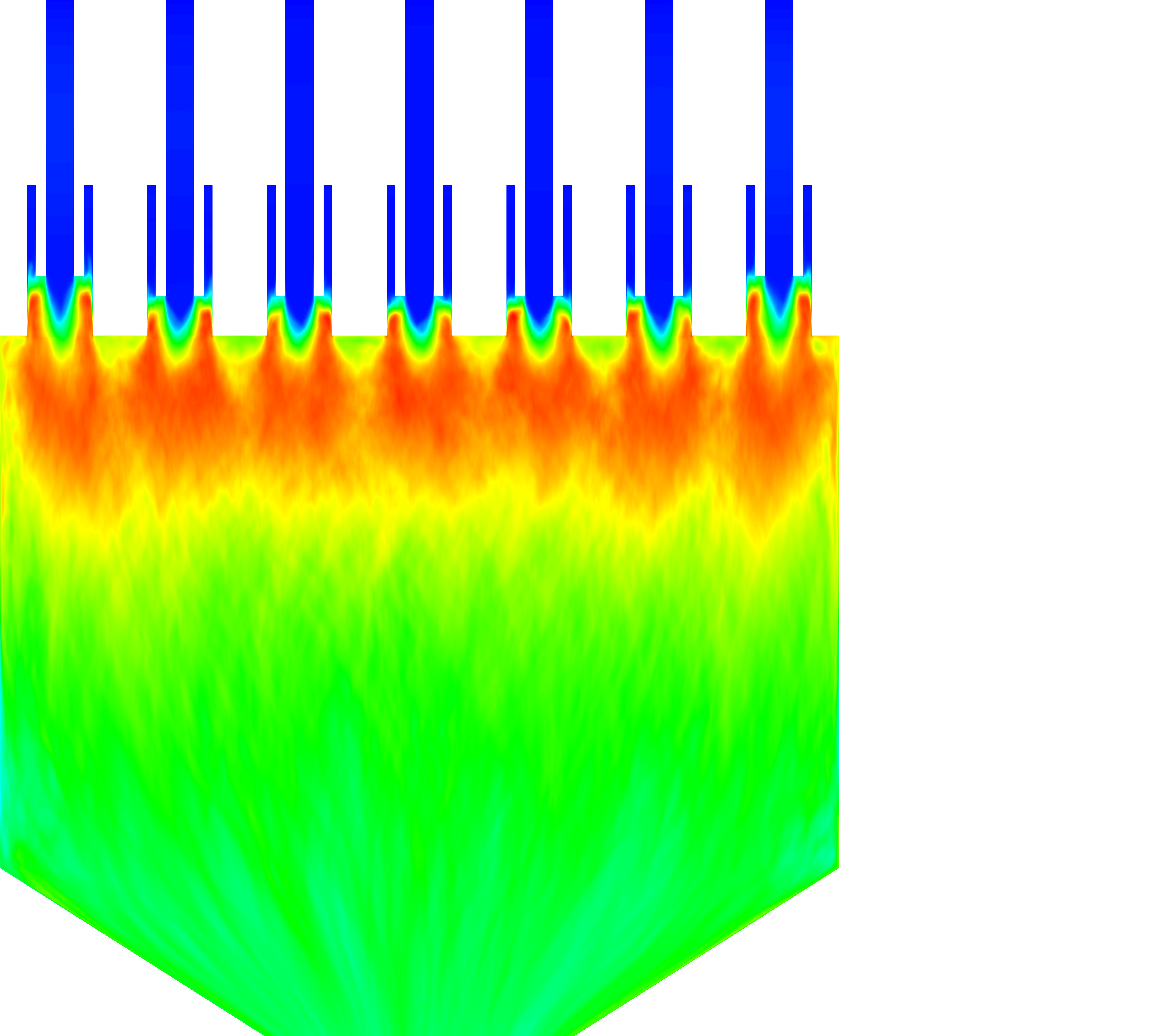}\\
        \small\bfseries FOM
    \end{minipage}
    \hfill
    \begin{minipage}{0.48\linewidth}
        \centering
        \includegraphics[height=5.8cm, trim=0 0 0 0, clip]{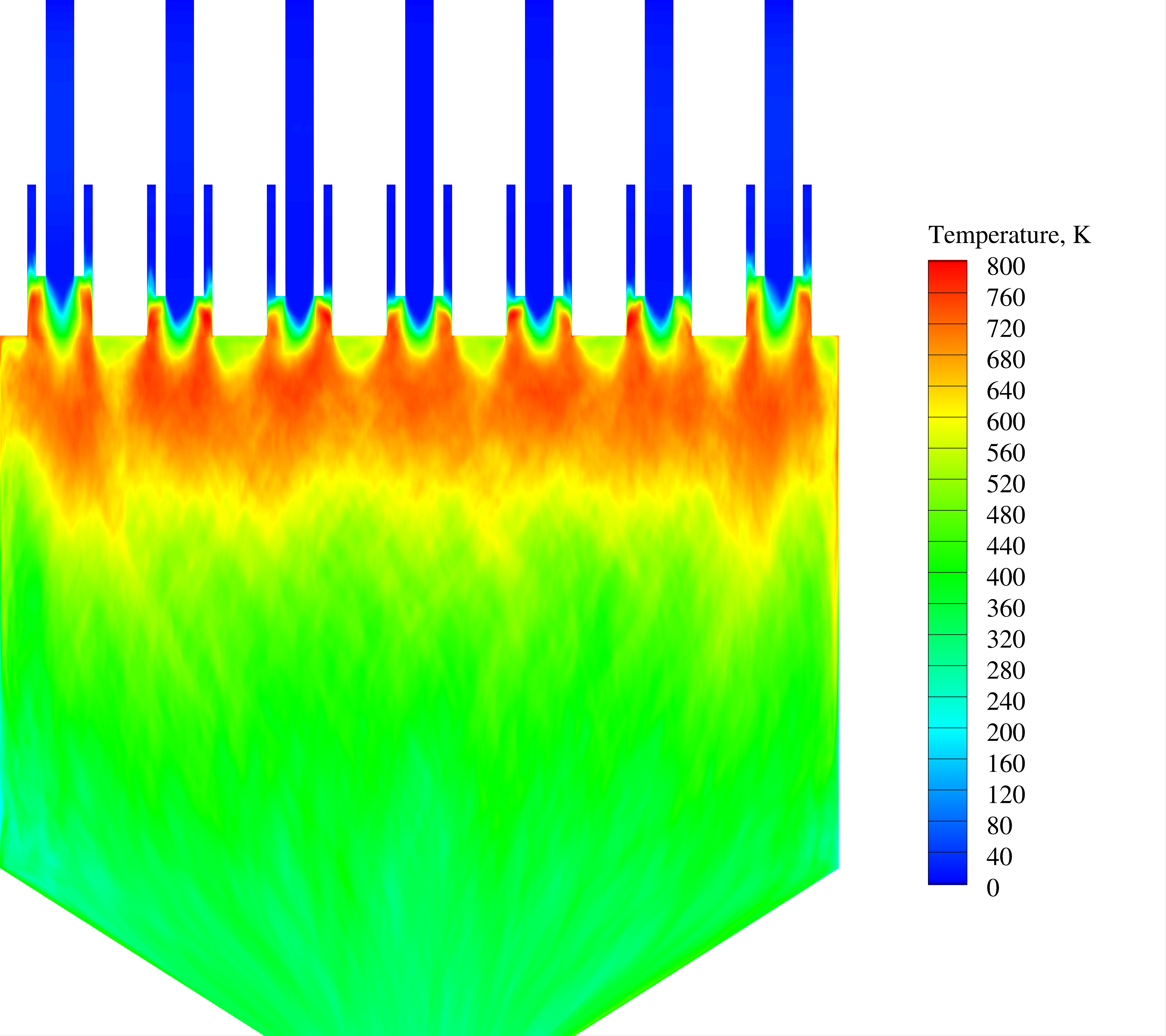}\\
        \hspace*{-1.4cm}\small\bfseries CBROM Framework
    \end{minipage}
\end{subfigure}
\caption{Comparisons of RMS temperature fields between FOM and CBROM framework for injector geometry variations}
\label{rms_geom}
\end{figure}

We first examine the instantaneous temperature fields predicted by the FOM and the CBROM framework as shown in Fig. \ref{bwlr_T}. Overall, the combustion dynamics remain similar to those of the baseline case, with the primary difference being the elongated high-temperature regions within the wall-injector recess. Figure~\ref{bwlr_T} demonstrates strong qualitative agreement between the FOM and CBROM results, particularly in capturing the enhanced reactant mixing inside the extended wall-injector recesses, as evidenced by the stretched high-temperature zones in these regions. Following the same procedure in Sec. \ref{sect5C2}, we further evaluate the CBROM framework based on key statistical QoIs. Figure~\ref{geom_DMD} compares the DMD spectra of pressure and temperature between the FOM and the CBROM framework for the wall injectors with elongated recess lengths, which shows significant magnitude reduction of the dominant acoustic modes relative the baseline case in Fig. \ref{baseline_DMD}, with the second mode in temperature spectrum becoming near indiscernible. We remark that such substantial changes in DMD spectra are accurately captured by the CBROM framework, which demonstrates its viability as a modeling tool for simulating combustion dynamics under geometric variations. Moreover, we examine and compare the time-averaged and RMS temperature fields in Figs.~\ref{mean_geom} and \ref{rms_geom} respectively, which exhibit overall similar spatial patterns as the baseline case in Figs. \ref{baseline_mean_para} and \ref{baseline_rms_para}. The major difference from the baseline case is that the mixing regions downstream of the two wall injectors with elongated recesses are inclined toward the center of the combustor and exhibit further expansion downstream into the combustor, which is not observed in the baseline case. Such changes in local temperature fields indicate the entrainment of reactants from the wall injectors toward the interior injectors, and more importantly, are all accurately captured by the CBROM framework. 

The last evaluation of the CBROM framework focuses on the computational acceleration it can achieve for all the test cases, which is quantified as the ratio of the FOM runtime over the CBROM framework runtime, the results of which are summarized in Table \ref{table1}. It can readily seen that the CBROM framework consistently delivers a computational acceleration of approximately 7.7x for all four test cases, which is a significant reduction in computational cost. 

\begin{table}[hbt!]
\caption{\label{table1} Summary of Computational Accelerations}
\centering
\begin{tabular}{lcccccc}
\hline\\[-2ex]
& Baseline 
& \shortstack{Center\\ Fuel Injector Shutdown} 
& \shortstack{Center \& Right-Wall\\ Fuel Injector Shutdown} 
& \shortstack{Elongated Wall Injector\\Recess Length} \\
\hline
Computational Acceleration & 7.6x & 7.7x& 7.7x& 7.7x&\\
\hline
\end{tabular}
\end{table}

\section{Conclusion}\label{sect6}
A component-based reduced-order modeling (CBROM) framework is established to simulate combustion dynamics in large-scale rocket combustors. The framework decomposes the full-scale rocket combustor into components with identical geometric features, which allows for the training and construction of component-based reduced-order models (ROMs) for each component. A component-based training strategy has been adopted to generate the training data for the component-based ROMs, which requires high-fidelity simulations only for small component domains, thus significantly reducing the computational cost compared to training with a full-domain FOM. In addition, the component-based ROM construction is achieved through an adaptive model-order reduction (MOR) technique, which leverages the state-of-the-art hyper-reduced model-form preserving least-squares with variable transformation (MP-LSVT) ROM formulation and seeks to update the basis and sampling points to tailor them toward the crucial physics. The resulting component-based ROMs are coupled together via a direct-flux-matching interfacing method to enable full-scale combustor simulations. 

The performance of the CBROM framework is assessed in detail using a 2D seven-injector rocket combustor exhibiting distinct self-excited combustion dynamics with parametric variations in operating conditions and geometries. Specifically, four test cases are considered for the framework evaluations, which include a baseline case, two parametric cases with the mass flow rates of selected fuel injectors set to zero, and one geometric case with variations of injector geometries. The predictions from the CBROM framework are compared directly with the FOM solutions based on three major metrics: analysis via the dynamic mode decomposition (DMD), time-averaged fields, and RMS fields. Overall, the CBROM framework provides accurate predictions of the combustion dynamics in all four test cases with all the major coherent structures captured in the unsteady temperature fields. In addition, the DMD analysis shows that the CBROM framework accurately captures the dominant modes in pressure and temperature present in the FOM results. More importantly, the CBROM framework's predictions show excellent agreement with the FOM in the time-averaged and RMS fields of temperature for all four test cases, especially in capturing the substantial variations in the temperature fields in the two parametric cases with the mass flow rates of the selected fuel injectors are set to zero. Moreover, the CBROM framework achieves a factor of 7.7 acceleration in computational time compared to the FOM. The success of the CBROM framework in the current work serves as a stepping stone towards modeling different types of full-scale propulsion devices, such as rocket engines \cite{fedorov_chamber_2006} and rotating detonation engines \cite{prakash_three-dimensional_2024}. Future work will mainly focus on two aspects. First, we will investigate the use of a recently developed feature-guided sampling strategy  to further enhance the efficiency of the framework. Second, we will focus on demonstrating the scalability of the CBROM framework using a large-scale 3D nine-injector rocket combustor configuration \cite{harvazinski_modeling_2019} which has been studied both numerically and experimentally.

\section*{Acknowledgments}
The authors would like to acknowledge the Madison and Lila Self Graduate Fellowship and the support from the Air Force Center of Excellence under Grant FA9550-17-1-0195, titled “Multi-Fidelity Modeling of Rocket Combustor Dynamics” (Technical Monitors: Fariba Fahroo, Justin Koo, and Ramakanth Munipalli).

\bibliography{citations}

\end{document}